\documentclass[prd,nofootinbib,twocolumn,superscriptaddress,preprintnumbers,longbibliography]{revtex4-1} %showpacs

\usepackage{lipsum}
\usepackage{amssymb,amsmath,latexsym,graphics, graphicx,epsfig,multirow,comment,appendix,feyn,slashed,xcolor,afterpage, makecell} 
\usepackage{booktabs}
\usepackage{tabularx}
\usepackage[colorlinks=true
,urlcolor=blue
,anchorcolor=blue
,citecolor=blue
,filecolor=blue
,linkcolor=blue
,menucolor=blue
,linktocpage=true
,pdfproducer=medialab
,pdfa=true
]{hyperref}
\usepackage[utf8]{inputenc}
\usepackage{lipsum}

\newcommand {\be} {\begin {equation}}
\newcommand {\ee} {\end {equation}} 

\newcommand {\bes} {\begin {equation*}}
\newcommand {\ees} {\end {equation*}}

\newcommand{\es}[2] {\begin{equation} \label{#1} \begin{split} #2 \end{split} \end{equation}}

\usepackage{csvsimple}

\newcommand{\beq}{\begin{equation}}
\newcommand{\eeq}{\end{equation}}

\begin{document}

\title{Mapping Extragalactic Dark Matter Annihilation with Galaxy Surveys: \\
A Systematic Study of Stacked Group Searches }

\preprint{MIT/CTP-4930, PUPT-2533, MCTP-17-14}

\author{Mariangela Lisanti}
\affiliation{Department of Physics, Princeton University, Princeton, NJ 08544}

\author{Siddharth Mishra-Sharma}
\affiliation{Department of Physics, Princeton University, Princeton, NJ 08544}

\author{Nicholas L. Rodd}
\affiliation{Center for Theoretical Physics, Massachusetts Institute of Technology, Cambridge, MA 02139}

\author{Benjamin R. Safdi}
\affiliation{Center for Theoretical Physics, Massachusetts Institute of Technology, Cambridge, MA 02139}
\affiliation{Michigan Center for Theoretical Physics, Department of Physics, University of Michigan, Ann Arbor, MI 48109}

\author{Risa H.~Wechsler}
\affiliation{Kavli Institute for Particle Astrophysics and Cosmology \& Physics Department, Stanford University, Stanford, CA 94305}
\affiliation{SLAC National Accelerator Laboratory, Menlo Park, CA 94025}

\date{\today}

\begin{abstract}
Dark matter in the halos surrounding galaxy groups and clusters can annihilate to high-energy photons.  Recent advancements in the construction of galaxy group catalogs provide many thousands of potential extragalactic targets for dark matter.  In this paper, we outline a procedure to infer the dark matter signal associated with a given galaxy group.  Applying this procedure to a catalog of sources, one can create a full-sky map of the brightest extragalactic dark matter targets in the nearby Universe ($z\lesssim 0.03$), supplementing sources of dark matter annihilation from within the Local Group.  As with searches for dark matter in dwarf galaxies, these extragalactic targets can be stacked together to enhance the signals associated with dark matter.  We validate this procedure on mock {\it Fermi} gamma-ray data sets using a galaxy catalog constructed from the \texttt{DarkSky} $N$-body cosmological simulation
and demonstrate that the limits are robust, at $\mathcal{O}(1)$ levels, 
to systematic uncertainties on halo mass and concentration.  We also quantify other sources of systematic uncertainty arising from the analysis and modeling assumptions.  Our results suggest that a stacking analysis using galaxy group catalogs provides a powerful opportunity to discover extragalactic dark matter and complements existing studies of Milky Way dwarf galaxies.
 \end{abstract}
\maketitle

\section{Introduction}

Dark matter (DM) annihilation into visible final states remains one of the most promising avenues for discovering non-gravitational interactions in the dark sector.  While an individual annihilation event is rare, the probability of observing it can be maximized by searching for excess photons in regions of high dark matter density.  The center of the Milky Way is potentially one of the brightest regions of DM annihilation as seen from Earth, but the astrophysical uncertainties associated with the baryonic physics at the heart of our Galaxy motivate exploring other targets.  Gamma-ray studies of DM-dominated dwarf galaxies in the Local Group currently provide some of the most robust constraints on the annihilation cross section~\cite{Fermi-LAT:2016uux, Ackermann:2015zua}.  However, many more potential targets are available beyond the Local Group.  This paper proposes a new analysis strategy to search for DM emission from hundreds more DM halos identified in galaxy group catalogs.  

A variety of methods have been used to study gamma-ray signatures of extragalactic DM annihilation, including modeling potential contributions to the Isotropic Gamma-Ray Background~\cite{Bengtsson:1990xf,Bergstrom:2001jj,Ullio:2002pj,Bottino:2004qi,Bertone:2004pz,Bringmann:2012ez,Ajello:2015mfa, DiMauro:2015tfa, Ackermann:2015tah, Feng:2016fkl}, measuring the \emph{Fermi} auto-correlation power spectrum~\cite{Ackermann:2012uf,Fornasa:2016ohl,Ando:2006cr,Ando:2013ff}, and cross-correlating the \emph{Fermi} data with galaxy counts~\cite{Branchini:2016glc, Xia:2011ax,Ando:2014aoa,Ando:2013xwa,Xia:2015wka,Regis:2015zka,Cuoco:2015rfa,Ando:2016ang}, cosmic shear~\cite{Camera:2014rja,Troster:2016sgf,Choi:2015mnp,Camera:2012cj,Shirasaki:2015nqp,Shirasaki:2014noa,Shirasaki:2016kol} and lensing of the Cosmic Microwave Background~\cite{Fornengo:2014cya, Feng:2016fkl}.  These methods typically rely on using a probabilistic distribution of the DM annihilation signal on the sky.  Our approach is more deterministic in nature.  In particular, we treat a collection of known galaxies as seeds for DM halos.  The properties of each galaxy---such as its luminosity and redshift---enable one to deduce the characteristics of its associated halo and the expected DM-induced gamma-ray flux from that particular direction in the sky.  In this way, we can build a map of the expected DM annihilation flux that traces  the observed distribution of galaxy groups. 

In certain ways, our approach resembles that used in previous studies  of DM annihilation from individual galaxy clusters.  For example, most recently the Andromeda galaxy~\cite{Ackermann:2017nya} and Virgo cluster~\cite{Ackermann:2015fdi} have been the subject of dedicated study by the \emph{Fermi} Collaboration.  Other work has inferred the properties of the DM halos associated with galaxy clusters detected in X-rays~\cite{Ackermann:2010rg, Ando:2012vu,Ackermann:2013iaq,Anderson:2015dpc,Rephaeli:2015nca,2016A&A...589A..33A,Liang:2016pvm}.  Most of these studies focused on a small number of galaxy clusters and obtained DM sensitivities weaker than those from dwarf galaxies.  

Recent advancements in the development of galaxy group catalogs allow us to now build a full-sky map of the nearby galaxies that should be the brightest DM gamma-ray emitters.  Catalogs based primarily on the 2MASS Redshift Survey (2MRS)~\cite{ Huchra:2011ii} provide an unprecedented amount of information regarding a group's constituents and halo properties~\cite{Tully:2015opa,2017ApJ...843...16K,Lu:2016vmu}.  This information allows us to build a list of the brightest extragalactic DM targets on the sky and to perform a stacked analysis for gamma-ray emission from their centers.  A gamma-ray line search using this methodology was recently performed by Ref.~\cite{Adams:2016alz}.  Our focus is on continuum DM signatures, which carry considerably more complications in terms of the treatment of astrophysical backgrounds.  

\begin{figure*}[htbp]
   \centering
   \includegraphics[width=0.9\textwidth]{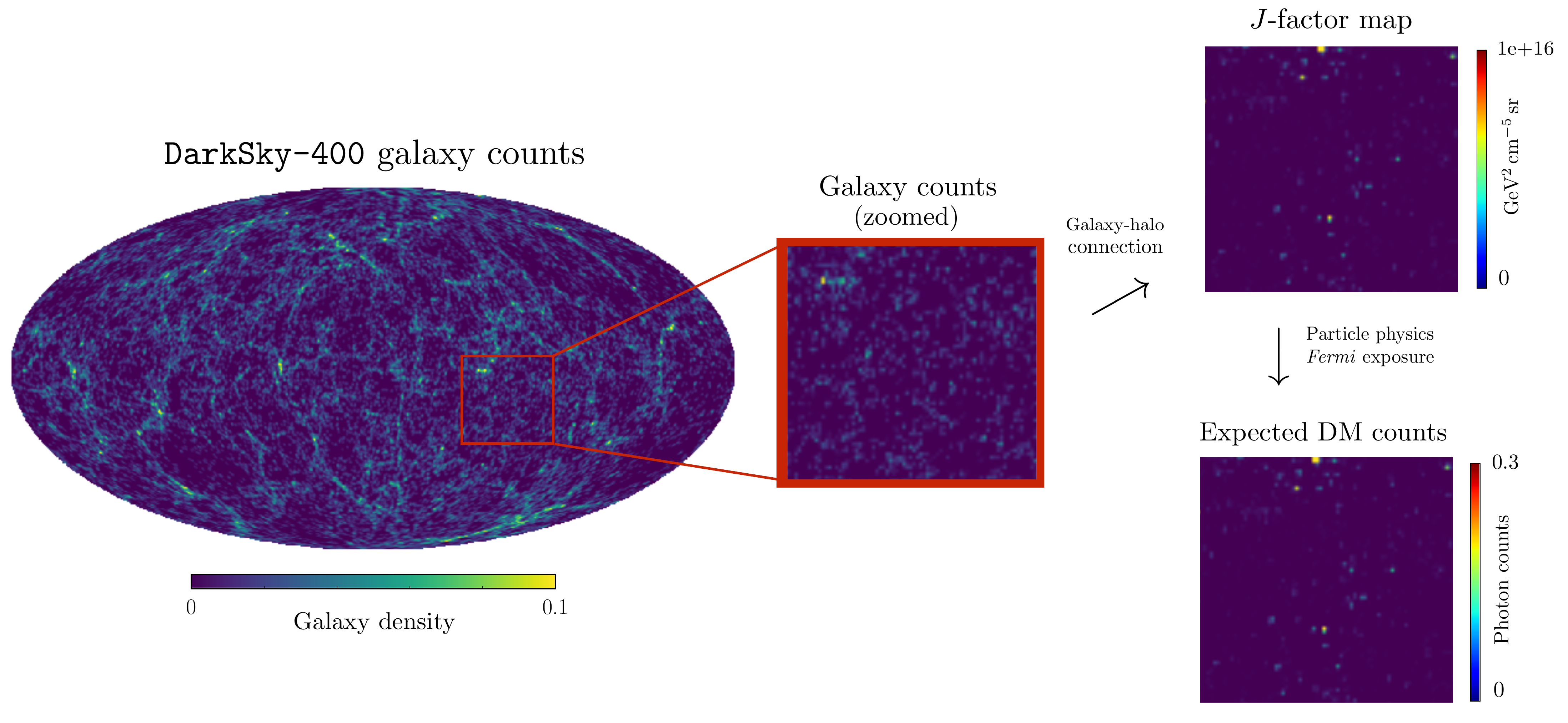}
   \caption{A schematic illustration of the analysis procedure as applied to \texttt{DarkSky}.   We begin with a sky map of galaxy counts (center left).  The \texttt{DarkSky} group catalog categorizes the galaxies into groups, which likely share a common DM halo.  From the \texttt{DarkSky} group catalog, we build a map of the $J$-factors for the host halos, as shown in the top right.  In reality, the properties of the halos surrounding each group of galaxies must be inferred from its total luminosity.  For a given DM model (here, a 100~GeV particle annihilating to $b\bar{b}$ with cross section $\langle \sigma v \rangle \approx 10^{-24}$ cm$^3$s$^{-1}$) and detector energy range (here, $\sim0.9-1.4$ GeV) the DM annihilation flux can be obtained (bottom right).  Going from the map of $J$-factors to that of DM counts also requires knowledge of the \emph{Fermi} exposure. Note that the full sky map has been subjected to 2$^{\circ}$ Gaussian smoothing.}
   \label{fig:DarkSkycounts}
\end{figure*}

In a companion paper~\cite{companion}, we present results implementing a stacked analysis of the group catalogs from Ref.~\cite{Tully:2015opa,2017ApJ...843...16K} on \emph{Fermi} data and show explicitly that this method yields competitive sensitivity to the dwarf searches.  Here, we present the full details of the analysis method and a thorough discussion of the systematic uncertainties involved in deducing the DM-induced flux associated with a given galaxy group.  To fully understand these uncertainties, we apply these methods on mock data where it is possible to compare the inferred DM properties to their true values.  For this purpose, we use the \texttt{DarkSky} cosmological $N$-body simulation~\cite{Skillman:2014qca,Lehmann:2015ioa} and an associated galaxy catalog from Ref.~\cite{Lehmann:2015ioa}.  We emphasize that, while we illustrate the analysis method on gamma-ray data, it can also be applied to other wavelengths and even other messengers, such as neutrinos.  

This paper is organized as follows. In Sec.~\ref{sec:galaxyfilteringpipeline}, we describe how to build DM annihilation flux maps starting from a galaxy group catalog and discuss the associated systematic uncertainties.  Sec.~\ref{sec:stats} presents a detailed description of the statistical methods that we follow to implement the stacking. We show the results of applying the limit-setting and signal recovery procedures on mock data in Sec.~\ref{sec:smallrois} and conclude in Sec.~\ref{sec:conclusions}.  Appendix~\ref{app:JDrelations} provides a  detailed discussion of the $J$-factor expressions used in the main text.  Appendix~\ref{app:energyrange} discusses the validity of approximations made in the profile likelihood procedure.   Appendix~\ref{app:inflimitimpact} shows the results of performing a full-sky template study, as a contrast to the stacking results presented in the main text.

\newpage

\section{Tracing Dark Matter Flux with Galaxy Surveys}
\label{sec:galaxyfilteringpipeline}

In this Section, we describe how to construct catalogs of extragalactic DM targets starting from a list of galaxy groups.  We begin by reviewing the properties of the galaxy group catalogs and then describe how to predict the DM signal from a given galaxy group and quantify the systematic uncertainties of this extrapolation.  

\subsection{Galaxy and Halo Catalogs}

The approach that we use throughout this work relies on galaxy surveys as an input.  Different galaxy catalogs span a range of redshifts and luminosities. Optimal catalogs for DM searches should cover as much of the sky as possible (to increase statistics) and sample low redshifts.  The strength of the DM signal increases at lower redshifts due to accretion of mass at late times, affecting both the halo mass distribution and substructure~\cite{Ando:2014aoa}.  In contrast, the integrated gamma-ray flux of standard astrophysical sources, such as Active Galactic Nuclei and star-forming galaxies, is expected to peak at higher redshifts between $\sim$0.1 and $\sim$2 depending on the specific source class and model for its unresolved contribution~\cite{Ando:2014aoa, Xia:2015wka}.  

The Two Micron All-Sky Survey Extended Sources Catalog (2MASS XSC)~\cite{Bilicki:2013sza,Huchra:2011ii}  satisfies the criteria listed above and has been used extensively in past cross-correlation studies~\cite{Ando:2013xwa,Ando:2014aoa,Ando:2016ang,Cuoco:2015rfa,Regis:2015zka,Xia:2011ax,Xia:2015wka}. The XSC is an all-sky infrared survey that consists of approximately one million galaxies up to a limiting magnitude of $K = 13.5$~mag. Several redshift surveys based on the 2MASS XSCmap the redshifts associated with these galaxies. The 2MRS~\cite{Huchra:2011ii}, for example, samples about 45,000 galaxies in the 2MASS XSC with redshifts to a limiting magnitude of $K=11.75$ mag. This corresponds to a nearly complete galaxy sample up to redshifts of $z=0.03$, which is ideal for DM studies.

Galaxies from large surveys such as 2MASS can be organized into group catalogs.  A group of gravitationally-bound galaxies shares a DM host halo.  The  brightest galaxy in the group is referred to as the central galaxy; the additional galaxies are bound satellites surrounded by their own subhalos.  The total luminosity of the galaxies in the group is a good predictor of the mass of the DM host halo.  A variety of group finders have been developed and applied to the 2MASS data set~\cite{Tully:2015opa,Lu:2016vmu,2017ApJ...843...16K}, using the 2MRS which adds information in the redshift dimension.  The groups in these catalogs range from cluster scales with $\sim$190 members and associated halo masses of $\sim$10$^{15}$~M$_\odot$, down to much smaller systems with only a single member. Galaxy group catalogs are especially relevant for the present study, since halo properties tend to be correlated with properties of galaxy groups rather than those of individual galaxies.

While in our companion paper~\cite{companion} we use information from the 2MASS group catalogs in the analysis of \emph{Fermi} data, we focus on a catalog of simulated galaxies and halos here.  We use the \texttt{DarkSky-400} cosmological $N$-body simulation (version \texttt{ds14\_i})~\cite{Skillman:2014qca,Lehmann:2015ioa} and an associated $r$-band galaxy catalog.  Using the code \texttt{2hot}~\cite{Warren:2013vma}, \texttt{DarkSky-400}  follows the evolution of $4096^3$ particles of mass $7.63 \times 10^7$~M$_\odot$ in a box 400~Mpc$\,h^{-1}$ per side.  Initial perturbations are tracked from $z=93$ to today, assuming $(\Omega_M, n_s, \sigma_8, h) = (0.295, 0.968, 0.834, 0.688)$.   The halo catalog was generated using the \texttt{Rockstar} halo finder~\cite{Behroozi:2011ju, Lehmann:2015ioa}.  Crucially, the simulation covers the relevant redshift space for DM studies.\footnote{The snapshot of the simulation analyzed in this work is taken at $z = 0$, but we will refer to distance using redshift because that is the more appropriate language when applied to real data.}  In particular, an observer at the center of the simulation box has a complete sample of galaxies out to \mbox{$z \sim 0.045$}, with the furthest galaxies extending out to \mbox{$z \sim 0.067$}.  In our work, we only consider groups located within $z\lesssim 0.03$, which is the approximate redshift cutoff of the catalogs in Ref.~\cite{Tully:2015opa,2017ApJ...843...16K,Lu:2016vmu}. We include only well-resolved halos in our analysis by imposing a lower cut-off of $5\times10^{11}$~M$_\odot$ on the mass of included host halos. The associated galaxy catalog is generated using the abundance matching technique following Ref.~\cite{Behroozi:2010rx,Reddick:2012qy} with luminosity function and two-point correlation measurements from the Sloan Digital Sky Survey (SDSS).  Specifically, the $\alpha=0.5$ model from Ref.~\cite{Lehmann:2015ioa} is used, which was shown to provide the best fit to SDSS two-point clustering.  The \texttt{DarkSky} galaxy catalog contains the same information that would be found in, \emph{e.g.} the 2MASS galaxy catalog and associated group catalogs, such as individual galaxy luminosities and sky locations. 

Figure~\ref{fig:DarkSkycounts} shows a sky map of the galaxy counts in \texttt{DarkSky} up to $z=0.03$ for an observer at the center of the simulation box.  It is a \texttt{HEALPix}~\cite{Gorski:2004by} map with resolution \texttt{nside=128}. To first approximation, the galaxies are isotropically distributed throughout the sky.  However, regions of higher and lower galaxy density are clearly visible. Note that this is shown for a particular sky realization and placing the observer in different parts of the \texttt{DarkSky} box would change the regions of contrasting galaxy density.

\subsection{Dark Matter Annihilation Flux Map}
\label{sec:dmflux}

One can predict the DM annihilation flux associated with a halo that surrounds a given galaxy group.  This requires knowing the halo's properties, including its mass and concentration.  In this subsection, we discuss how to determine the flux when the halo's properties are known exactly.  Then, in the following subsection, we consider how to generalize the results to the more realistic scenario where the halo properties have to be inferred.

Each halo in \texttt{DarkSky} is fit by the \texttt{Rockstar} halo finder with 
a Navarro-Frenk-White (NFW) distribution~\cite{Navarro:1995iw} of the form
\begin{equation}
\rho_\text{NFW}(r)=\frac{\rho_{s}}{r/r_{s}\,(1+r/r_{s})^{2}}\, ,
\label{eq:NFW}
\end{equation}
where $r_s$ is the scale radius and $\rho_s$ is the normalization.  The NFW parameters are determined from the parameters that are provided for each DM halo---specifically, its redshift $z$, virial mass $M_\text{vir}$, virial radius $r_\text{vir}$, and virial concentration parameter $c_{\text{vir}}=r_\text{vir}/r_{s}$.

In the simplest scenarios, the annihilation flux factorizes as 
\begin{equation}
\frac{d\Phi}{dE_{\gamma}} = \frac{d\Phi_\text{pp}}{dE_{\gamma}}\times J \, ,
\label{eq:flux}
\end{equation}
where $E_\gamma$ is the photon energy and $\Phi_\text{pp}$  ($J$) encodes the particle physics (astrophysical) dependence.
The particle physics contribution is given by
\begin{equation}
\frac{d\Phi_\text{pp}}{dE_{\gamma}}=\frac{\langle\sigma v\rangle}{8\pi m_{\chi}^{2}}\sum_i \text{Br}_{i}\, \left. \frac{dN_{i}}{dE_{\gamma}'} \right|_{E'_{\gamma} = (1+z) E_{\gamma}},
\end{equation}
where $m_\chi$ is the DM mass, $\langle \sigma v\rangle $ is its annihilation cross section, $\text{Br}_{i}$ is its branching fraction to the $i^\text{th}$ annihilation channel, $ dN_{i}/dE_{\gamma}$ is the photon energy distribution in this channel, which is modeled using  PPPC4DMID~\cite{Cirelli:2010xx}, and $z$ is the redshift.  We consider the case of annihilation into the $b \bar{b}$ channel as a generic example of a continuum spectrum. Of course, the exact limits will vary for different spectra, and one should  consider a range of final states when applying the method to data, or use model independent-approaches (see, {\it e.g.}, Ref.~\cite{Elor:2015tva,Elor:2015bho}).
\begin{figure}[t]
   \centering
     \includegraphics[width=0.5\textwidth]{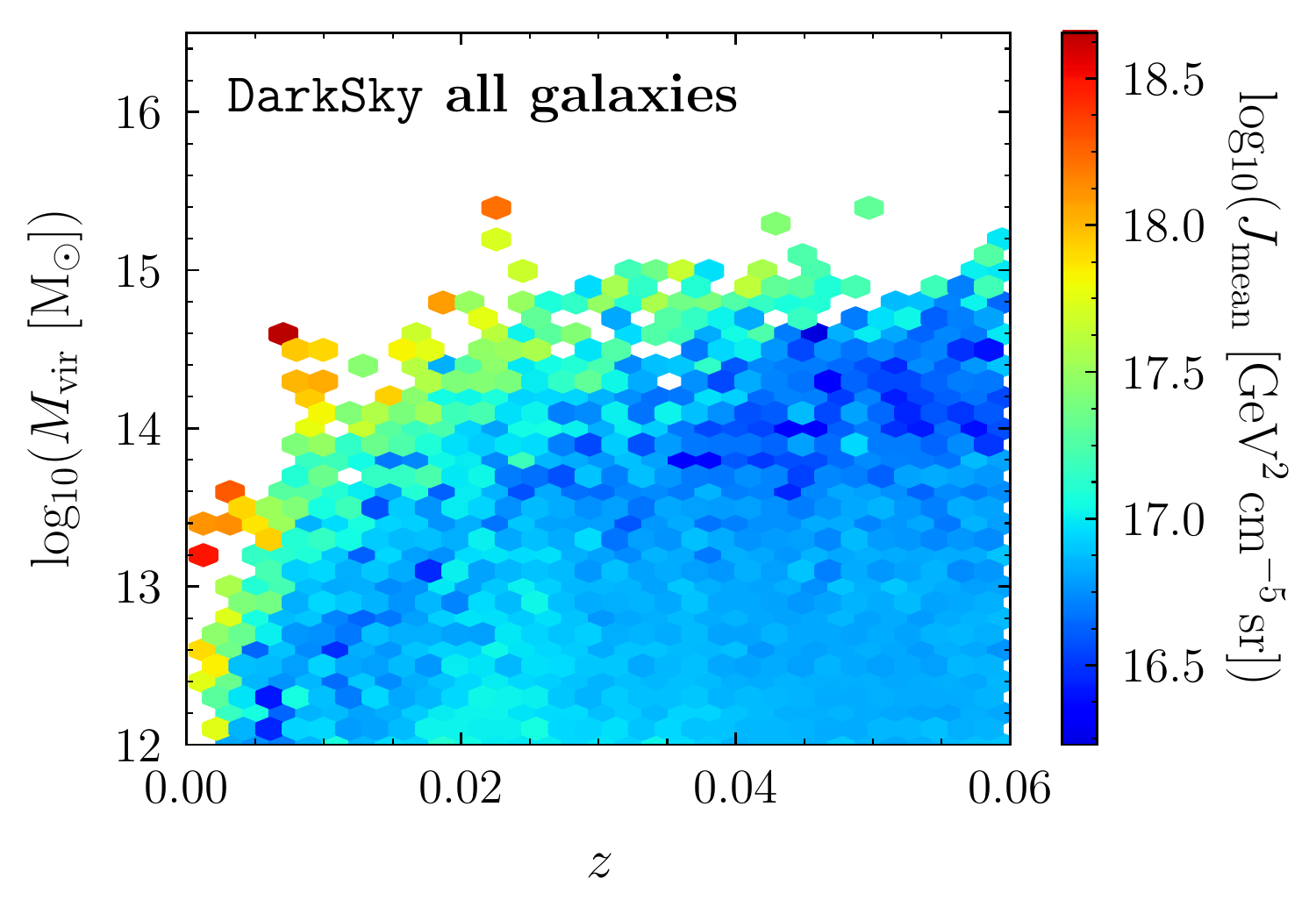}
   \caption{Heatmap of $J$-factors for the halos associated with all the galaxy groups in \texttt{DarkSky}, as a function of redshift and virial mass.  For this example, the observer is placed in the center of the simulation box. }
   \label{fig:heatmap}
\end{figure}

The $J$-factor is defined as the integral along the line-of-sight of the squared DM density of the observed object:\footnote{As defined, the $J$-factor has units of [${\rm GeV}^2 \cdot {\rm cm}^{-5} \cdot {\rm sr}$]. This definition is convenient for extragalactic objects, but beware because another common definition of the $J$-factor involves dividing out by a solid angle to remove the units of $[{\rm sr}]$. A detailed discussion of the units is provided in Appendix~\ref{app:JDrelations}.}
\begin{equation}
J = \left(1+b_\text{sh}[M_\text{vir}]\right)\,  \int ds \, d\Omega\,\rho^{2}_\text{NFW}(s,\Omega)\,,
\label{eq:jfactor}
\end{equation}
where $s$ is the line-of-sight distance and $b_\text{sh}[M_\text{vir}]$ is the so-called boost factor.  The boost factor accounts for the enhancement in the flux due to the annihilation in subhalos.   For the case of extragalactic objects, one can obtain a closed form solution that is an excellent approximation to the integral in Eq.~\ref{eq:jfactor}, which is proportional to
\begin{equation}
J \propto \left(1+b_\text{sh}[M_\text{vir}]\right) \frac{M_{\rm vir} \, c_{\rm vir}^3\, \rho_c}{d_c^2[z]}\,,
\label{eq:jscale}
\end{equation}
where $d_c$ is the comoving distance (a function of redshift, $z$), $\rho_c$ is the critical density, and $c_\text{vir}$ is the concentration.  In our analysis, we calculate the $J$-factor exactly, but the scaling illustrated in Eq.~\ref{eq:jscale} is useful for understanding the dependence of $J$ on the halo mass and concentration. The derivation of the $J$-factor expression is reviewed in detail in Appendix~\ref{app:JDrelations}, where we also show the result for the Burkert profile.

\begin{figure*}
   \centering
   \includegraphics[width=0.45\textwidth]{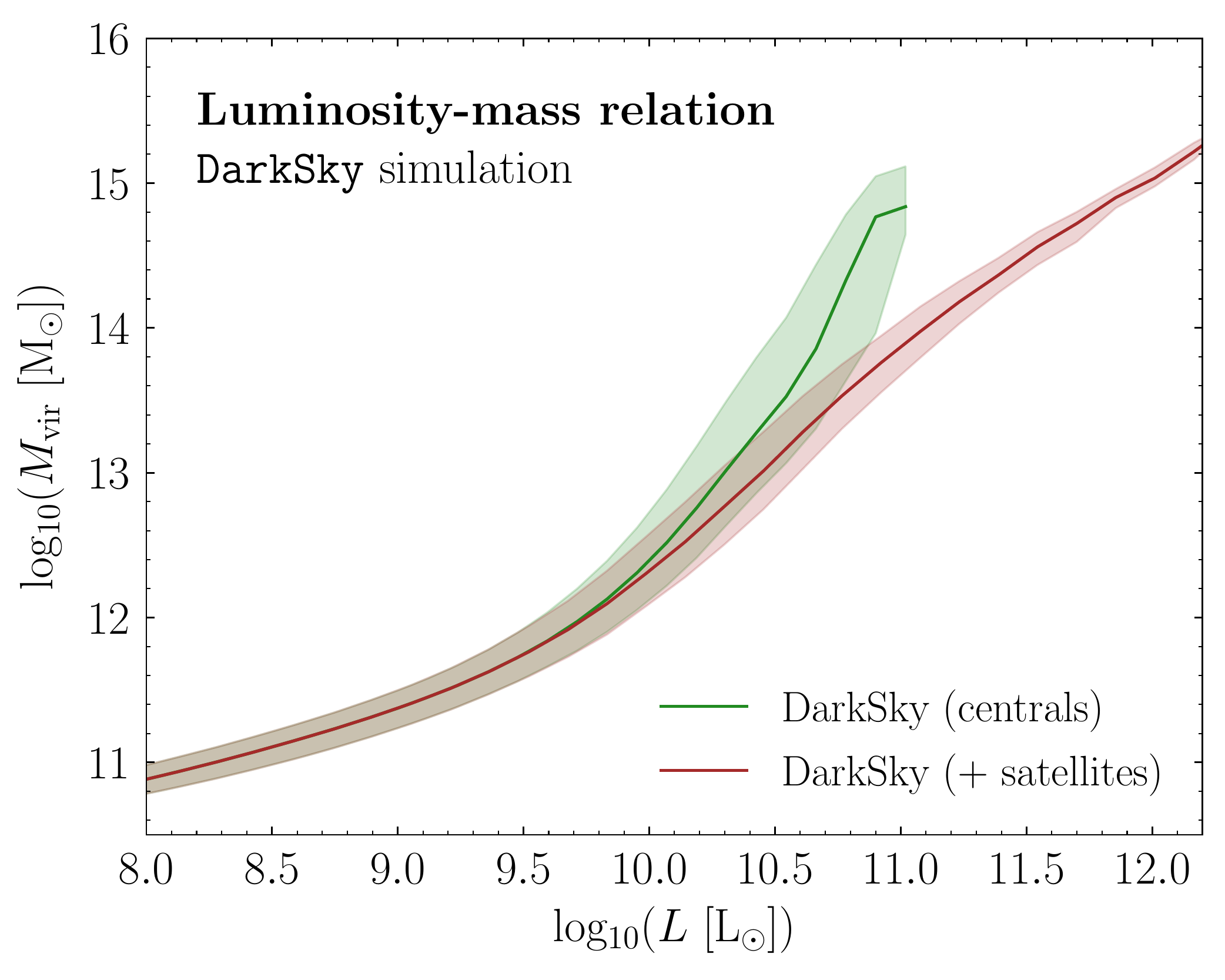}
   \includegraphics[width=0.45\textwidth]{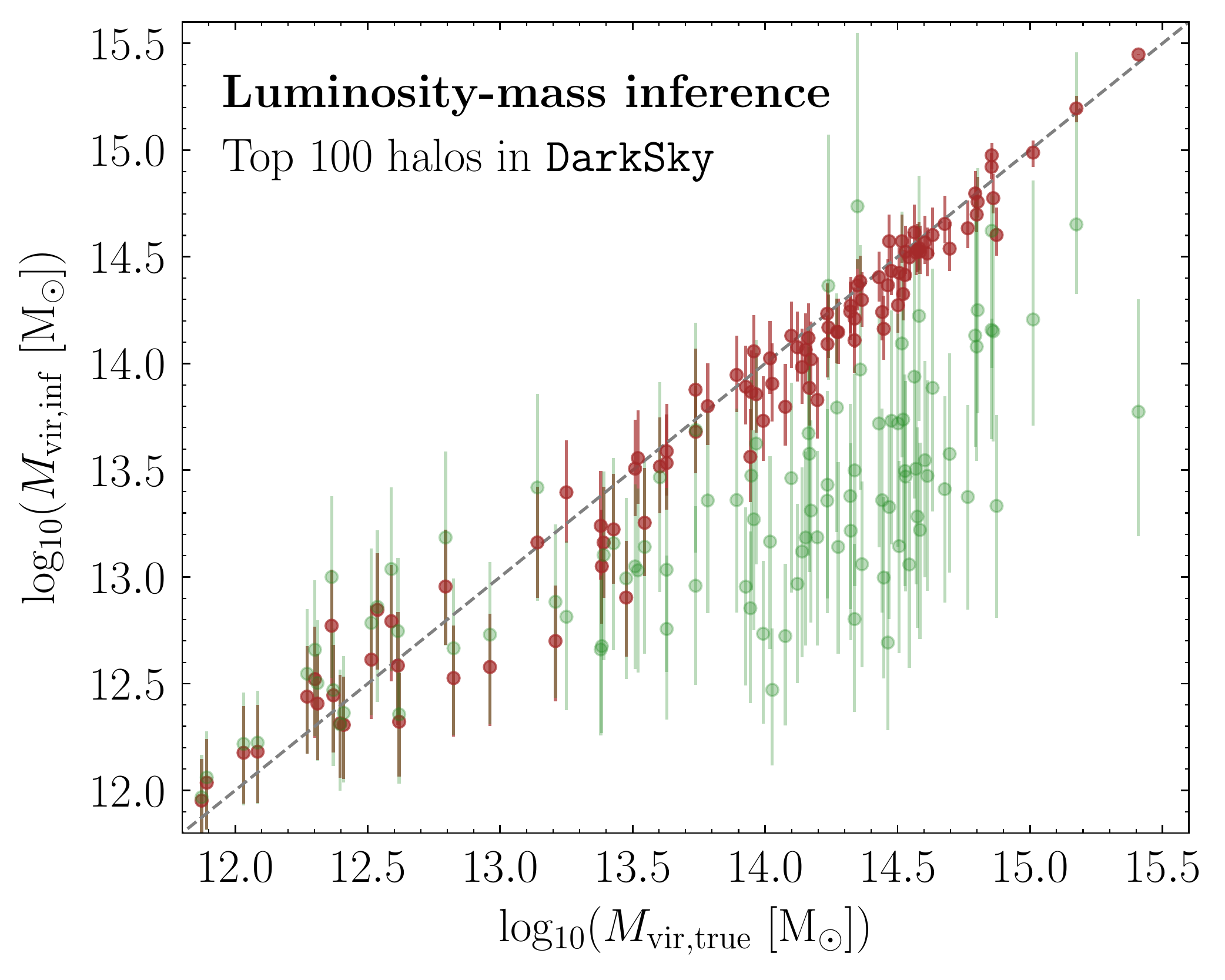}
   \caption{(Left) From \texttt{DarkSky}, we obtain the host halo mass as a function of absolute luminosity.  The green line represents the best-fit $M(L)$ relation when the central galaxy luminosity $(L_\mathrm{cen})$ is used to infer the host halo mass, while the red line uses the total luminosity $L_\mathrm{tot}$ (central + satellite).  The shaded region denotes the 68\% containment region in each case. (Right) Halo masses and uncertainties, inferred using the $M(L_\mathrm{cen})$ relation (green) and the $M(L_\mathrm{tot})$ relation (red).  The inclusion of the satellite luminosity allows one to better recover the halo mass.}
   \label{fig:ltom}
\end{figure*}

Figure~\ref{fig:DarkSkycounts} illustrates the truth $J$-factor map associated with \texttt{DarkSky}, obtained by putting the observer in the center of the simulation box.  This map is constructed by applying Eq.~\ref{eq:jfactor} to all host halos in the \texttt{DarkSky} catalog and using the boost model from Ref.~\cite{Bartels:2015uba} to describe the contribution from substructure.  Once the $J$-factors are known, the expected photon counts per pixel can be determined using Eq.~\ref{eq:flux} and \emph{Fermi}'s exposure map.  This is also shown in Fig.~\ref{fig:DarkSkycounts}, assuming a DM particle with $m_\chi = 100$~GeV that annihilates to $b \bar{b}$ with $\langle \sigma v \rangle \approx 10^{-24}$~cm$^3\,$s$^{-1}$.  Not all the pixels that contain one or more galaxies correspond to significant regions of DM annihilation.  The DM annihilation flux is largest for the most massive, concentrated, and/or closest galaxy groups.  

Note that when constructing Fig.~\ref{fig:DarkSkycounts}, we perform the angular integrals in Eq.~\ref{eq:jfactor} as a function of angular extent, $\Omega$.  In doing so, we implicitly assume that the boost factor is simply a multiplicative factor.  In reality, the boost factor likely broadens the angular profile, because the subhalo annihilation should extend further away from the halo center. However, since the angular extent of the annihilation in most halos is small compared to the instrument point-spread function (PSF), we do not model this extension here. Some nearby halos may have significantly larger angular extent, as would be expected for the Andromeda galaxy. Nevertheless, such considerations need to be made case by case and are discussed in detail in our companion paper~\cite{companion}, where we choose to exclude Andromeda due to its size.

Figure~\ref{fig:heatmap} is a heatmap representing the average $J$-factor, for a given $M_\text{vir}$ and $z$, of the \texttt{DarkSky} halos in the above configuration.  The halos span a wide range of masses and redshifts, with $J$-factors averaging over several orders of magnitude from $\sim$~10$^{16.5-18.5}~{\rm GeV}^2\,{\rm cm}^{-5}\,{\rm sr}$.     
The largest $J$-factors are observed for the most massive, cluster-sized halos at $z \sim $~0.01--0.02, as well as for less-massive halos at smaller redshifts ($z \lesssim 0.01$). 

\subsection{Uncertainties in Halo Modeling}
\label{sec:uncertainties}

Now, we consider more carefully the systematic uncertainties associated with modeling the halo properties.  
A halo with an NFW density profile has a $J$-factor dictated by its parameters as given in Eq.~\ref{eq:jscale}.  In addition to the distance, the $J$-factor also depends on the virial mass and concentration.\footnote{Note that uncertainties on the halo redshift also feed into the $J$-factor.  However, we consider this uncertainty to be subdominant for spectroscopically determined redshifts. For nearby halos, where the relation between distance and redshift is nontrivial, the uncertainty on the distance can be noticeably larger, and as high as $\sim$5\%~\cite{Tully:2016ppz}. Nonetheless, even such uncertainties are considerably smaller than those associated with the mass and concentration, and so we do not consider them.}  
  Therefore, any uncertainty in the determination of these halo properties is propagated through to the uncertainty on the DM annihilation flux.  Up until now, we have taken the halo mass and concentration directly from \texttt{DarkSky}, but in practice these parameters need to be inferred from properties of the observed galaxy groups.
 
Within \texttt{DarkSky}, the halo mass can be inferred from the absolute luminosity of its associated galaxy group.  We obtain a deterministic $M(L)$ relation following a procedure similar to that in Ref.~\cite{Vale:2005mw}, which derived a relation between the $K$-band galaxy luminosity and the mass of its DM halo.  The left panel of Fig.~\ref{fig:ltom} shows the true masses for the \texttt{DarkSky} halos, as a function of central galaxy luminosity (green) or the total luminosity, which includes the luminosity of the satellite galaxies (red).  The \texttt{DarkSky} catalog provides the associations for all galaxies, central and satellite, so we include all satellites that are associated to the group when calculating the total absolute luminosity. This is similar to what is done in published group catalogs~\cite{Tully:2015opa,2017ApJ...843...16K,Lu:2016vmu}, where they account for the loss in luminosity of satellite galaxies that are farther away.  

From Fig.~\ref{fig:ltom}, we see that the spread in the associated halo mass increases above $\sim10^{10}$~L$_\odot$, up to the brightest galaxy at $\sim10^{11}$~L$_\odot$, when the central galaxy luminosity is used.  In contrast, the spread is significantly smaller when the total luminosity is used, making it a better predictor for the halo mass.  As demonstrated in the right panel of Fig.~\ref{fig:ltom}, including the satellite luminosities allows one to better reconstruct the halo mass.  Therefore, we use the median $M(L)$ relation thus obtained as our fiducial case to infer the central mass estimate, and we use the spread in the $M(L)$ relation to infer the uncertainty on the mass.  Note that the $M(L)$ relation shown in Fig.~\ref{fig:ltom} is constructed by binning the \texttt{DarkSky} data in luminosity and calculating the 16, 50, and 84 percentiles in $M_\text{vir}$; different results would be obtained by binning in $M_\text{vir}$ and then constructing the percentiles from the luminosity distributions.   This procedure is similar to that adopted by galaxy group catalogs to infer the halo mass~\cite{Tully:2015opa,Lu:2016vmu,2017ApJ...843...16K}. Using this $M(L)$ relation, we can infer the halo mass and uncertainty  for each galaxy-group host halo in \texttt{DarkSky}. 

DM halos of the same mass can have very different characteristics, usually reflecting their distinct formation history and environment.  One such characteristic is the halo's virial concentration $c_\text{vir} = r_\text{vir}/r_s$. 
The scale radius is the relevant quantity to compare to as it indicates an isothermal slope for the density profile, which is required for a flat rotation curve.  The virial radius corresponds to the spherical volume within which the mean density is $\Delta_{c}$ times the critical density of the Universe at that redshift. We use $\Delta_c(z) = 18\pi^2 +82x-39x^2$ with $x = \Omega_{m}(1+z)^3/[\Omega_{m}(1+z)^3 + \Omega_{\Lambda}]-1$ in accordance with Ref.~\cite{Bryan:1997dn}.  The cosmology associated with the \texttt{DarkSky} simulation is used throughout, with $\Omega_\Lambda = 0.705$,  $\Omega_m = 0.295$ and $h = 0.688$. 

\begin{figure*}[t]
   \centering
  \includegraphics[width=0.49\textwidth]{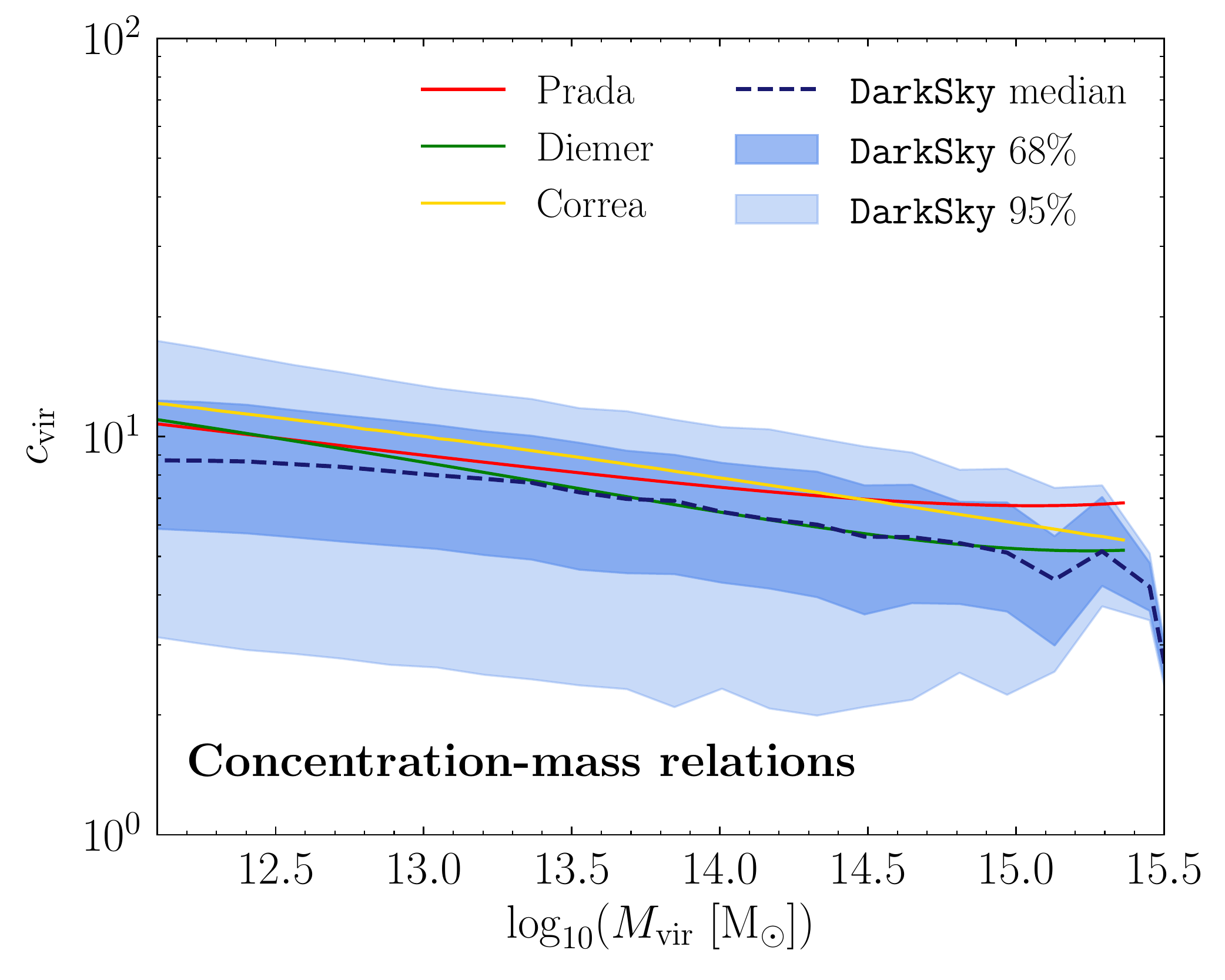} 
   \includegraphics[width=0.49\textwidth]{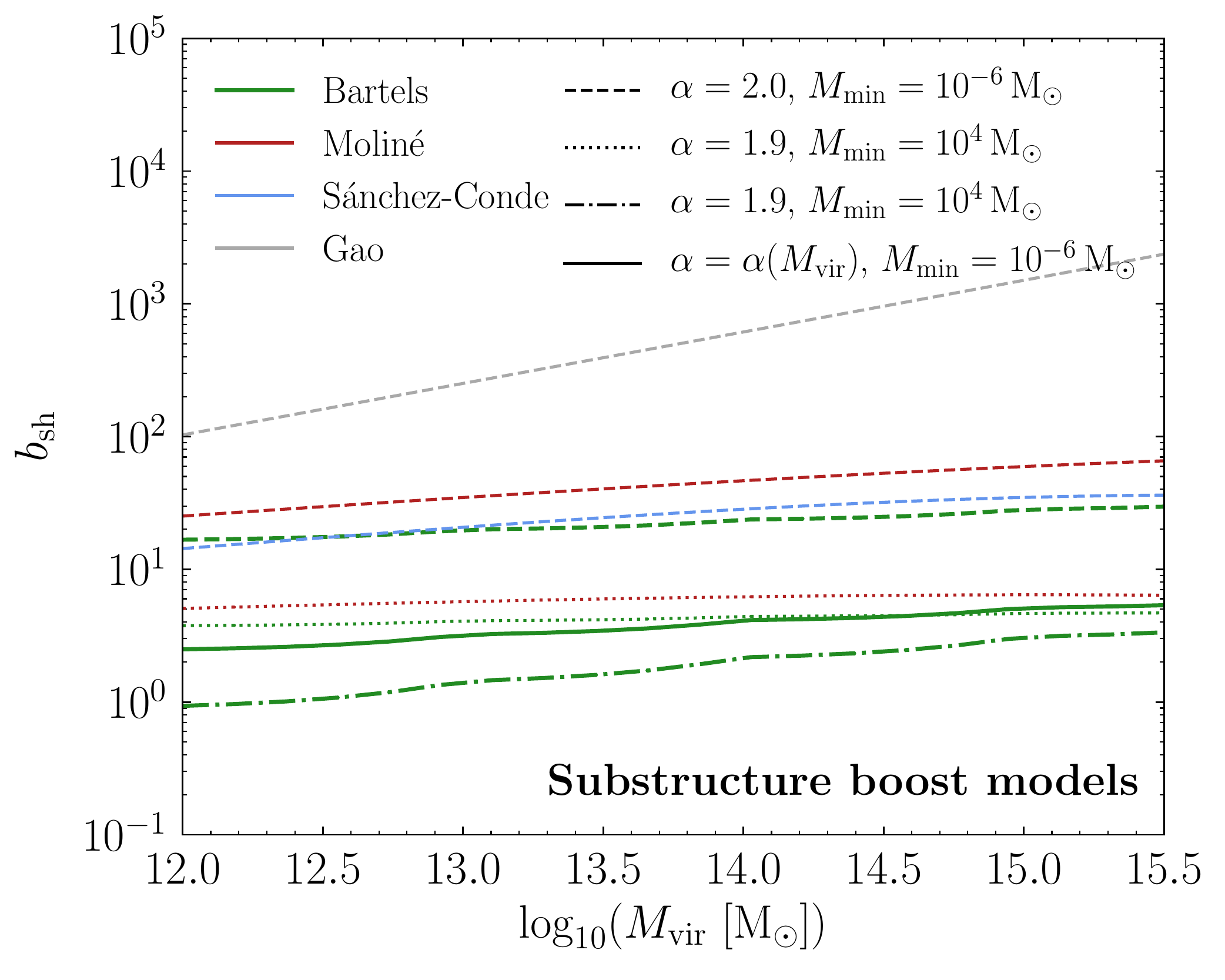} 
         \caption{(Left) The median concentration-mass relation in \texttt{DarkSky} (dashed black) along with the middle 68 and 95\% spread (blue regions) compared with models found in the literature.  For comparison, we also show the models of Correa \emph{et al.} (yellow)~\cite{Correa:2015dva}, Diemer and Kravtsov (green)~\cite{Diemer:2014gba}, and Prada \emph{et al.} (red) \cite{Prada:2011jf}.  All concentration models are evaluated for the \texttt{DarkSky} cosmology. (Right)  Boost models found in the literature as a function of host halo mass. As a conservative choice, we select the Bartels and Ando model~\cite{Bartels:2015uba} shown in thick solid green. In blue, red and gray, we compare this to the boost models of S{\'a}nchez-Conde \emph{et al.}~\cite{Sanchez-Conde:2013yxa}, Molin{\'e} \emph{et al.}~\cite{Moline:2016pbm}, and Gao \emph{et al.}~\cite{Gao:2011rf}, respectively. The line type (dashed, dotted, dot-dashed, and solid) denotes the assumption being made on the slope of the subhalo mass function, $\alpha$, and the mass cutoff, M$_\text{min}$.  }
    \label{fig:concentrations}
\end{figure*}

In general, the concentration correlates strongly with halo mass due to the dependence of halo formation time on mass---on average, lower mass halos tend to be more concentrated because they collapsed earlier, when the Universe was denser.  For the same reason, the concentration is sensitive to the cosmology, which determines how early halos start to assemble. 
The concentration of field halos has been extensively studied and several concentration-mass relations have been proposed in the literature, usually based on $N$-body simulations or physically motivated analytic approaches~\cite{Duffy:2008pz,Prada:2011jf,Sanchez-Conde:2013yxa,Diemer:2014gba,Correa:2015dva,Moline:2016pbm,Klypin:2014kpa,Dutton:2014xda}. In the left panel of Fig.~\ref{fig:concentrations}, we show the median value of the concentration-mass relation derived directly from the \texttt{DarkSky} simulation, as well as the middle 68 and 95\% spread. The middle 68\% scatter in the relation is typically in the range 0.14-0.19 across the halo mass range considered. For comparison, we also show several concentration models that are commonly used in the literature.  As is standard in the literature~\cite{Hutten:2016jko,Sanchez-Conde:2013yxa}, we model the uncertainty in the concentration, for a given virial mass, as a log-normal distribution around its median value.  

To summarize, it is possible to infer the halo mass from the luminosity of the galaxy group and to then obtain the concentration.  The final remaining property that is needed to solve for the $J$-factor in Eq.~\ref{eq:jfactor} is the boost factor, which depends on the distribution and minimum cutoff of the subhalos' mass.  The boost factor encapsulates the complicated dependence of the subhalo mass distribution on both the particle physics assumptions of the DM model as well as the dynamics of the host halo formation.  A variety of different boost models typically used in the literature are illustrated in the right panel of Fig.~\ref{fig:concentrations}.  As our fiducial case, we adopt the boost model of Ref.~\cite{Bartels:2015uba} (labeled as `Bartels Boost Model'), which self-consistently accounts for  the concentration-mass relation of subhalos (compared to field halos) as well as the effects of tidal stripping. Specifically, in the subhalo mass function $dn/dM_\text{sh}\propto M_\text{sh}^{-\alpha}$, we use a minimum subhalo mass cutoff of $M_\text{min}=10^{-6}$~M$_{\odot}$ and slope $\alpha$ that varies self-consistently with host halo mass while accounting for evolution effects (see Ref.~\cite{Bartels:2015uba} for details).

We have now built up a framework that allows us to determine the expected DM annihilation flux map associated with a catalog of galaxy groups.  Next, we show how to use this information to search for signals of DM  from hundreds of galaxy groups.  

\section{Statistical Methods}
\label{sec:stats}

In this work, we introduce and study a statistical procedure to search for gamma-ray signals from DM by stacking galaxy groups.  All analyses discussed here are run on mock data, which is based on the expected astrophysical contributions to the real \emph{Fermi} data set. When building this mock data set, we include contributions from (1) the diffuse  emission, for which we use the \textit{Fermi} Collaboration's \texttt{p7v6} model; (2) isotropic emission;  (3) emission from the \textit{Fermi} Bubbles \cite{Su:2010qj}; and (4) emission from point sources in the \textit{Fermi} 3FGL catalog \cite{Acero:2015hja}. The overall flux normalization  for each component must be known \emph{a priori} to create the mock data. To obtain this, we fit spatial maps of (1)--(4) above to the actual \emph{Fermi} data.  We use 413 weeks of UltracleanVeto (all PSF quartile) Pass 8 data collected between August 4, 2008 and July 7, 2016. We break the data into 40 equally log-spaced energy bins between 200~MeV and 2~TeV, applying the recommended data cuts: zenith angle $< 90^{\circ}$, \texttt{DATA\_QUAL} $> 0$, and \texttt{LAT\_CONFIG} $=1$.  To minimize the Galactic contamination in this initial fit, we mask the region $|b|<30^{\circ}$ as well as the 68\% containment radius for the 300 brightest and most variable sources in the 3FGL catalog. We emphasize that these masks are only used when creating the mock data and not in the stacked analysis. The fitting procedure described here provides the expected astrophysical background contribution from the real data.  Monte Carlo (MC) is then generated by summing up these contributions and taking a Poisson draw from the resulting map.  In the following discussion, we will show how results vary over different MC realizations of the mock data as a demonstration of Poisson fluctuations in the photon distribution. 

We now describe in detail the statistical procedure we employ to implement the stacking analysis on the mock data.  We perform a template-fitting profile likelihood analysis in a 10$^{\circ}$ region-of-interest (ROI) around each group.  Template studies amount to describing the sky by a series of spatial maps (called templates).  The normalization of each template is proportional to its relative gamma-ray flux.  We use five templates in our study.  The first four are associated with the known astrophysical sources (1)--(4) described above. Within $10^\circ$ of the halo center, we independently float the normalization of each 3FGL source.\footnote{The results do not change when floating all the point sources together as one combined template. This can potentially cause problems when implemented on data, however, because the 3FGL normalizations can be erroneous in certain energy bins.  Allowing the normalizations of the sources to float separately helps to mitigate this potential problem.}  Sources outside this region may potentially contribute within the ROI because of the tails of the {\it Fermi} PSF.  Therefore, between 10$^{\circ}$ and 18$^{\circ}$ of the halo center,  we float the sources as a single template. The fifth and final template that we include is associated with the expected DM annihilation flux for the halo, which is effectively a map of the $J$-factor and is described in Sec.~\ref{sec:galaxyfilteringpipeline}.  Note that all templates have been carefully smoothed using the \textit{Fermi} PSF. The diffuse model is smoothed with the \textit{Fermi} Science Tools, whereas other templates are smoothed according to the instrument response function using custom routines.  Mismodeling the smoothing of either the point sources or individual halos can potentially impact the  results.

A given mock data set, $d$, is divided into 40 log-spaced energy bins indexed by $i$. Each energy bin is then spatially binned using \texttt{HEALPix}~\cite{Gorski:2004by} with \texttt{nside}=128 and individual pixels indexed by $p$.  In this way, the full data set is reduced to a two-dimensional array of integers $n_i^p$ describing the number of photons in energy bin $i$ and pixel $p$.  For a given halo, indexed by $r$, only a subset of all the pixels in its vicinity are relevant.  In particular, the relevant pixels are those with centers within 10$^{\circ}$ of the object.  Restricting to these pixels leaves a subset of the data, which we denote by $n_i^{p,r}$. Template fitting dictates that this data is described with a set of spatial templates binned in the same way as the data, which we label as $T^{p,\ell}_i$, where $\ell$ indexes the different templates considered. The number of counts in a given pixel, energy bin, and region consists of a combination of these templates:
\begin{equation}
\mu_i^{p,r}({\boldsymbol \theta_i^r}) = \sum_{\ell} A^{r, \ell}_i\,T^{p,\ell}_i\,.
\end{equation}
Here, ${\boldsymbol \theta_i^r}$ represents the set of model parameters. For Poissonian template fitting, these are given by the normalizations of the templates $A^{r,\ell}_i$, \emph{i.e.}, ${\boldsymbol \theta_i^r} = \{ A^{r,\ell}_i \}$. Note that the template normalizations have an energy but not a spatial index, as the templates have an independent degree of freedom in each energy bin as written, but the spatial distribution of the model is fixed by the shapes of the templates themselves.  In principle, we could also remove this freedom in the relative emission across energy bins, because we have models for the spectra of the various background components, and in particular DM. Nevertheless, we still allow the template normalizations to float independently in each energy bin for the various backgrounds. This is more conservative than assuming a model for the background spectra, and in particular we can use the shape of the derived spectra as a check that the dominant background components are being correctly modeled. The spectral shape of the DM forms part of our model prediction, however, and once we pick a final state such as annihilation to two $b$-quarks, we fix the relative emission between the energy bins.

As we assume that the data comes from a Poisson draw of the model, the appropriate likelihood in energy bin $i$ and ROI $r$ is
\begin{equation}
\mathcal{L}_i^r(d_i^r | {\boldsymbol \theta}_i^r) = \prod_p \frac{\mu_i^{p,r}({\boldsymbol \theta}_i^r)^{n_i^{p,r}} e^{-\mu_i^{p,r}({\boldsymbol \theta}_i^r)}}{n_i^{p,r}!}\,.
\label{eq:pi}
\end{equation}
Of the templates that enter this likelihood, there are some we are more interested in than others. In particular, we care about the the DM annihilation intensity, which we denote as $\psi_i$.  We treat the normalizations of the templates associated with the known astrophysical emission as nuisance parameters, $\lambda_i^r$. Below, we will describe how to remove the nuisance parameters to reduce Eq.~\ref{eq:pi} to a likelihood profile that depends only on the DM annihilation intensity, but for now we have ${\boldsymbol \theta}_i^r = \{ \psi_i, \lambda_i^r \}$.

Importantly, the nuisance parameters have different values between ROIs, but the DM parameters do not.  This is because the DM parameters, such as the DM mass, annihilation rate, and set of final states, are universal, while the parameters that describe the astrophysical emission can vary from region to region.
We do, however, profile over the $J$-factor uncertainty in each ROI.
Explicitly, each halo is given a model parameter $J^r$, which is described by a
log-normal distribution around the central value $\log_{10} J_{\rm c}^r$ with width $\sigma_r = \log_{10} J_{\rm err}^r$, both of which depend on the object and hence ROI considered.  The $J$-factor error, $J_{\rm err}^r$, is determined by propagating the errors associated with the mass and concentration of a given halo. To account for this, we append the following addition onto our likelihood as follows:
\begin{equation}\begin{aligned}
&\mathcal{L}_i^r \left( d_i^r | \boldsymbol{\theta}_i^r \right) \to \mathcal{L}_i^r \left( d_i^r | \boldsymbol{\theta}_i^r \right) \\
&\times \frac{1}{\ln(10) J_{\rm c}^r\sqrt{2\pi} \sigma_r} \exp \left[ - \frac{(\log_{10} J^r - \log_{10} J_{\rm c}^r)^2}{2\sigma_r^2} \right]\,.
\label{eq:Jlognormal}
\end{aligned}\end{equation}
A detailed justification for this choice of the likelihood is provided in Appendix~\ref{app:Juncertainties}.  Note that this procedure does not account for any systematic uncertainties that can bias the determination of the $J$-factor.

The nuisance parameter $J^r$ can now be eliminated via the profile likelihood---see Ref.~\cite{Rolke:2004mj} for a review. Unlike for the other nuisance parameters, the value of $J^r$ does not depend on energy and so we eliminate the energy-dependent parameters first:
\begin{equation}
\mathcal{L}_i^r(d_i^r | \psi_i) = \max_{\{\lambda_i^r\}}\,\mathcal{L}_i^r \left( d_i^r | \boldsymbol{\theta}_i^r \right)\,.
\label{eq:firstprofile}
\end{equation}
The full implementation of the profile likelihood method as suggested by this equation requires determining the maximum likelihood for the $\lambda_i^r$ template coefficients, for every value of $\psi_i$.  Nevertheless, an excellent approximation to the profile likelihood, which is computationally more tractable, is simply to set the nuisance parameters to their maximum value obtained in an initial scan where all templates are floated.\footnote{The DM template is only included for energy bins above 1 GeV.  At lower energies, the large \textit{Fermi} PSF leads to confusion between the DM, isotropic and point source templates, which can introduce a spurious preference for the DM template.} 
 We follow this procedure throughout in calculating likelihoods and discuss its validity in Appendix~\ref{app:energyrange}. 

Using this approach to determine the likelihood in Eq.~\ref{eq:firstprofile}, we can build a total likelihood by combining the energy bins. Once this is done, the likelihood  depends on the full set of DM intensities $\psi_i$, which are specified by a DM model $\mathcal{M}$, cross section $\langle \sigma v \rangle$, mass $m_{\chi}$, and $J$-factor via Eq.~\ref{eq:flux}. Explicitly:
\begin{equation}
\mathcal{L}^r(d_r | \mathcal{M}, \langle \sigma v \rangle, m_\chi, J^r) = \prod_i \mathcal{L}_i^r(d_i^r | \psi_i)\,,
\end{equation}
and recall that unlike the other parameters on the left hand side, the $J$-factor not only determines the $\psi_i$, but also enters the likelihood through the expression in Eq.~\ref{eq:Jlognormal}. We emphasize that in this equation, the DM model and mass specify the spectra, and thereby the relative weightings of the $\psi_i$, whereas the cross section and $J$-factor set the overall scale of the emission.

The remaining step to get the complete likelihood for a given halo $r$ is to remove $J^r$, again using profile-likelihood:
\begin{equation}
\mathcal{L}^r(d_r | \mathcal{M}, \langle \sigma v \rangle, m_\chi) = \max_{J^r} \mathcal{L}^r(d_r | \mathcal{M}, \langle \sigma v \rangle, m_\chi, J^r)\,.
\end{equation}
This provides the full likelihood for this object as a function of the DM model parameters.  The likelihood for the full stacked catalog is then simply a product over the individual likelihoods:
\begin{equation}
\mathcal{L}(d | \mathcal{M}, \langle \sigma v \rangle, m_\chi) = \prod_r \mathcal{L}^r(d_r | \mathcal{M}, \langle \sigma v \rangle, m_\chi)\,.
\label{eq:likelihoodobjprod}
\end{equation}
Using this likelihood, we define a test statistic (TS) profile as follows:
\begin{equation}\begin{aligned}
{\rm TS}(\mathcal{M}, \langle\sigma v\rangle, m_\chi) \equiv 2 &\left[ \log \mathcal{L}(d | \mathcal{M}, \langle\sigma v\rangle, m_\chi ) \right.\\
&\left.- \log \mathcal{L}(d | \mathcal{M}, \widehat{\langle\sigma v\rangle}, m_\chi ) \right]\,,
\label{eq:TSdef}
\end{aligned}\end{equation}
where $\widehat{\langle\sigma v\rangle}$ is the cross section that maximizes the likelihood for that DM model and mass. From here, we can use this TS, which is always nonpositive by definition, to set a threshold for limits on the cross-section.  When searching for evidence for a signal, we use an alternate definition of the test statistic defined as 
\es{maxTS}{
{\rm TS}_\text{max}(\mathcal{M}, m_\chi) \equiv &2 \left[ \log \mathcal{L}(d | \mathcal{M}, \widehat{\langle\sigma v\rangle}, m_\chi ) \right. \\
 &\left.- \log \mathcal{L}(d | \mathcal{M}, \langle\sigma v\rangle =0, m_\chi ) \right] \,.
}

We implement template fitting with the package \texttt{NPTFit} \cite{Mishra-Sharma:2016gis}, which uses \texttt{MultiNest}~\cite{Feroz:2008xx,Buchner:2014nha} by default, but we have employed \texttt{Minuit}~\cite{James:1975dr} in our analysis.

\section{Analysis Results}
\label{sec:smallrois}

In this Section, we present the results of our analysis on mock data using the \texttt{DarkSky} galaxy catalog. We begin by describing the sensitivity estimates associated with this study, commenting on the impact of statistical as well as systematic uncertainties and studying the effect of stacking a progressively larger number of halos.  Then, we justify the halo selection criteria that are used by showing that we can recover injected signals on mock data. 

\subsection{Halo Selection and Limits}
\label{sec:limits}

We now discuss the results obtained by applying the halo inference pipeline described in Sec.~\ref{sec:galaxyfilteringpipeline} and the  statistical analysis described in Sec.~\ref{sec:stats} to  mock gamma-ray data.  We focus on the top 1000 galaxy groups in the \texttt{DarkSky} catalog, as ranked by the inferred $J$-factors of their associated halos, placing ourselves at the center of the simulation box. In addition, we mask regions of the sky associated with seven large-scale structures that are challenging to model accurately: the Large and Small Magellanic Clouds, the Orion molecular clouds, the galaxy NGC5090, the blazar 3C454.3, and the pulsars J1836+5925 and Geminga. 

While we start from an initial list of 1000 galaxy groups, we do not include all of them in the stacking procedure.  A galaxy group is excluded if: 
\begin{enumerate}
\item it is located within $|b| \leq 20^\circ$;
\item it is located less than $2^\circ$ from the center of another brighter group in the catalog;
\item it has TS$_\text{max}$ $> 9$ and $\langle \sigma v \rangle_\text{best}  > 10.0 \times \langle \sigma v \rangle_\text{lim}^*$\, ,
\end{enumerate}
where $\langle \sigma v \rangle_\text{best}$ is the best-fit cross section at any mass and $\langle \sigma v \rangle^*_\text{lim}$ is the best-fit limit set by \emph{any} halo at the specified DM mass.  Note that the second requirement is applied sequentially to the ranked list of halos, ordered by $J$-factor.  
We now explain the motivation for each of these requirements separately.  The first requirement listed above removes groups that are located close to the Galactic plane to reduce contamination from regions of high diffuse emission and the associated uncertainties in modeling these. The second requirement demands that the halos be reasonably well-separated, which avoids issues having to do with overlapping halos and accounting for multiple DM parameters in the same ROI. The non-overlap criterion of 2$^\circ$ is chosen based on the \emph{Fermi} PSF containment in the lowest energy bins used and on the largest spatial extent of gamma-ray emission associated with the extended halos, which collectively drive the possible overlap between nearby halos.

\begin{figure*}[t]
   \centering
   \includegraphics[width=0.45\textwidth]{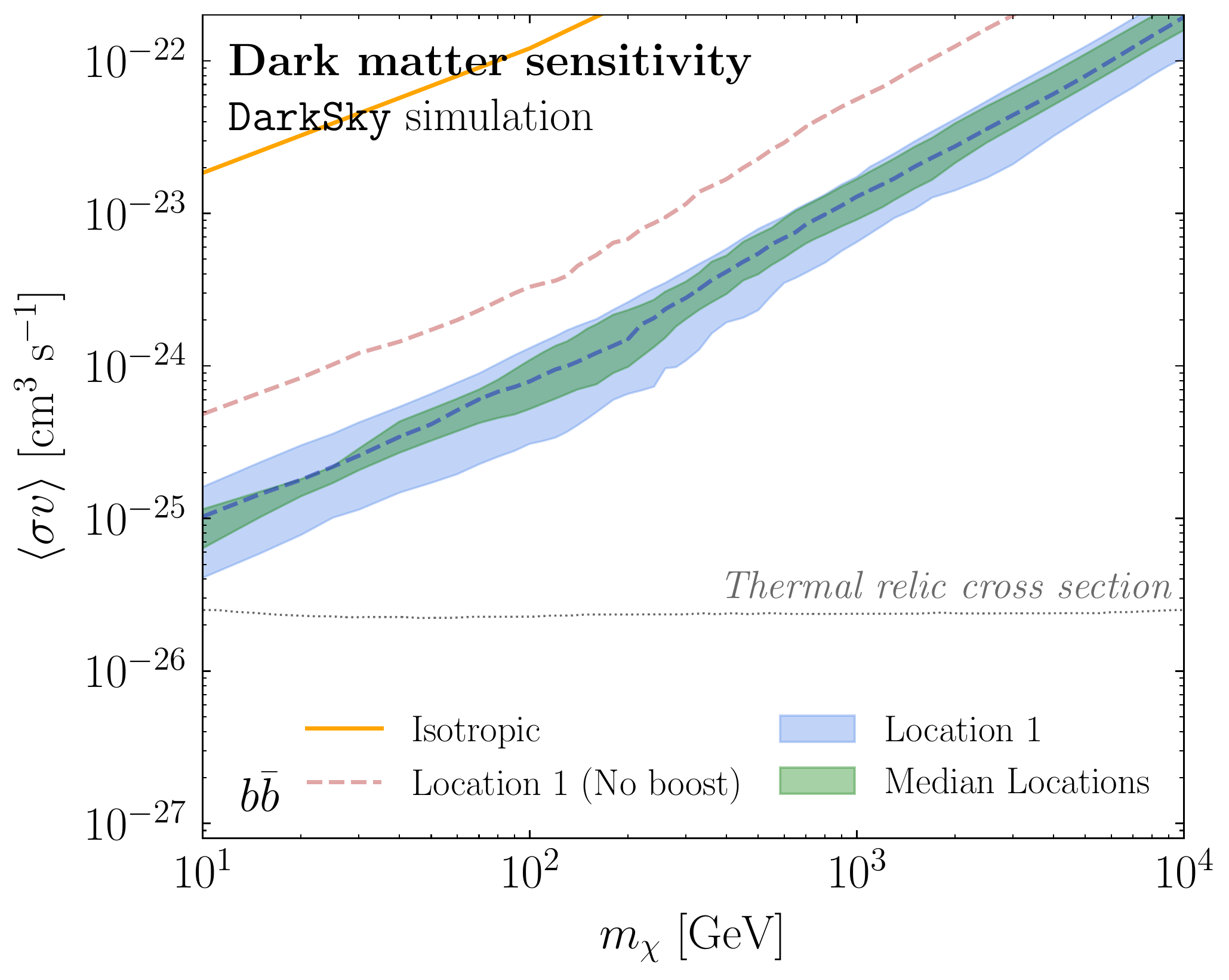} 
   \includegraphics[width=0.45\textwidth]{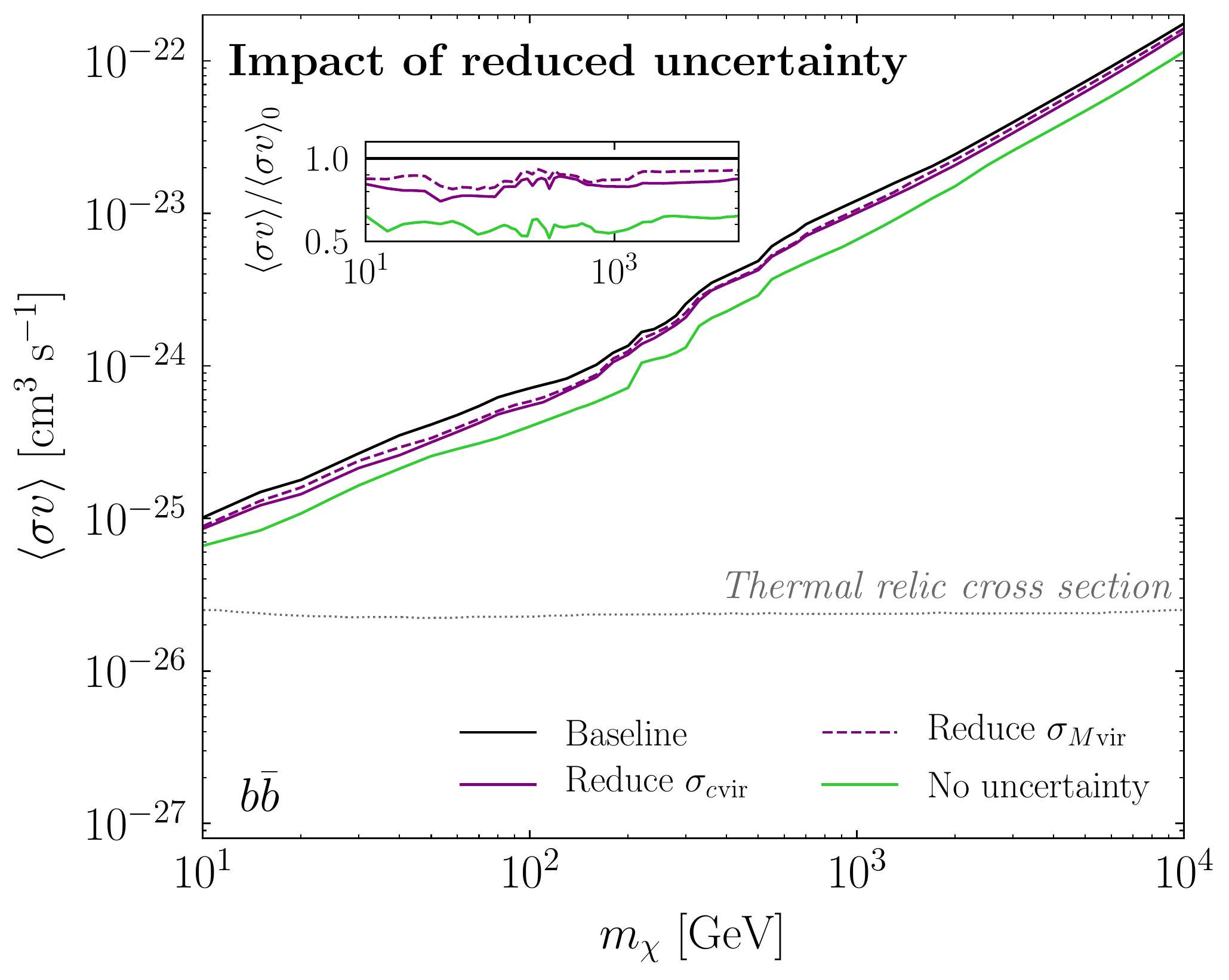} 
   \caption{(Left) The 95\% confidence limit on the DM annihilation cross section, $\langle \sigma v \rangle$, as a function of the DM mass, $m_\chi$, for the $b\bar{b}$ final state, assuming the fiducial boost-factor model from Ref.~\cite{Bartels:2015uba} (dashed blue); the corresponding result with no boost factor is shown in dashed red.  These limits correspond to the default position where the observer is placed in the center of the \texttt{DarkSky} simulation box (`Location 1').  The blue band shows the middle 68\% spread in the median limits obtained from 100 Monte Carlo realizations of the mock data.  The green band shows the same spread on the median limits obtained from nine random observer locations within the \texttt{DarkSky} simulation box.  The orange line shows the limit obtained by requiring that DM emission not overproduce the observed isotropic gamma-ray intensity and highlights how the sensitivity improves when one resolves the DM structure.  The thermal relic cross section for a generic weakly interacting massive particle~\cite{Steigman:2012nb} is indicated by the thin dotted line. (Right) The effect of reducing the uncertainty on virial mass, $M_\text{vir}$, and concentration, $c_\text{vir}$, in the stacking analysis.  The case where no uncertainty on the $J$-factor is assumed (green) is compared with the baseline analysis (black). We also show the impact of individually reducing the uncertainty on the concentration (solid purple) or mass (dashed purple) by 50\% for each halo. The inset shows the ratio of the improved cross section limit to the baseline case.}
   \label{fig:DSLimits}
\end{figure*}

The final requirement excludes a galaxy group if it has an excess of at least 3$\sigma$ significance associated with the DM template that is simultaneously excluded by the other galaxy groups in the sample.  This selection is necessary because we expect that some galaxy groups will have true cosmic-ray-induced gamma-ray emission from conventional astrophysics in the real data, unrelated to DM.  To identify these groups, we take advantage of the fact that we are starting from a large population of halos that are all expected to be bright DM sources in the presence of a signal.  Thus, if one halo sets a strong limit on the annihilation rate and another halo, at the same time, has a large excess that is severely in conflict with the limit, then most likely the large excess is not due to DM.  The worry here is that we could have mis-constructed the $J$-factor of the halo that gave the strong limit, so that the real limit is not as strong as we think it is.  However, with the TS$_\text{max}$ and $\langle \sigma  v \rangle$ criteria outlined above, this does not appear to be the case.  In particular, we find that the criteria very rarely rejects halos due to statistical fluctuations.  For example, over 50 MC iterations of the mock data, $966\pm8$ halos (out of 1000) remain after applying the TS$_\text{max}$ and cross section cuts alone, and the excluded halos tend to have lower $J$-factors, since there the $\langle \sigma v\rangle$ requirement is more readily satisfied.  

We expect that this selection criteria will be very important on real data, however, where real excesses can abound.  In addition, as we will describe in the next subsection, injected signals are not excluded when the analysis pipeline is run on mock data.  In an ideal scenario, we would attempt to understand the origin of these excesses by correlating their emission to known astrophysics either individually or statistically. In the present analysis, however, we take the conservative approach of removing halos that are robustly inconsistent with a DM signal and leave a deeper understanding of the underlying astrophysics to future work.

 \begin{figure*}[t]
   \centering
   \includegraphics[width=1.0\textwidth]{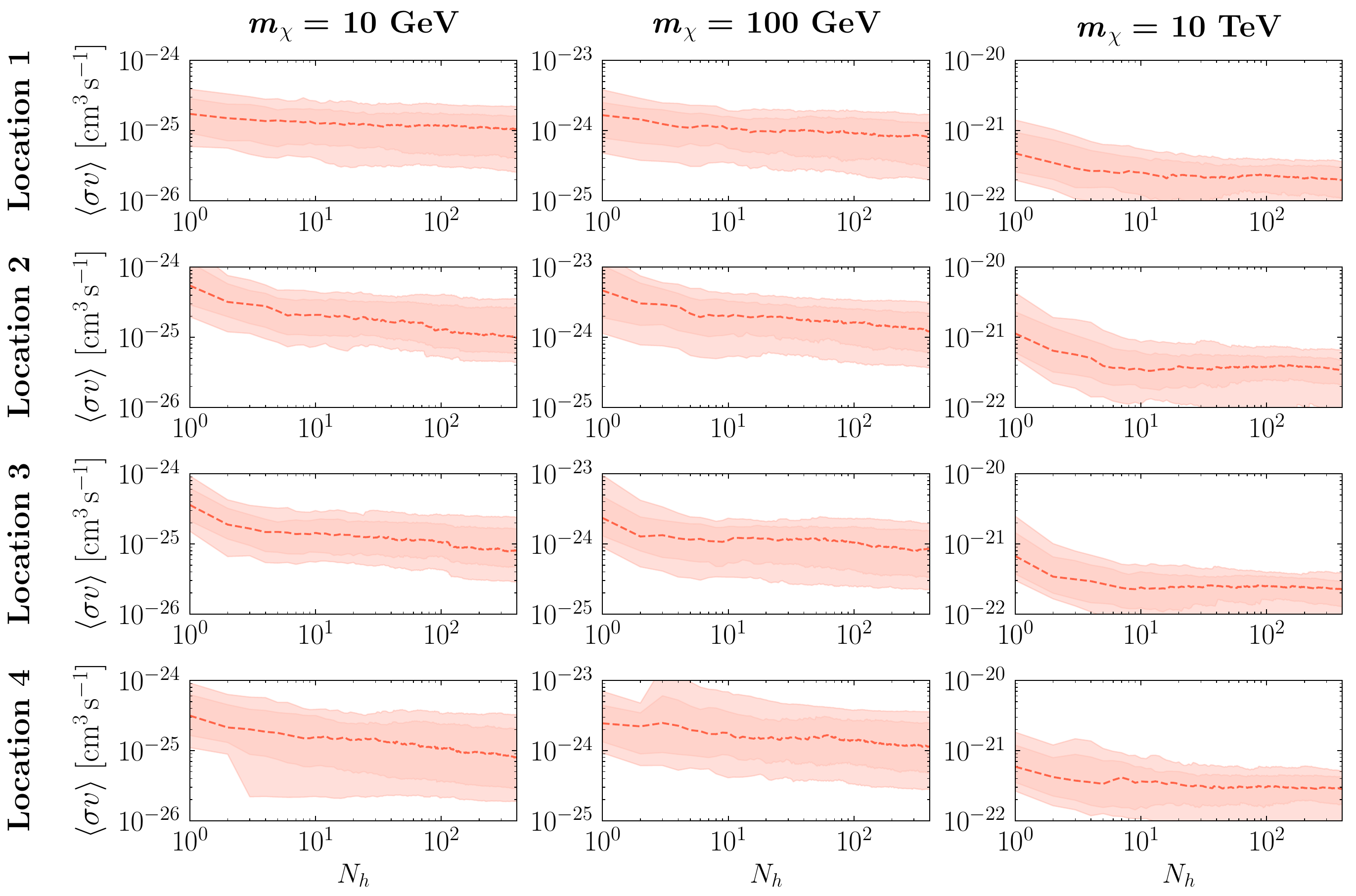}
   \caption{Variation of the limits as the number of galaxy groups (ranked by $J$-factor) included in the stacking, $N_h$, increases. The left, center, and right columns correspond to masses of 10~GeV, 100~GeV, and 10~TeV, respectively. Note that the scale of the y-axis varies between masses.  The four rows show how the limits vary for four different observer locations within the \texttt{DarkSky} simulation box.  }
   \label{fig:DSelephants}
\end{figure*}
We apply the procedure outlined in Secs.~\ref{sec:galaxyfilteringpipeline} and~\ref{sec:stats} to  the mock data to infer the 95\% confidence limit on the DM annihilation cross section.  The resulting sensitivity is shown by the blue dashed line in the left panel of Fig.~\ref{fig:DSLimits}, which uses the boost factor from Ref.~\cite{Bartels:2015uba}.  For comparison, we also show the limit assuming no boost factor (red dashed line); note that the boost-factor model that we use provides a modest $\mathcal{O}(1)$ improvement to the limit.  Because the limit can potentially vary over different MC realizations of the mock data, we repeat the procedure for 100 MCs (associated with different Poisson realizations of the map); the blue band indicates the middle 68\% spread in the limit associated with this statistical variation.

To see how the limit depends on the observer's location within the \texttt{DarkSky} simulation box, we repeat the procedure described above over nine different locations.\footnote{The nine locations we used are at the following coordinates $(x,y,z)$ Mpc/h in \texttt{DarkSky}: $(200,200,200)$, $(100,100,100)$, $(100,100,300)$, $(100,300,100)$, $(300,100,100)$, $(300,300,100)$, $(100,300,300)$, $(300,100,300)$, $(300,300,300)$. The first listed location is our default position, and any time we use more than one location they are selected in order from this list.}  At each location, we perform 20 MCs and obtain the median DM limit.  The green band in the left panel of Fig.~\ref{fig:DSLimits} denotes the middle 68\% spread on the median bounds for each of the different sky locations. 
In general, we find that the results obtained by an observer at the center of the \texttt{DarkSky} box are fairly representative, compared to random locations.  Note, however, that this bound does not necessarily reflect the sensitivity reach one would expect to get with actual \emph{Fermi} data. The reason for this is that the locations probed in \texttt{DarkSky} do not resemble that of the Local Group in detail.
We will come back to this point below, when we compare the $J$-factors of the \texttt{DarkSky} halos to those from galaxy catalogs that map the local Universe.

The orange line in the left panel of Fig.~\ref{fig:DSLimits} shows the limit obtained by requiring that the DM emission from the groups not overproduce the measured isotropic gamma-ray component~\cite{Ackermann:2014usa}. This should not be compared to the published DM bounds obtained with the \emph{Fermi} Isotropic Gamma-Ray Background~\cite{Ackermann:2015tah} because that study accounts for the integrated effect of the DM annihilation flux from halos much deeper than those we consider here.  The inclusion of these halos results in a total flux that can be greater than those from our sample by over an order of magnitude. Nevertheless, this gives an idea of how much we gain by resolving the spatial structure of the local DM population and knowing the locations of the individual galaxy groups.

The right panel of Fig.~\ref{fig:DSLimits} shows the effect of propagating uncertainties associated with inferring the halo properties. The green line indicates how the limit improves when no uncertainties are assumed, \emph{i.e.}, we can perfectly reconstruct the virial mass and concentration of the halos. The sensitivity reach improves by roughly a factor of two in this case. We further show the effect of individually reducing the error on $M_\text{vir}$ (dashed purple line) and $c_\text{vir}$ (purple line) by 50\%. The reductions in the uncertainties provide only marginal improvements to the overall sensitivity, still far below the level of systematic uncertainty associated with extragalactic analyses in general.

\begin{figure}[t]
   \centering
   \includegraphics[width=0.45\textwidth]{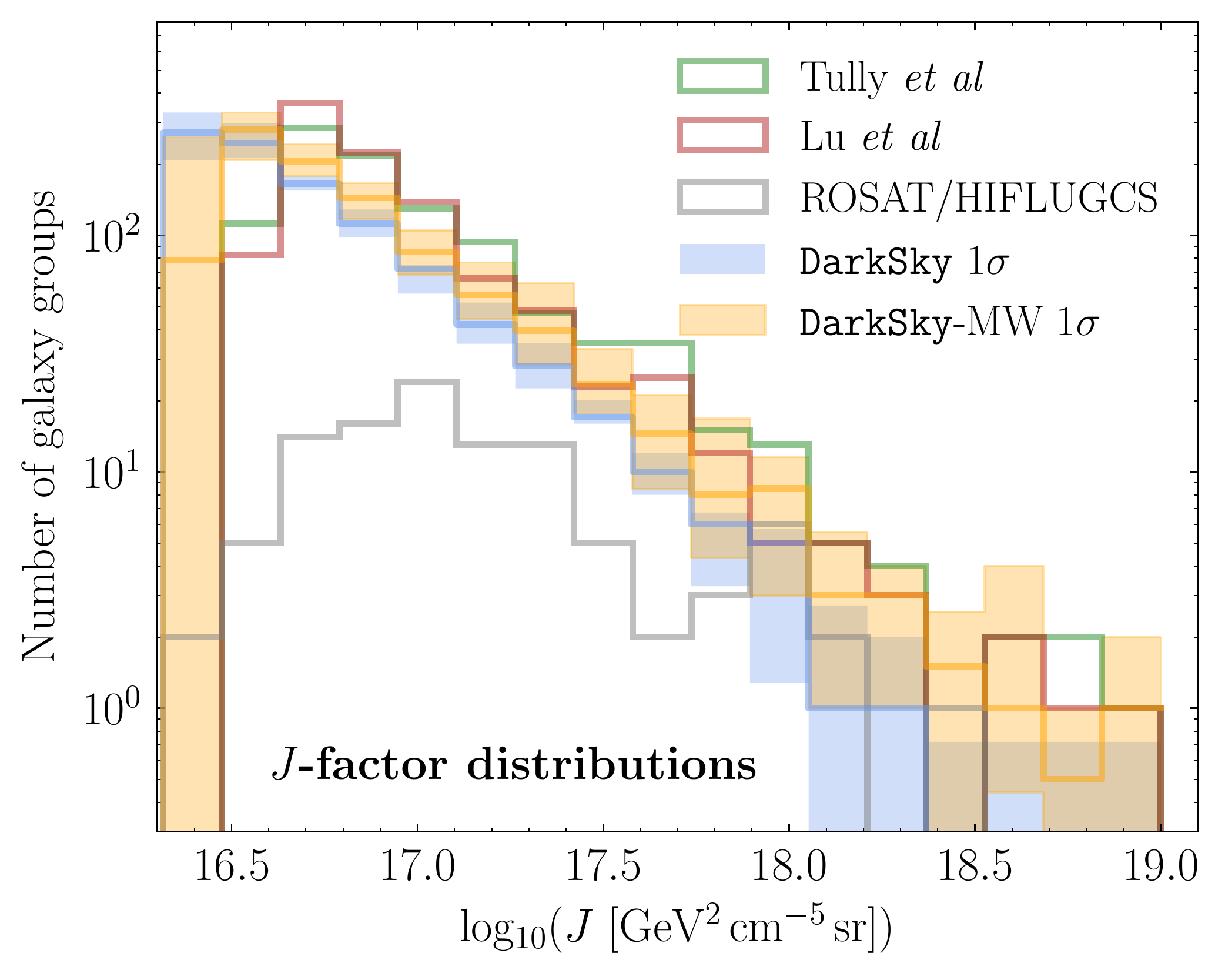} 
   \caption{Distribution of the top 1000 $J$-factors from the \texttt{DarkSky} catalog; the blue line  indicates the median distribution over nine random observer locations within the simulation box, with the blue band denoting the 68\% containment.  The orange line and band are the same, except for observers placed at ten random Milky~Way--like halos of mass  $\sim10^{12}$~M$_\odot$ in the box.  The distributions for the top 1000 $J$-factors in 2MRS galaxy-group catalogs are also shown; the green and red lines correspond to the Tully \emph{et al.}~\cite{Tully:2015opa,2017ApJ...843...16K} and the Lu \emph{et al.}~\cite{Lu:2016vmu} catalogs, respectively.  We also show the distribution (gray line) for the 106 galaxy clusters from the extended HIFGLUGCS catalog~\cite{Reiprich:2001zv,Chen:2007sz}, which is based on X-ray observations.  
   The $J$-factors for the real-world catalogs use the concentration model from Ref.~\cite{Correa:2015dva} and assume the \emph{Planck} 2015 cosmology~\cite{Ade:2015xua}, which is very similar to that used in \texttt{DarkSky}.
   }
   \label{fig:GroupCats}
\end{figure}

\begin{figure*}[htb]
   \centering
   \includegraphics[width=0.45\textwidth]{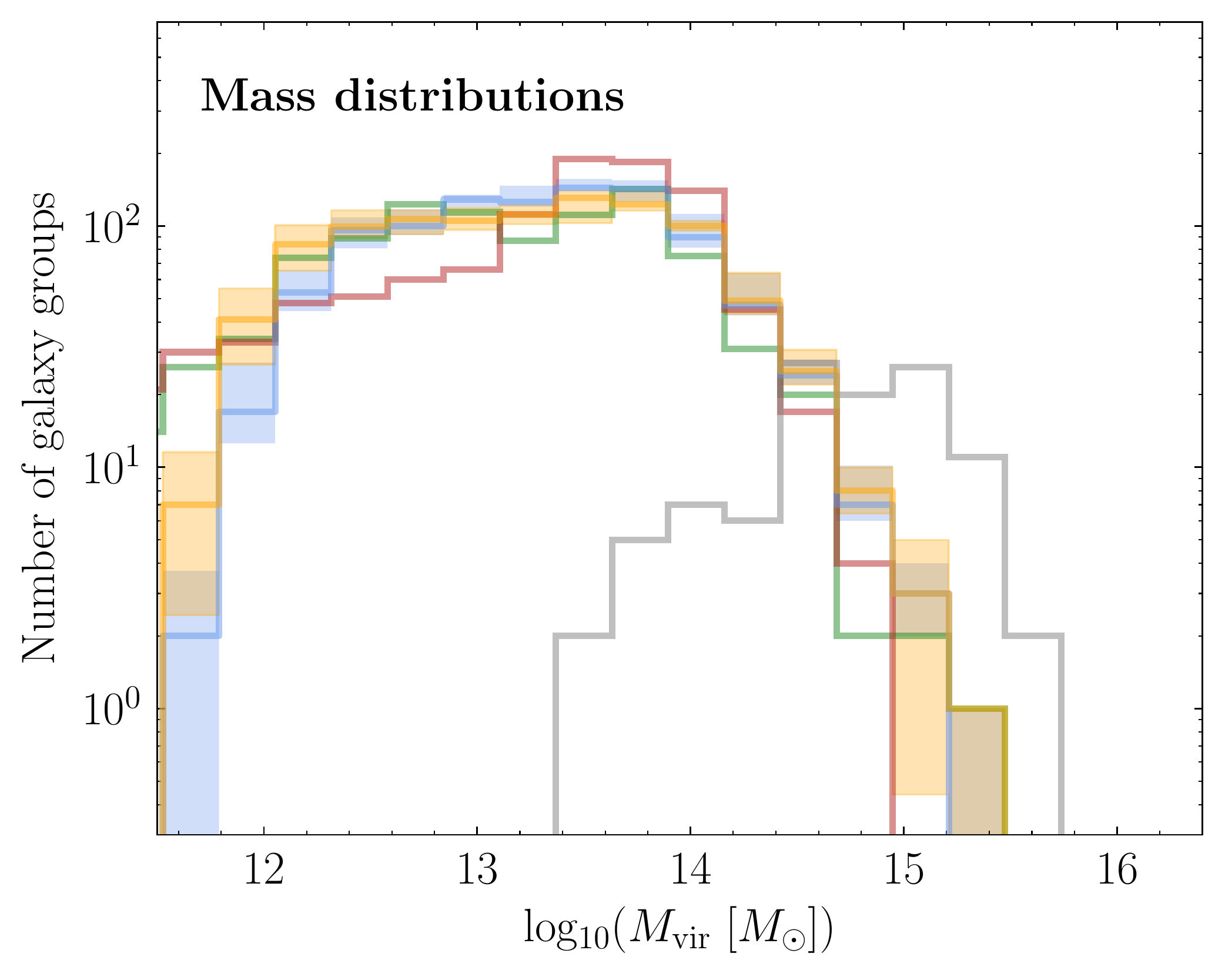}
     \includegraphics[width=0.45\textwidth]{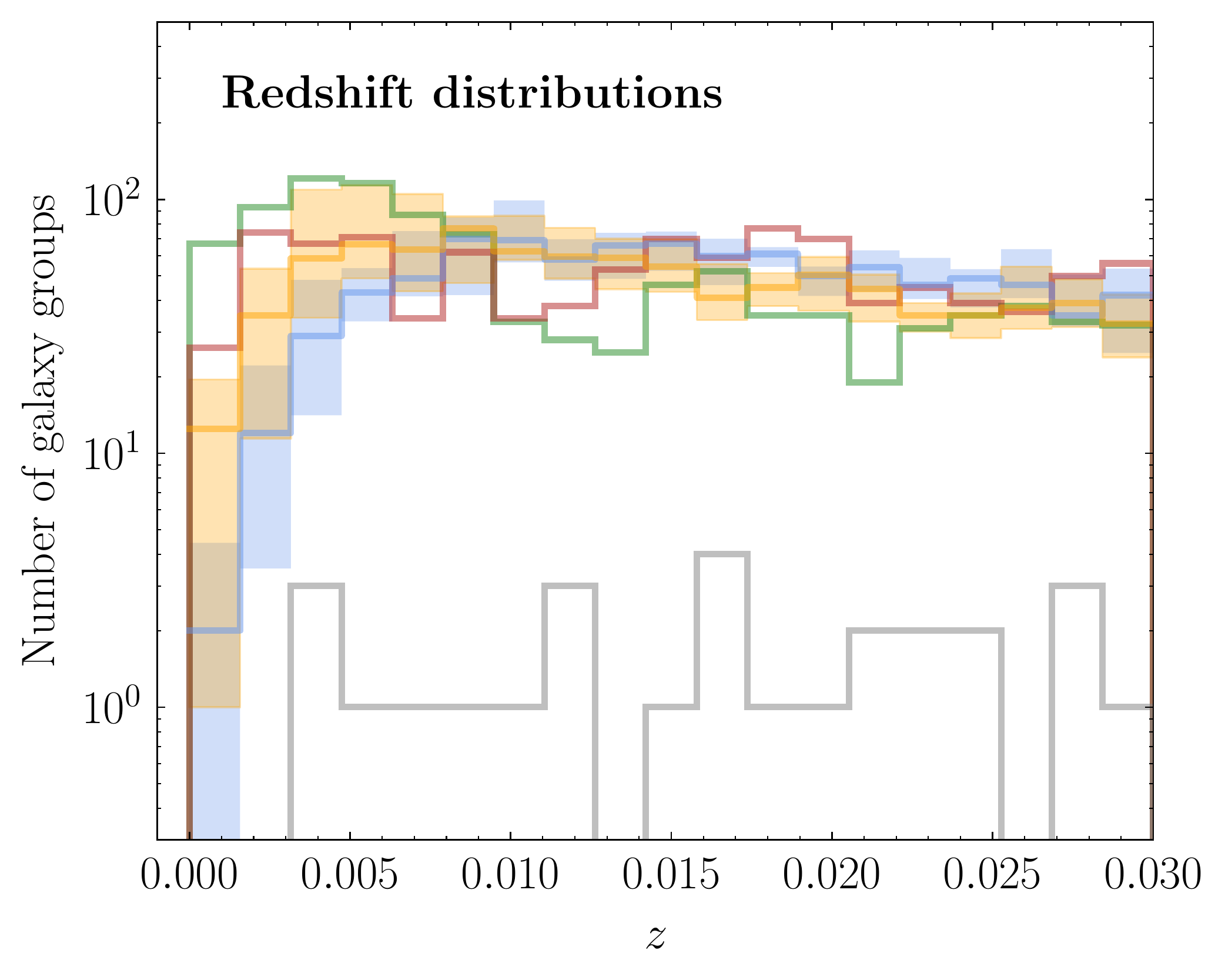}    
   \caption{Same as Fig.~\ref{fig:GroupCats}, except showing the mass function (left) and redshift distribution (right).  Note that the redshift distribution for the HIFLUGCS clusters extends above $z\sim0.03$, even though these are not shown in the right panel.     }
   \label{fig:GroupCatsMZ}
\end{figure*}
  
It is interesting to study how the limit scales with the number of halos, $N_h$, included in the stacking procedure.  This result is shown in Fig.~\ref{fig:DSelephants} for $m_\chi = 10,$ $100,$ and $10^4$~GeV, for four different observer locations in the simulation box.  The dashed red line indicates the median 95\% confidence limit. The red bands are the 2.5, 16, 84 and 97.5 percentiles on the limit, obtained from 100 MC realizations of the mock data.  We observe that the limit typically improves continuously for the first $\sim$10 halos.  As more halos are included in the stacking, the gains diminish.  For some sky locations, the limit simply remains flat; for others we see some marginal improvements in the bounds.  These results are consistent, within uncertainties, between the DM masses and the different sky locations of the observer. 

We emphasize that the scaling on $N_h$ can be very different on applicaton to real data, because the distribution of $J$-factors in the random \texttt{DarkSky} locations is not representative of our own environment in the Local Group and also some halos can have residuals that are not related to DM but rather to mismodeling or real cosmic-ray--induced emission from the galaxy groups.  The former point is demonstrated in Fig.~\ref{fig:GroupCats}, where we histogram the top 1000 $J$-factors associated with the baseline \texttt{DarkSky} analysis (blue line/band).  For  comparison, we also show the distributions corresponding to 2MRS  galaxy group catalogs, specifically the Tully~\emph{et al.}~\cite{Tully:2015opa,2017ApJ...843...16K} (green line) and the Lu 
~\emph{et al.}~\cite{Lu:2016vmu} (red line) catalogs.  We see that the distribution of $J$-factors for the 2MRS catalogs is skewed towards higher values compared to that from \texttt{DarkSky}.  (Note that the cut-off at low $J$-factors is artificial and is simply a result of including 1000 halos for each catalog.)

The differences in the $J$-factor distributions can be traced to the redshift distribution of the galaxy groups, as illustrated in Fig.~\ref{fig:GroupCatsMZ}.  We see specifically that the mass function of the top 1000 \texttt{DarkSky} halos in each of the random sky locations sampled is roughly consistent with that observed in the 2MRS catalogs.  In contrast, the actual catalogs have more groups at lower $z$ than observed in the random \texttt{DarkSky} locations.  

While a random location in the \texttt{DarkSky} box does not resemble our own Local Group, we can try to find specific locations in the simulation box that do.  Therefore, we place the observer at ten random Milky Way--like halos in the simulation box, which have a mass $\sim10^{12}$~M$_\odot$.  More specifically, we select halos with mass $\log_{10}(M/\mathrm{M}_\odot) \in [11.8,12.2]$ and at least 100 Mpc\,$h^{-1}$ from the box boundaries.  The distribution of the top 1000 $J$-factors is indicated by the orange line/band in Fig.~\ref{fig:GroupCats}, while the corresponding mass and redshift distributions are shown in Fig.~\ref{fig:GroupCatsMZ}.  We see that the redshift---and, consequently, $J$-factor---distributions approach the observations, though the correspondence is still not exact.  A more thorough study could be done assessing the likelihood that an observer in \texttt{DarkSky} is located at a position that closely resembles the Local Group.  However, as our primary goal here is to outline an analysis procedure that we can apply to actual data, we simply conclude that our own local Universe appears to be a richer environment compared to a random location within the \texttt{DarkSky} simulation box, which bodes well for studying the actual \emph{Fermi} data.

\subsection{Signal Recovery Tests}
\label{sec:siginj}
 
\begin{figure*}[t]
   \centering
   \includegraphics[width=0.35\textwidth]{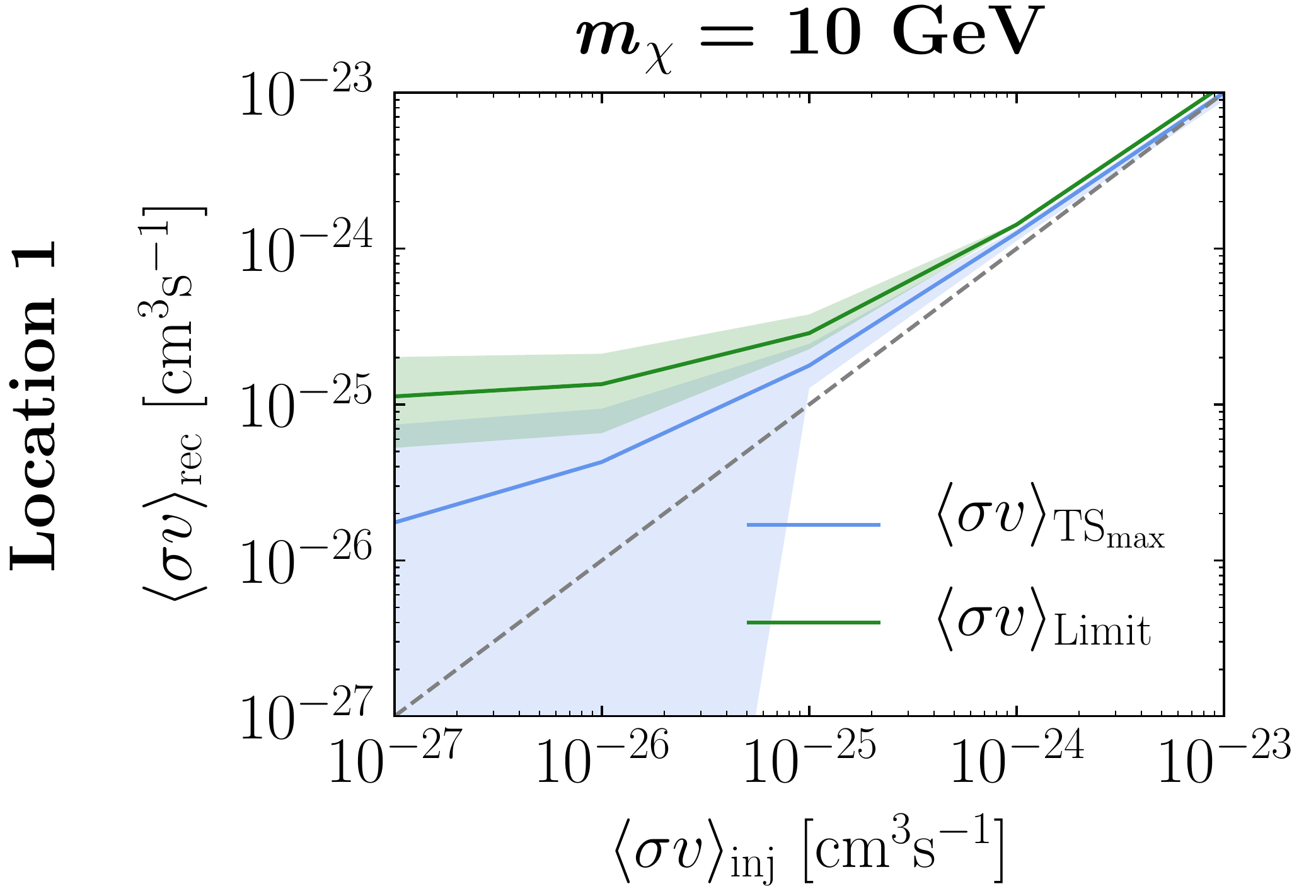} \hspace{0.01cm}
   \includegraphics[width=0.30\textwidth]{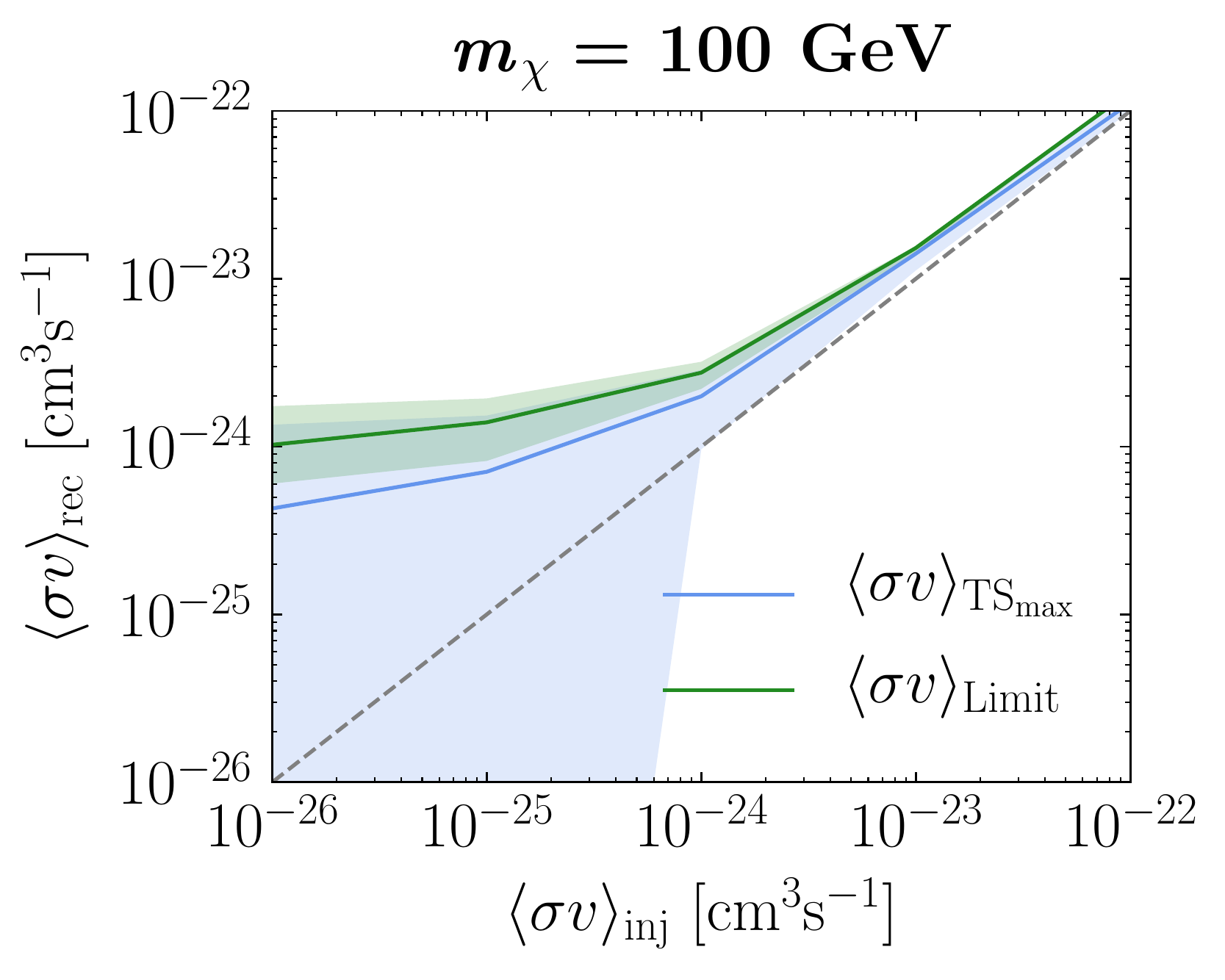} \hspace{0.01cm} 
   \includegraphics[width=0.30\textwidth]{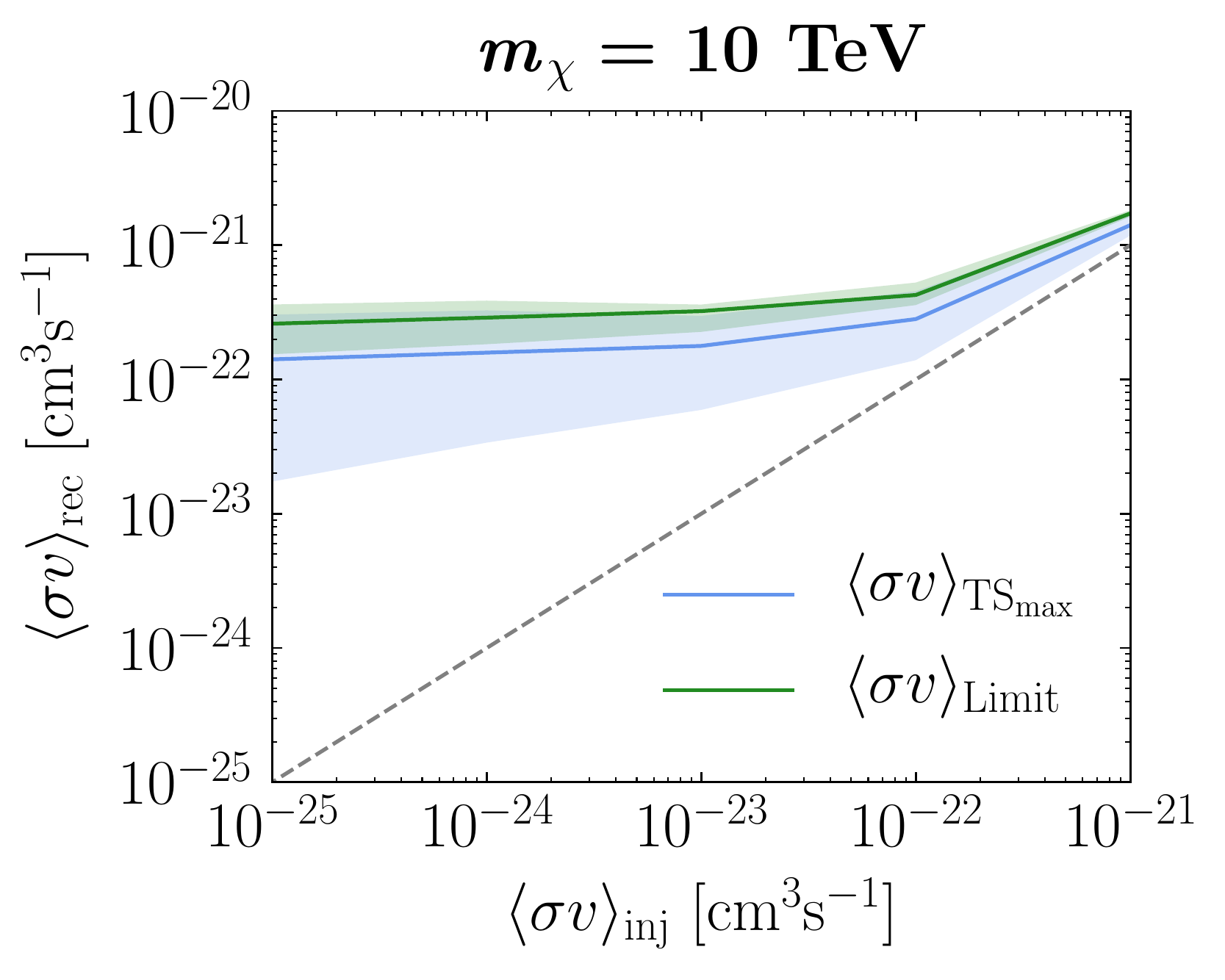} \\ \vspace{0.01cm}  
   \includegraphics[width=0.34\textwidth]{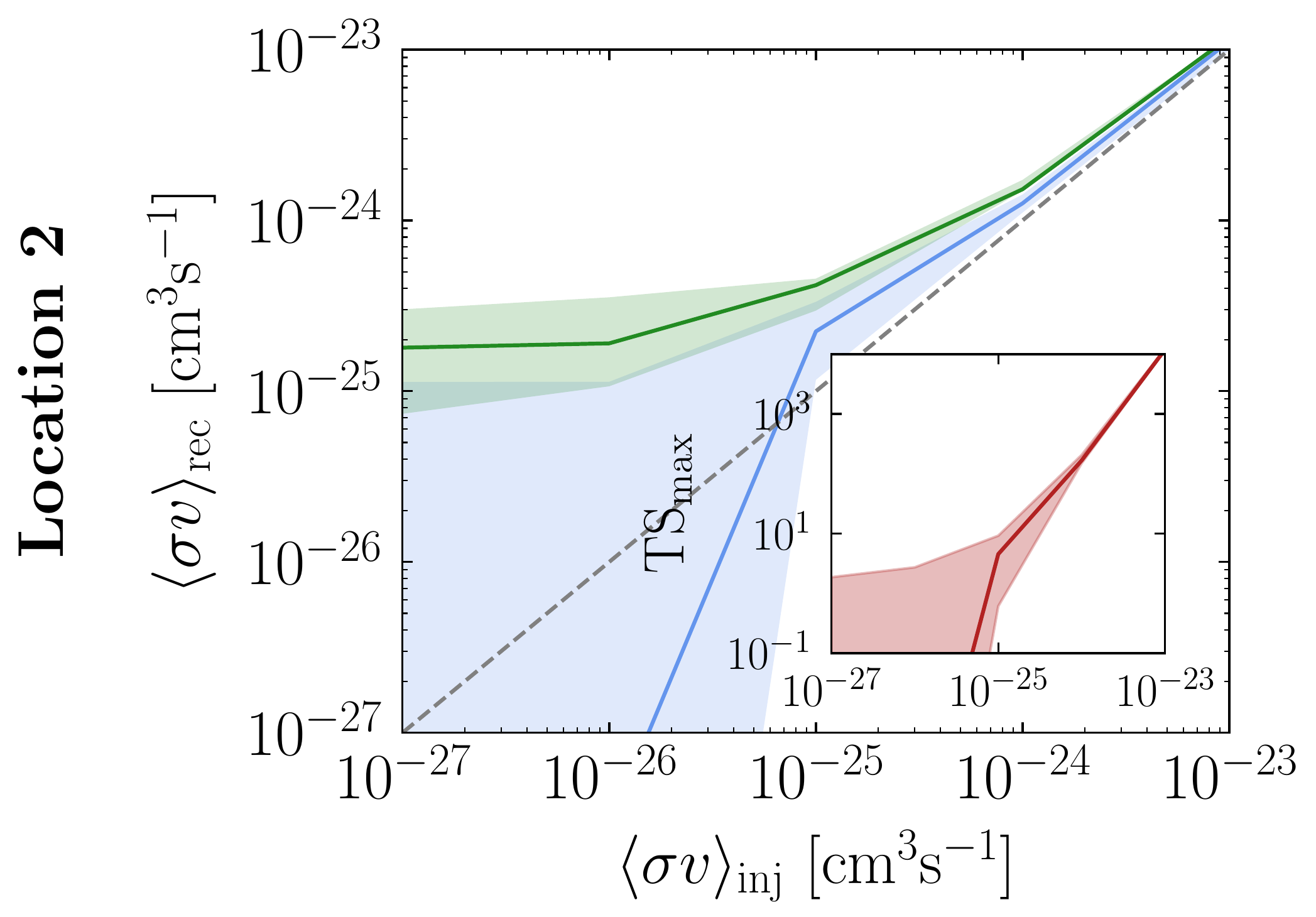} \hspace{0.01cm}
   \includegraphics[width=0.30\textwidth]{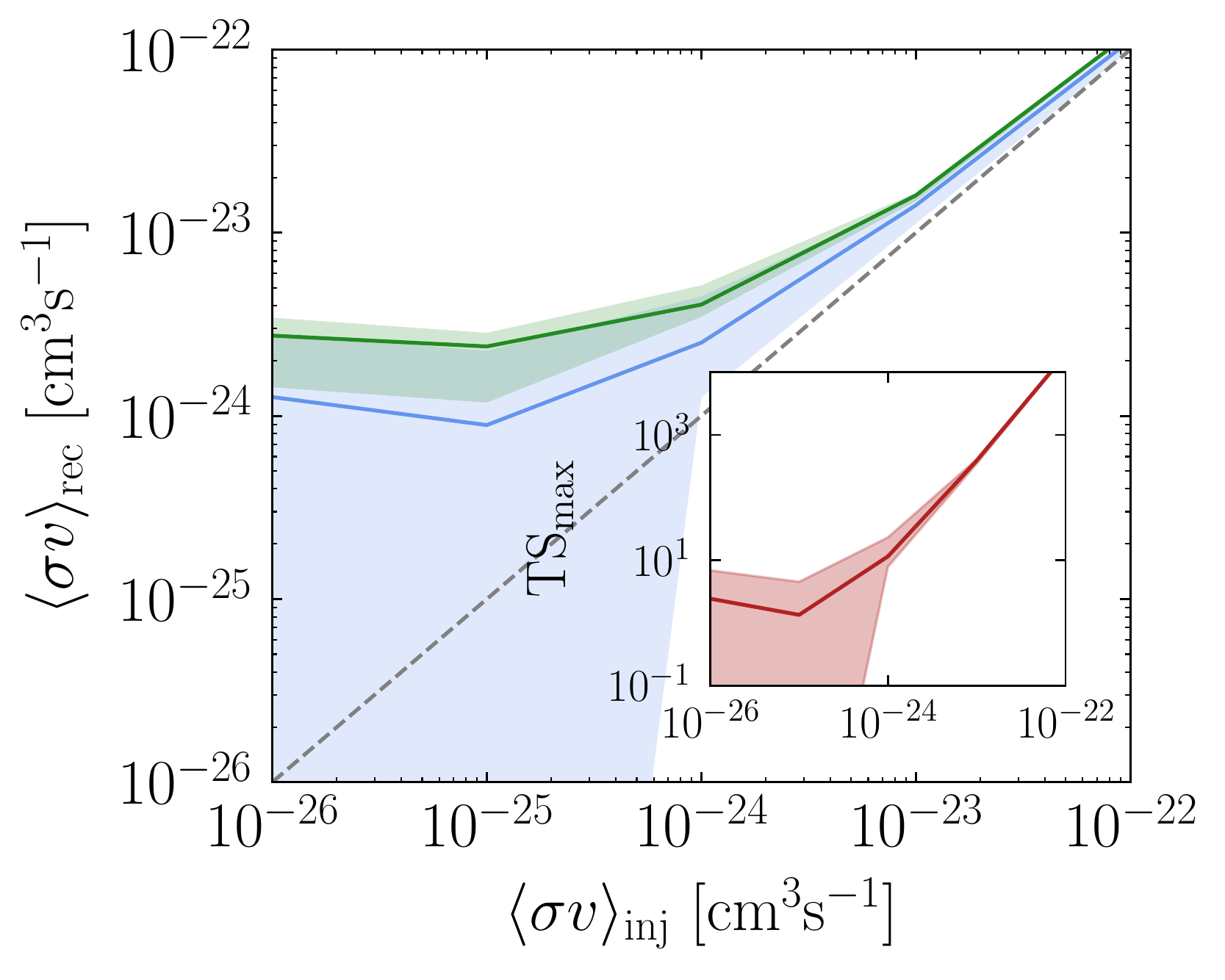} \hspace{0.01cm} 
   \includegraphics[width=0.30\textwidth]{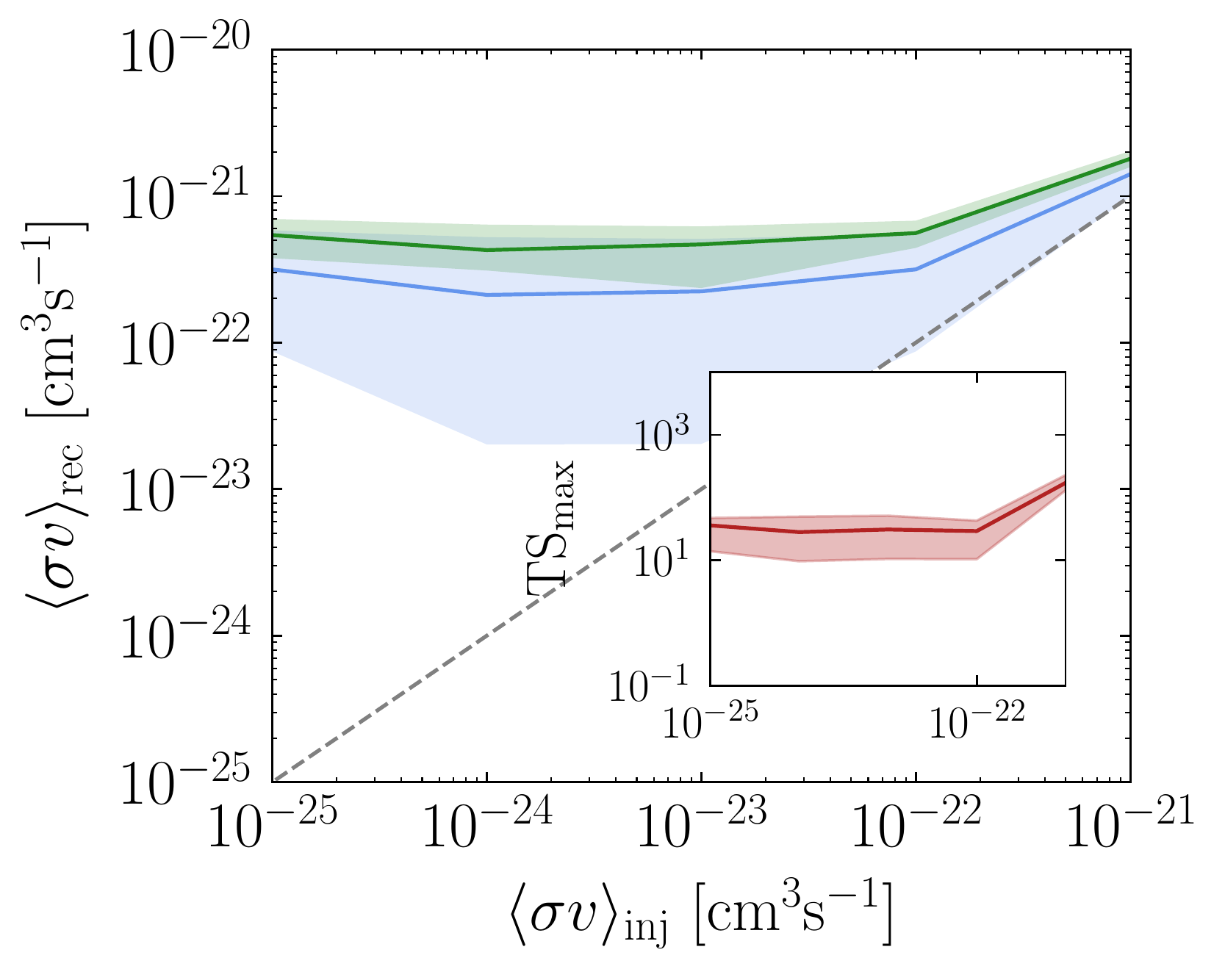}
   \caption{The results of injecting a DM signal with cross section $\langle \sigma v \rangle_\text{inj}$ into the mock data and studying the recovered cross section, $\langle \sigma v \rangle_\text{rec}$.  Each column shows the result for a different DM mass ($m_\chi = 10, 100, 10^4$~GeV), while each row shows a different observer location within the \texttt{DarkSky} simulation box.  The green line shows the 95\% confidence limit, with the green band denoting the 68\% containment region over twenty different Monte Carlo (MC) realizations of the mock data.  Critically, the limit never rules out an injected signal.  The blue line shows the median value of $\langle \sigma v\rangle_{\text{TS}_\text{max}}$, the cross section associated with the maximum test statistic (TS$_\text{max}$), over twenty MCs of the data. The blue band spans the median cross sections associated with TS$_\text{max} \pm 1$.  The maximum test statistic for each mass (with the band denoting the 68\% spread over MC realizations) is shown as an inset for each mass.  See Appendix~\ref{app:energyrange} for a more detailed discussion of these signal injection tests.  }
   \label{fig:DSinjsiglocs}
\end{figure*}

 It is critical that the halo selection criteria described in the previous section do not exclude a potential DM signal if one were present. To verify this, we have conducted extensive tests where we inject a signal into the mock data, pass it through the analysis pipeline and test our ability to accurately recover its cross section in the presence of the selection cuts.  Figure~\ref{fig:DSinjsiglocs} summarizes the results of the signal injection tests for two different observer locations in the \texttt{DarkSky} simulation box (top and bottom rows, respectively).  We inject a signal in the mock data that is associated with $b\bar{b}$ annihilation for three different masses ($m_\chi = 10, 100, 10^4$~GeV) that traces the DM annihilation flux map associated with \texttt{DarkSky}. The dashed line in each panel delineates where the injected cross section, $\langle \sigma v \rangle_\text{inj}$, matches the recovered cross section, $\langle \sigma v \rangle_\text{rec}$.  

The green line shows the 95\% one-sided limit on the cross section $\langle \sigma v \rangle_\text{rec}$ found using Eq.~\ref{eq:TSdef}, with a TS threshold corresponding to $\text{TS} = -2.71$.  The green band shows the 68\% containment region on this limit, constructed from twenty different MC realizations of the mock data set.  Importantly, the limit on $\langle \sigma v \rangle_\text{rec}$ roughly follows---but is slightly weaker than---the injected signal, up until the maximum sensitivity is reached and smaller cross sections can no longer be probed.  This behavior is generally consistent between the three DM masses tested and both sky locations.  We clearly see that the limit obtained by the statistical procedure never excludes an injected signal over the entire cross section range. 

Next, we consider the recovered cross section that is associated with the maximum test statistic, TS$_\text{max}$, in the total likelihood. The blue line in each panel of Fig.~\ref{fig:DSinjsiglocs} shows the median value of $\langle \sigma v\rangle_{\text{TS}_\text{max}}$ over 20 MCs of the mock data.  The blue band spans the median cross sections associated with $\text{TS}_\text{max}\pm1$.  The inset plots show the median and 68\% containment region for TS$_\text{max}$ as a function of the injected cross section.  The maximum test statistic is an indicator for the significance of the DM model and as such the $\langle \sigma v\rangle_{\text{TS}_\text{max}}$ distributions are only influenced by the data at high injected cross sections where TS$_\text{max}$ has begun to increase.  At lower injected cross sections, the distributions for $\langle \sigma v\rangle_{\text{TS}_\text{max}}$ are not meaningful.  

Two issues are visible in Fig.~\ref{fig:DSinjsiglocs}: (i) at high injected cross sections, the best-fit recovered cross sections are systematically around 1$\sigma$ too high, and (ii) at high and low DM masses and near-zero injected cross sections, the distribution of TS$_\text{max}$ deviates from the chi-square distribution.  The first issue stems from the way we model the $J$-factor contribution to the likelihood, while the second arises from the approximations we make to perform the profile likelihood in a computationally efficient manner.  Appendix~\ref{app:energyrange} discusses these issues and ways they may be mitigated, in more detail.

\section{Conclusions}
\label{sec:conclusions}

In this paper, we introduced a procedure to build a full-sky map of extragalactic DM targets based on galaxy surveys and demonstrated this methodology using the \texttt{DarkSky} cosmological simulation.   Starting from the galaxies in the \texttt{DarkSky} catalog, we inferred the properties of their respective DM halos using the galaxy-halo connection.  In so doing, we identified the halos that are the brightest sources of extragalactic DM annihilation and which act as the best annihilation targets.  This procedure allows us to account for the fact that not all galaxy groups are expected to be bright DM emitters; the most massive, concentrated, and/or most nearby galaxies dominate the signals.  By building a map of extragalactic DM targets, we can focus our search for DM annihilation on the most relevant regions of sky.  This philosophy contrasts with that of cross-correlation studies, which treat all galaxies as equally good targets for DM.   

With a list of extragalactic DM halos in hand, as well as their inferred $J$-factors, we performed a stacked analysis to search for gamma-ray signatures of DM annihilation in mock data.  We described the likelihood procedure for the stacking analysis in detail.  There are two clear advantages to this approach over, say,  a full-sky template study.\footnote{ Appendix~\ref{app:inflimitimpact} includes a more detailed discussion of using full-sky DM annihilation flux templates.}  First, focusing on smaller regions around each halo significantly reduces the sensitivity to mis-modeling of the foregrounds.  Second, uncertainties on the predicted DM annihilation flux can be straightforwardly included in the likelihood function.  In particular, we outlined how uncertainties in the $J$-factors, which arise from the determination of the virial mass and concentration, are marginalized over in the analysis. 

We presented limits on the DM annihilation cross section for mock data and, most importantly, demonstrated that the analysis procedure robustly recovers injected signals.  We found that the sensitivity improves by nearly two orders of magnitude when the structure of extragalactic DM emission on the sky is accounted for, rather than simply assuming an isotropic distribution.  Typically, the limit is dominated by the brightest $\mathcal{O}(10)$ halos in the stacking, though this varies depending on the location in the simulation box.  The $J$-factor distribution of nearby groups in our own Galaxy differs from the random locations sampled in the \texttt{DarkSky} box, which can change the number of halos that dominate the limit.  In actuality, one would want to continue adding halos to the analysis---ranked starting from the brightest $J$-factors---until the gains in the limit are observed to level off.

One advantage of using the \texttt{DarkSky} simulation in this initial study is that the truth information for all the halos is known.  We can therefore study how the DM limits improve when the virial mass and  concentration of the halos are known precisely.  For this ideal scenario, we find that that the limits improve by roughly 50\% over those obtained by marginalizing over uncertainties.  This suggests that a concrete way to improve the bounds on DM annihilation is to reduce the uncertainties on $M_\text{vir}$ and $c_\text{vir}$ for the brightest halos in the catalog.

The substructure boost factor remains one of the most difficult systematics to handle.  In this work, we use recent boost-factor models that account for tidal stripping of subhalos.  This boost factor changes the limit by an $\mathcal{O}(1)$ factor, which is more conservative than other models sometimes used in extragalactic DM studies.  While the boost-factor enhancement is fairly modest, it is still the dominant systematic uncertainty over the halo mass and  concentration.   

The analysis outlined in this paper can be repeated on \emph{Fermi} data using published galaxy group catalogs.  In particular, the Tully \emph{et al.} catalogs~\cite{Tully:2015opa,2017ApJ...843...16K} and the Lu \emph{et al.} catalog~\cite{Lu:2016vmu} provide a map of the galaxy groups in the local Universe within $z\lesssim0.03$.  Both catalogs are based primarily on 2MRS, but use different clustering algorithms and halo mass determinations.  
Taken together, they provide a way to estimate the systematic uncertainties associated with the galaxy to halo mapping procedure.  Previous cluster studies on $\emph{Fermi}$ data~\cite{Ackermann:2010rg, Ando:2012vu,Ackermann:2013iaq,Anderson:2015dpc,Liang:2016pvm} used the extended HIghest X-ray FLUx Galaxy Cluster Sample (HIFLUGCS)~\cite{Reiprich:2001zv,Chen:2007sz}, which includes 106 of the brightest clusters observed in X-ray with the ROSAT all-sky survey.  These clusters cover redshifts from $0.0037 \lesssim z\lesssim 0.2$; the distribution of their $J$-factors, masses, and redshifts are shown in Fig.~\ref{fig:GroupCats} and~\ref{fig:GroupCatsMZ}.  In general, the 2MRS catalogs provide a larger number of groups that should be brighter in DM annihilation flux, so we expect a corresponding improvement in the sensitivity to annihilation signatures.  

The recent advancement of galaxy catalogs based on 2MRS and other nearby group catalogs allows us for the first time to map out the most important extragalactic DM targets in the nearby Universe.  This, in turn, enables us to perform a search that focuses on regions of sky where we expect the DM signals to be the brightest outside the Local Group.  We present the complete results of such an analysis, as applied to data, in our companion paper~\cite{companion}.

\section*{Acknowledgments}
We thank S.~Ando, N.~Bahcall, R.~Bartels, J.~Beacom, P.~Behroozi, F.~Calore, W.~Coulton, A.~Drlica-Wagner, D.~Hooper, S.~Horiuchi, A.~Kravtsov, T.~Linden, Y.~Mao, K.~Murase, L.~Necib, J.~Ostriker, A.~Peter, T.~Slatyer, B.~Tully, C.~Weniger, and S.~Zimmer for helpful conversations. This research made use of the \texttt{DarkSky} Simulations, which were produced using an INCITE 2014 allocation on the Oak Ridge Leadership Computing Facility at Oak Ridge National Laboratory.  We thank S.~Skillman, M.~Warren, M.~Turk, and the \texttt{DarkSky} collaboration for their efforts in creating these simulations and for providing access to them, and Y. Mao for providing the galaxy catalogs used herein.
 This research made use of the \texttt{Astropy}~\cite{2013A&A...558A..33A}, \texttt{IPython}~\cite{PER-GRA:2007}, \texttt{Minuit}~\cite{James:1975dr} and \texttt{NPTFit}~\cite{Mishra-Sharma:2016gis}  software packages.
ML is supported by the DOE under contract DESC0007968, the Alfred P.~Sloan Foundation and the Cottrell Scholar Program through the Research Corporation for Science Advancement.  NLR is supported by the DOE under contracts DESC00012567 and DESC0013999.  BRS is supported by a Pappalardo Fellowship in Physics at MIT. RHW received support from the U.S. Department of Energy under contract number DE-AC02-76SF00515. This work was performed in part at Aspen Center for Physics, which is supported by NSF grant PHY-1607611.

\newpage
\appendix
\setcounter{equation}{0}
\setcounter{figure}{0}
\setcounter{table}{0}
\setcounter{section}{0}
\makeatletter
\renewcommand{\theequation}{A\arabic{equation}}
\renewcommand{\thefigure}{A\arabic{figure}}
\renewcommand{\thetable}{A\arabic{table}}
\newcommand\ptwiddle[1]{\mathord{\mathop{#1}\limits^{\scriptscriptstyle(\sim)}}}

\section{$J$- and $D$-factors for Extragalactic Sources}
\label{app:JDrelations}

In this Appendix, we derive the $J$-factor relations used in the main text. We also derive the corresponding $D$-factor relations, which apply to the case of decaying DM.  Although we do not make use of the decay results in the main text, we include these results for completeness because much of our main analysis can be extended to the decaying case.
This Appendix is broken into three subsections. In the first of these, we detail the units and conventions used in our definition of the $J$- and $D$-factors.  After this, we derive an approximate form of the astrophysics factors for different DM density profiles and discuss the accuracy of the approximations made.  We conclude with a discussion of error propagation in the $J$-factors.  Note that several of the details presented in these appendices have been discussed elsewhere, see \emph{e.g.}, Ref.~\cite{Abdo:2010ex,Charbonnier:2011ft,Charbonnier:2012gf,Evans:2016xwx}. 

\subsection{Units and Conventions}

\subsubsection{Dark Matter Flux}

We begin by carefully outlining the units associated with the $J$- and $D$-factors. 
The flux, $\Phi$, associated with either DM annihilation or decay factorizes into two parts:
\begin{equation}\begin{aligned}
\frac{d\Phi^{\rm ann.}}{dE_{\gamma}} &= \frac{d\Phi_\text{pp}^{\rm ann.}}{dE_{\gamma}}\times J \, , \\
\frac{d\Phi^{\rm dec.}}{dE_{\gamma}} &= \frac{d\Phi_\text{pp}^{\rm dec.}}{dE_{\gamma}}\times D \, ,
\end{aligned}\end{equation}
where $E_\gamma$ is the photon energy and the `ann.'~(`dec.') superscripts denote annihilation~(decay).  
The particle physics factors are given by:
\begin{equation}\begin{aligned}
\frac{d\Phi_\text{pp}^{\rm ann.}}{dE_{\gamma}} &=\frac{\langle\sigma v\rangle}{8\pi m_{\chi}^{2}}\sum_i \text{Br}_{i}\, \frac{dN_{i}}{dE_{\gamma}}\,, \\
\frac{d\Phi_\text{pp}^{\rm dec.}}{dE_{\gamma}} &=\frac{1}{4\pi m_{\chi} \tau}\sum_i \text{Br}_{i}\, \frac{dN_{i}}{dE_{\gamma}}\,,
\end{aligned}\end{equation}
where $\langle \sigma v \rangle$ is the velocity-averaged annihilation cross section, $m_\chi$ is the DM mass, Br$_i$ is the branching fraction into the $i^\text{th}$ channel, $dN_i/dE_\gamma$ is the photon energy distribution associated with this channel, and $\tau$ is the DM lifetime.  The annihilation factor assumes that the DM is its own antiparticle; if this were not the case, and assuming no asymmetry in the dark sector, then the factor would be half as large.  The particle physics factors carry the following dimensions:
\begin{equation}\begin{aligned}
\left[ \frac{d\Phi_\text{pp}^{\rm ann.}}{dE_{\gamma}} \right] &= {\rm counts} \cdot {\rm cm}^3 \cdot {\rm s}^{-1} \cdot {\rm GeV}^{-3} \cdot {\rm sr}^{-1} \,, \\
\left[ \frac{d\Phi_\text{pp}^{\rm dec.}}{dE_{\gamma}} \right] &= {\rm counts} \cdot {\rm s}^{-1} \cdot {\rm GeV}^{-2} \cdot {\rm sr}^{-1}\,,
\label{eq:ppunits}
\end{aligned}\end{equation}
where `counts' refers to the number of gamma-rays produced in the interaction and the ${\rm sr}^{-1}$ is associated with the $1/4\pi$ in the particle physics factors.  Note that some references include this $4\pi$ in the definition of the $J$- or $D$-factors, but this is not the convention that we follow here.

The $J$- and $D$-factors are defined as follows:
\begin{equation}\begin{aligned}
J &= \left(1+b_\text{sh}[M_\text{vir}]\right)\,\int ds \, d\Omega \,\rho_{\rm DM}^2(s,\Omega) \,, \\
D &= \int ds\, d\Omega\, \rho_{\rm DM}(s,\Omega)\,,
\label{eq:JDdef}
\end{aligned}\end{equation}
where $b_\text{sh}[M_\text{vir}]$ is the subhalo boost factor.  The $J$- and $D$-factors carry the following units:
\begin{equation}\begin{aligned}
\left[ J \right] &= {\rm GeV}^2 \cdot {\rm cm}^{-5} \cdot {\rm sr}\,, \\
\left[ D \right] &= {\rm GeV} \cdot {\rm cm}^{-2} \cdot {\rm sr}\,.
\end{aligned}\end{equation}
Combining these with Eq.~\ref{eq:ppunits}, we find that 
\begin{equation}
\left[ \frac{d\Phi}{dE_{\gamma}} \right] = {\rm counts} \cdot {\rm cm}^{-2} \cdot {\rm s}^{-1} \cdot {\rm GeV}^{-1}\,
\end{equation}
for both the annihilation and decay case.  This means that $\Phi$ is given in units of counts per experimental effective area [${\rm cm}^2$] per experimental run time [${\rm s}$].  
In this work, we study extragalactic objects with small angular extent.  So long as each object is centered on the region-of-interest (ROI), we expect that all of its flux will be contained within the ROI as well.  This means that the photon counts obtained by integrating Eq.~\ref{eq:JDdef} over the entire sky corresponds to the total counts expected from that object in the ROI.  The situation is different when treating objects with a large angular extent that exceeds the size of the ROI---\emph{e.g.}, when looking for emission from the halo of the Milky Way.  In such cases, it is more common to divide the $J$- and $D$-factors by the solid angle of the ROI ($\Delta \Omega$) such that both they, and consequently $\Phi$, are averages rather than totals.  

\subsubsection{Halo Mass and Concentration}

We briefly comment here on different mass and concentration definitions (virial and 200) as relevant to our analysis.   Boost-factor models, concentration-mass relations, and masses are often specified in terms of 200 quantities, which must be converted to virial ones. In order to do this, we use the fact that
\begin{equation}\begin{aligned}
\frac{\rho_\text{s}}{\rho_c} \equiv \delta_\mathrm{c} = \frac{\Delta_\text{c}}{3}\frac{c^3}{\log{(1+c)}-c/(1+c)}
\label{eq:CritOverdens}
\end{aligned}\end{equation}
for the NFW profile~\cite{Navarro:1995iw}, where $\rho_s$ is the normalization of the density profile, $\rho_c$ is the critical density, $c$ is the concentration parameter, and $\delta_\mathrm{c}$ is the critical overdensity.  For virial quantities, $\Delta_c(z) = 18\pi^2 +82x-39x^2$ with $x = \Omega_{m}(1+z)^3/[\Omega_{m}(1+z)^3 + \Omega_{\Lambda}]-1$ in accordance with Ref.~\cite{Bryan:1997dn}, while for 200 quantities, $\Delta_c = 200$.  Therefore, Eq.~\ref{eq:CritOverdens} can be equated between the 200 and virial quantities and solved numerically to convert between definitions of the concentration.

For different mass definitions, we have 
\begin{equation}\begin{aligned}
\frac{M_{200}}{M_\text{vir}} = \left(\frac{c_{200}[M_{200}]}{c_\text{vir}[M_\text{vir}]}\right)^3\frac{200}{\Delta_\text{c}} \, ,
\label{eq:MassConvert}
\end{aligned}\end{equation}
where the concentration definitions on the right-hand side depend on $M_{200}$ and $M_\text{vir}$ and may have to be converted between each other and we have suppressed the redshift dependence for clarity.  Solving this numerically, we can convert between the two mass definitions.

\subsection{Approximate $J$- and $D$-factors}

For an extragalactic DM halo, the astrophysical factors in Eq.~\ref{eq:JDdef} can be approximated as:
\begin{equation}\begin{aligned}
J &\approx \left(1+b_\text{sh}[M_\text{vir}]\right)\, \frac{1}{d_c^2[z]} \int_V dV' \rho_{\rm DM}^2(r') \,, \\
D &\approx \frac{1}{d_c^2[z]} \int_V dV' \rho_{\rm DM}(r')\,,
\vspace{0.5in}
\label{eq:JDdapprox}
\end{aligned}\end{equation}
where the integrals are performed in a coordinate system centered on the halo, and $d_c[z]$ is the comoving distance, which is a function of redshift for a given cosmology.  The aim of this subsection is to derive Eq.~\ref{eq:JDdapprox} from Eq.~\ref{eq:JDdef} and to quantify the error associated with this approximation. 

To handle the $J$- and $D$-factors simultaneously, we consider the following integral over all space:
\begin{equation}
\int ds \, d\Omega \,\rho_{\rm DM}^n(s,\Omega)\,,
\end{equation}
with $n \geq 1$. Here, $s$ is playing the role of a radius in a spherical coordinate system centered on the Earth.  Therefore, we can rewrite the measure as
\begin{equation}
\int s^2\, ds \, d\Omega \,\, \frac{\rho_{\rm DM}^n(s,\Omega)}{s^2} = \int dV\, \frac{\rho_{\rm DM}^n(s,\Omega)}{s^2}\,.
\end{equation}

Next, we transform to a coordinate system (denoted by primed quantities) that is centered at the origin of the halo described by $\rho_{\rm DM}$.  Because this change of coordinates is only a linear translation, it does not induce a Jacobian and $dV = dV'$.  Assuming that the Earth is located at a position $\mathbf{r}$ from the halo center and the DM interaction occurs at position $\mathbf{r'}$, then $s = | \mathbf{r} - \mathbf{r}'| $ and
\begin{equation}
\int dV\, \frac{\rho_{\rm DM}^n(s,\Omega)}{s^2} = \int dV'\, \frac{\rho_{\rm DM}^n(r',\Omega')}{r^{\prime 2} - 2 d_c r' \cos \theta' + d_c^2}\,,
\label{eq:haloframe}
\end{equation}
where we take $|\mathbf{r}| = d_c$ and $\mathbf{r} \cdot \mathbf{r}' = d_c\, r' \, \cos\theta'$.

Eq.~\ref{eq:haloframe} can be simplified by taking advantage of several properties of the halo density.  First, it is spherically symmetric about the origin of the primed coordinate system.  Second, it only has finite support in $r'$.  In particular, it does not make sense to integrate the object beyond the virial radius, $r_{\rm vir}$. This allows us to rewrite the integral as follows:
\vspace{0.1in}
\begin{widetext}
\begin{eqnarray}
\label{eq:multiline}
\int dV'\, \frac{\rho_{\rm DM}^n(r',\Omega')}{r^{\prime 2} - 2 d_c r' \cos \theta' + d_c^2}
&=& \int_0^{r_{\rm vir}} dr'\, \int d\Omega'\, \frac{\rho_{\rm DM}^n(r')}{r^{\prime 2} - 2 d_c r' \cos \theta' + d_c^2} \\
&=& \frac{2 \pi}{d_c^2} \, \int_0^{r_{\rm vir}} dr'\, \rho_{\rm DM}^n(r')  \int_0^{\pi} d\theta'\, \frac{\sin \theta'}{1 - 2 (r'/d_c) \cos \theta' + (r'/d_c)^2} \notag \\
&=& \frac{2\pi}{d_c^2} \, \int_0^{r_{\rm vir}} dr'\, \frac{ \rho_{\rm DM}^n(r')}{2\,(r'/d_c)} \ln \left[ \frac{((r'/d_c)+1)^2}{((r'/d_c)-1)^2} \right]\,. \notag
\end{eqnarray}
\end{widetext}
For extragalactic objects, $d_c \gg r_{\rm vir} \geq r'$.  As a result, we can take advantage of the following  expansion:
\begin{equation}
\frac{1}{2x} \ln \left[ \frac{(x+1)^2}{(x-1)^2} \right]  = 2 \left[ 1 + \frac{1}{3} x^2 + \mathcal{O} \left(x^4\right) \right] \, ,
\label{eq:logexpand}
\end{equation}
where $x= r'/d_c$.  It follows that the leading-order approximation to Eq.~\ref{eq:multiline} is \begin{equation}\begin{aligned}
\int ds \, d\Omega \,\rho_{\rm DM}^n(s,\Omega) = \frac{1}{d_c^2} \int dV' \rho_{\rm DM}^n(r') \,,
\end{aligned}\end{equation}
which when inserted into Eq.~\ref{eq:JDdef} gives Eq.~\ref{eq:JDdapprox}, as claimed. 

We can calculate the size of the neglected terms in Eq.~\ref{eq:logexpand} to quantify the accuracy of this approximation.  We take the parameters of the halo with the largest $J$-factor in the catalog to estimate the largest error possible amongst the \texttt{DarkSky} halos.  For this halo, the fractional correction to the $J$-factor of the first neglected term in the expansion is $\mathcal{O}(10^{-5})$ for either an NFW or Burkert profile (described below), whilst for the $D$-factor it is $\mathcal{O}(10^{-4})$. These values are significantly smaller than the other sources of uncertainty present in estimating these quantities and so we conclude that the approximations in Eq.~\ref{eq:JDdapprox} are sufficient for our purposes.

\subsection{Analytic Relations}

Starting from the approximate forms given in Eq.~\ref{eq:JDdapprox} and specifying a DM density profile $\rho_{\rm DM}$, the $J$- and $D$-factors can often be determined exactly.   We will now demonstrate that the final results only depend on the distance, mass, and concentration of the halo---for a given substructure boost model and cosmology.

As a starting point, consider the NFW profile:
\begin{equation}
\rho_{\rm NFW}(r) = \frac{\rho_s}{r/r_s(1+r/r_s)^2}\,.
\end{equation}
The parameter $r_s$ is the scale radius and dictates how sharply peaked the core of the DM distribution is.
Starting from this distribution, the volume integral in the $J$-factor evaluates to
\begin{equation}\begin{aligned}
\int dV'\, \rho_{\rm NFW}^2(r') &= 4\pi \rho_s^2 r_s^2 \int_0^{r_{\rm vir}} \frac{dr'}{(1+r'/r_s)^4} \\
&= \frac{4\pi}{3} \frac{\rho_s^2 r_{\rm vir}^3}{c_{\rm vir}^3} \left[ 1 - \frac{1}{(1+c_{\rm vir})^3} \right]\,,
\label{eq:NFWVolumeInt}
\end{aligned}\end{equation}
where $c_{\rm vir} = r_{\rm vir}/r_s$ is the virial concentration.  To remove the normalization factor $\rho_s$ from this equation, we can write the virial mass of the halo as
\begin{equation}\begin{aligned}
M_{\rm vir} &\equiv \int dV'\, \rho_{\rm NFW}(r') \\
&= 4\pi \rho_s \frac{r_{\rm vir}^3}{c_{\rm vir}^3} \left[ \ln \left( 1 + c_{\rm vir} \right) - \frac{c_{\rm vir}}{1+c_{\rm vir}} \right]\,,
\end{aligned}\end{equation}
which, when combined with Eq.~\ref{eq:NFWVolumeInt}, gives
\begin{equation}\begin{aligned}
\int dV'\, \rho_{\rm NFW}^2(r)
=\,&\frac{M_{\rm vir}^2 c_{\rm vir}^3}{12\pi r_{\rm vir}^3} \left[ 1 - \frac{1}{(1+c_{\rm vir})^3} \right] \\
\times &\left[ \ln \left( 1 + c_{\rm vir} \right) - \frac{c_{\rm vir}}{1+c_{\rm vir}} \right]^{-2}\,.
\end{aligned}\end{equation}
Stopping here, we would conclude that the $J$-factor scales as $M_{\rm vir}^2$.  However, for a given $M_{\rm vir}$ and cosmology, $r_{\rm vir}$ is not an independent parameter. Using the results of Ref.~\cite{Bryan:1997dn}, we can write:
\begin{equation}
\frac{3M_{\rm vir}}{4\pi r_{\rm vir}^3} = \rho_c \Delta_c[z]\,,
\end{equation}
where $\rho_c$ is the critical density and 
\begin{equation}\begin{aligned}
\Delta_c[z] &\equiv  18\pi^2 + 82\,x[z] - 39\,x[z]^2\,, \\
x[z] &\equiv \frac{\Omega_m \left( 1 + z \right)^3}{\Omega_m \left( 1 + z \right)^3 + \Omega_{\Lambda}} - 1\,.
\end{aligned}\end{equation}
This relation can then be used to remove $M_{\rm vir}/r_{\rm vir}^3$ from the volume integral and we conclude that
\begin{align}
J_{\rm NFW} \approx &\left(1+b_\text{sh}[M_\text{vir}]\right) \frac{M_{\rm vir} c_{\rm vir}^3 \rho_c \Delta_c[z]}{9 d_c^2[z]} \label{eq:JNFWfull} \\
\times &\left[ 1 - \frac{1}{(1+c_{\rm vir})^3} \right] \left[ \ln \left( 1 + c_{\rm vir} \right) - \frac{c_{\rm vir}}{1+c_{\rm vir}} \right]^{-2}\,. \notag
\end{align}
We see the additional mass dimension required from the fact this scales as $M_{\rm vir}$ not $M_{\rm vir}^2$ is carried by $\rho_c$. The $c_{\rm vir}^3$ dependence highlights that the annihilation flux is critically dependent upon how sharply peaked the halo is.   To summarize, Eq.~\ref{eq:JNFWfull} demonstrates that the $J$-factor is fully specified by three halo parameters for a given substructure boost model and cosmology: the redshift $z$, mass $M_{\rm vir}$, and concentration $c_{\rm vir}$.

\begin{figure*}[t]
   \centering
   \includegraphics[width=0.45\textwidth]{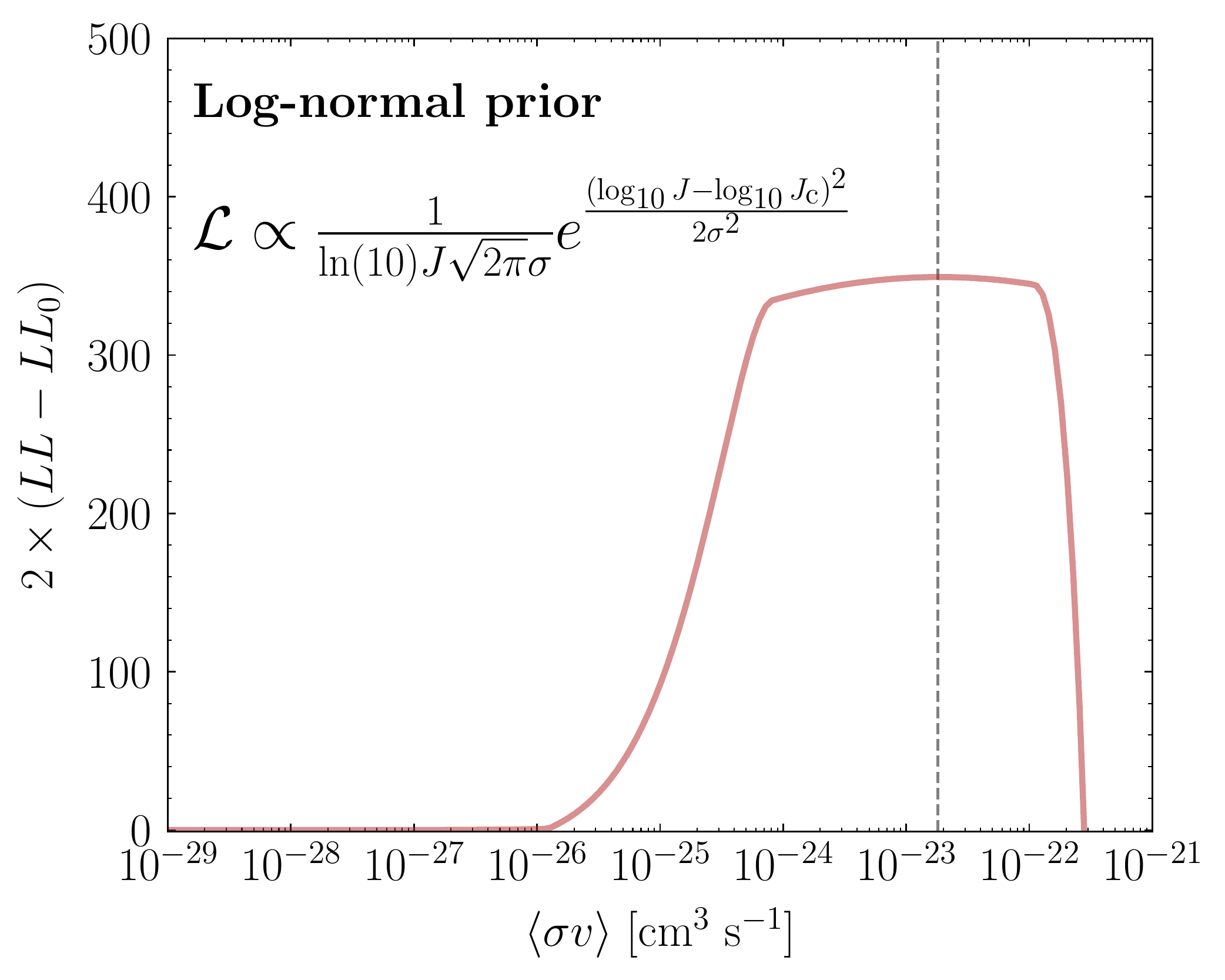}  ~\hspace{2mm}
   \includegraphics[width=0.45\textwidth]{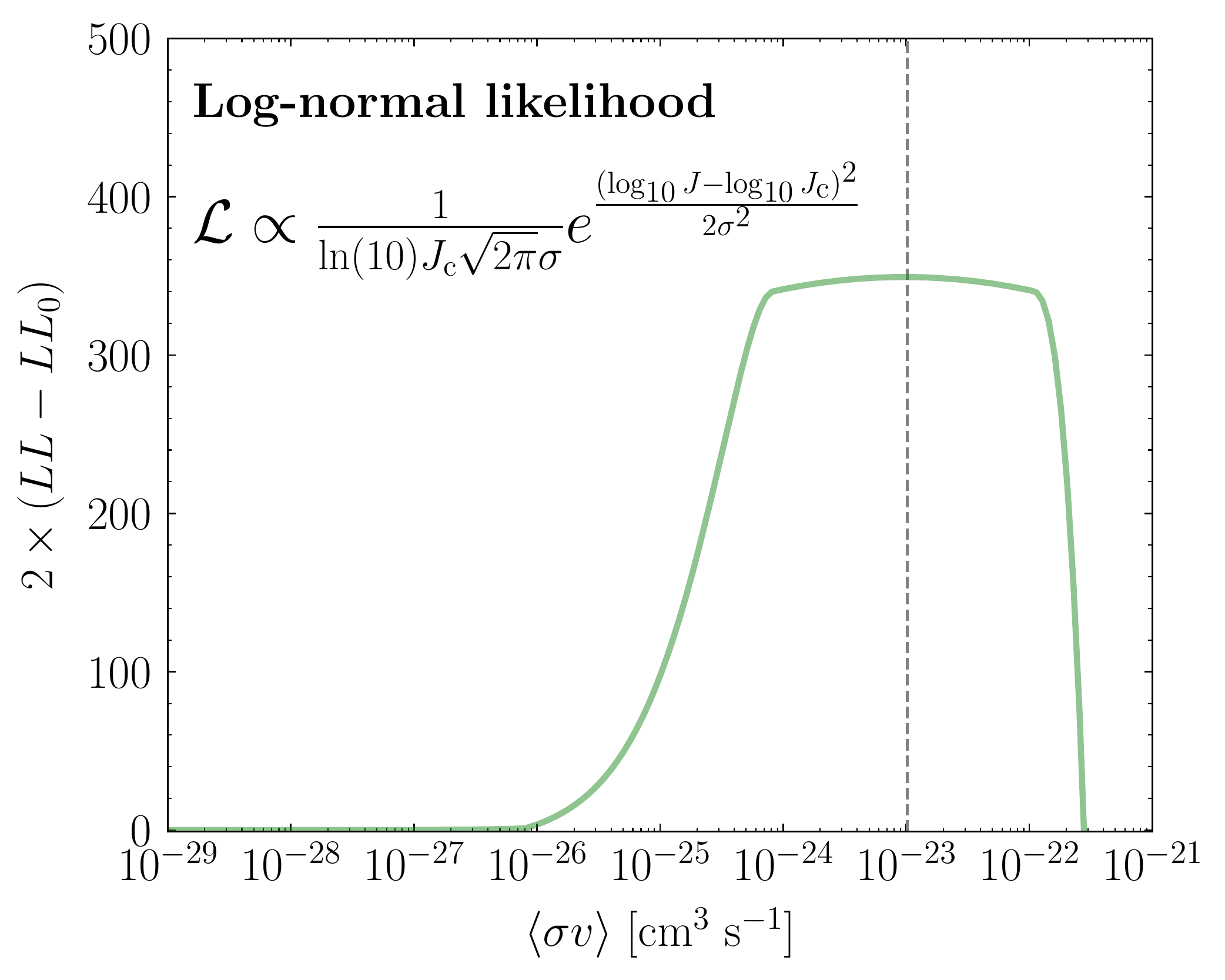} 
   \caption{Impact of using the log-normal prior for the $J$-factor (left) versus using the log-normal likelihood (right).  This is illustrated by showing the test statistic as a function of cross section for a single halo after injecting a DM signal of $\langle\sigma v \rangle= 10^{-23}\,{\rm cm}^3\,{\rm s}^{-1}$ at $m_\chi =100$ GeV. The recovered cross section is   associated with the maximum of the test statistic and is indicated by the vertical dashed line. The log-normal likelihood on the right correctly recovers the injected signal, while the log-normal prior on the left inevitably leads to an offset with respect to the underlying signal strength. Note that true, rather than inferred, $J$-factors are used in this case for illustration.}
   \label{fig:LLJfacForms}
\end{figure*}

The basic scalings and dependence shown above are not peculiar to the NFW profile, but are in fact more generic. To demonstrate this, we can repeat the above exercise for the cored Burkert profile~\cite{Burkert:1995yz}:
\begin{equation}
\rho_{\rm Burkert}(r) = \frac{\rho_B}{(1+r/r_B)(1+(r/r_B)^2)}\,,
\end{equation}
which is manifestly non-singular as $r \to 0$ unlike the NFW profile.  Here, $\rho_B$ and $r_B$ are the Burkert analogues of $\rho_s$ and $r_s$ in the NFW case, but they are not exactly the same.  Indeed, following \emph{e.g.}, Ref.~\cite{Bartels:2015uba}, by calculating physically measurable properties of halos such as the radius of maximum rotational velocity for both the NFW and Burkert cases and setting them equal, we find
\begin{equation}
r_B \simeq 0.7 r_s\,.
\end{equation}
We will replace $r_B$ with a concentration parameter $c_B = r_{\rm vir}/r_B$.  Following the same steps as for the NFW profile, we arrive at:
\begin{align}
J_{\rm Burkert} \approx &\left(1+b_\text{sh}[M_\text{vir}]\right) \frac{4M_{\rm vir} c_B^3 \rho_c \Delta_c[z]}{3 d_c^2[z]} \\
\times &\left[ \frac{c_B(1+c_B+2c_B^2)}{(1+c_B)(1+c_B^2)} - \arctan(c_B) \right] \notag \\
\times &\left[ \ln \left[ (1+c_B)^2 (1+c_B^2) \right] - 2 \arctan(c_B) \right]^{-2}\,, \notag
\end{align}
from which we see that $J \sim (1+b_\text{sh}) M_\text{vir} c_B^3\rho_c/d_c^2[z]$.

For the case of decaying DM, the approximate integral given in Eq.~\ref{eq:JDdapprox} can be evaluated independent of any choice for the halo profile.   Specifically:
\begin{equation}\begin{aligned}
D &\approx \frac{1}{d_c^2[z]} \int_V dV' \rho_{\rm DM}(r) = \frac{M_{\rm vir}}{d_c^2[z]}\,,
\label{eq:Dfactor}
\end{aligned}\end{equation}
where the second equality follows from the fact that the volume integral gives the virial mass exactly.
For DM decays in relatively nearby halos, the emission can be quite extended, as the flux is not as concentrated towards the center of the halo as in the annihilation case.  As such, it is often useful to have a version of the extragalactic $D$-factor where one only integrates out to some angle $\theta$ on the sky from the center of the halo, or equivalently to a distance $R = \theta \cdot d_c(z) < r_{\rm vir}$. In this case:\begin{align}
D &\approx \frac{M_{\rm vir}}{d_c^2(z)} \\
&\times \left[ \ln \left( 1 + \frac{c_{\rm vir} R}{r_{\rm vir}[M_{\rm vir}]} \right) - \frac{c_{\rm vir}}{r_{\rm vir}[M_{\rm vir}]/R + c_{\rm vir}} \right] \notag \\
&\times \left[ \ln (1 + c_{\rm vir}) - \frac{c_{\rm vir}}{1+c_{\rm vir}} \right]^{-1}\,, \notag
\end{align}
for the NFW profile, where we have made explicit the fact that $r_{\rm vir}$ is a function of $M_{\rm vir}$.  When $R=r_{\rm vir}$, this reduces to the simple result in Eq.~\ref{eq:Dfactor}.

\subsection{Propagating $J$-factor Uncertainties}
\label{app:Juncertainties}

We now discuss the propagation of uncertainties from inferred halo parameters into an overall uncertainty on the $J$-factor.  Specifically, we will justify the form of the log-normal distribution for the $J$-factor that was used in Eq.~\ref{eq:Jlognormal} of the main text.  While we focus our attention on the case of the $J$-factor for the NFW profile, the logic carries over straightforwardly to the Burkert profile, or even to the $D$-factor.

Our starting point is the approximate form of the NFW $J$-factor given in Eq.~\ref{eq:JNFWfull}, which shows that the $J$-factor is determined by the redshift  $z$, mass $M_{\rm vir}$, and concentration $c_{\rm vir}$, for a given substructure boost model and cosmology.  Therefore, the errors on these three parameters need to be propagated to determine the total error on the $J$-factor.  When the redshift $z$ is determined spectroscopically, the uncertainty on $d_c$ is  subdominant to the uncertainties on the $J$-factor that are induced by the mass and concentration. As such, we neglect the uncertainty in the redshift. If one were using photometric redshifts, however, these uncertainties would also need to be accounted for.  Additionally, for nearby halos in particular, the relation between redshift and distance can have further uncertainties, since this relation is affected by local peculiar velocities.  

From our studies of \texttt{DarkSky}, we see that the $M_{\rm vir}$ and $c_{\rm vir}$ distributions are well-approximated as log-normal. If the $J$-factor simply scaled as the product of several log-normal distributions ($J \sim M_{\rm vir} c_{\rm vir}^3$), then $J$ would also be log-normally distributed.  However, the dependence of the concentration parameter and boost factor on the virial mass mean that $J$ will not be exactly distributed in this way. Nevertheless, by explicitly calculating the full $J[M_{\rm vir}, c_{\rm vir}]$ distribution, we confirm that it is very accurately log-normally distributed. The reason for this is that the mass dependence of the boost factor in the Bartels substructure model~\cite{Bartels:2015uba} is very mild and additionally the $c_{\rm vir}$ dependence is subdominant in dictating the form of $J$, beyond the $c_{\rm vir}^3$ dependence.  Because the log-normal is  considerably simpler than the full distribution, we adopt it for the $J$-factor.

The form of the likelihood that we use for the $J$-factor in Eq.~\ref{eq:Jlognormal} is the same as that used in Ref.~\cite{Ackermann:2015zua}, but stands in contrast to Ref.~\cite{Ackermann:2011wa,Ackermann:2013yva} with the substitution of the nominal central $J$-factor in the denominator instead of the marginalized value. The interpretation of the $J$-factor as a log-normal likelihood~\cite{Ackermann:2015zua}  rather than a log-normal prior~\cite{Ackermann:2011wa,Ackermann:2013yva} ensures proper normalization for all values of $J$ and, when interpreted as a maximum likelihood estimator for signal recovery, coincides with the true underlying signal strength. This is illustrated in Fig.~\ref{fig:LLJfacForms}, where we show the test statistic as a function of DM cross section obtained for an injected signal of $\langle\sigma v \rangle= 10^{-23}\,{\rm cm}^3\,{\rm s}^{-1}$ at $m_\chi =100$ GeV after $J$-factor marginalization in the two different cases for a single halo. The left panel shows the traditional log-normal prior form of the likelihood.  The maximum of this, indicated by the vertical dotted line, is offset from the true value of the underlying cross section. This discrepancy can be substantially amplified when stacking a large number of objects. Using the log-normal likelihood form (right panel), we see that the maximum test statistic associated with the recovered signal coincides with the injected signal at the true value, $\langle\sigma v \rangle= 10^{-23}\,{\rm cm}^3\,{\rm s}^{-1}$.

\section{Validity of Likelihood Approximation}
\label{app:energyrange}

In this Appendix, we address issues pointed out in the main text where our analysis procedure appears to induce small systematic discrepancies.  First, we discuss the incorporation of the $J$-factor uncertainties, and then we discuss issues related to the profile likelihood procedure itself. 

\subsection{J-factor Likelihood}

\begin{figure}[t]
   \centering
   \includegraphics[width=0.48\textwidth]{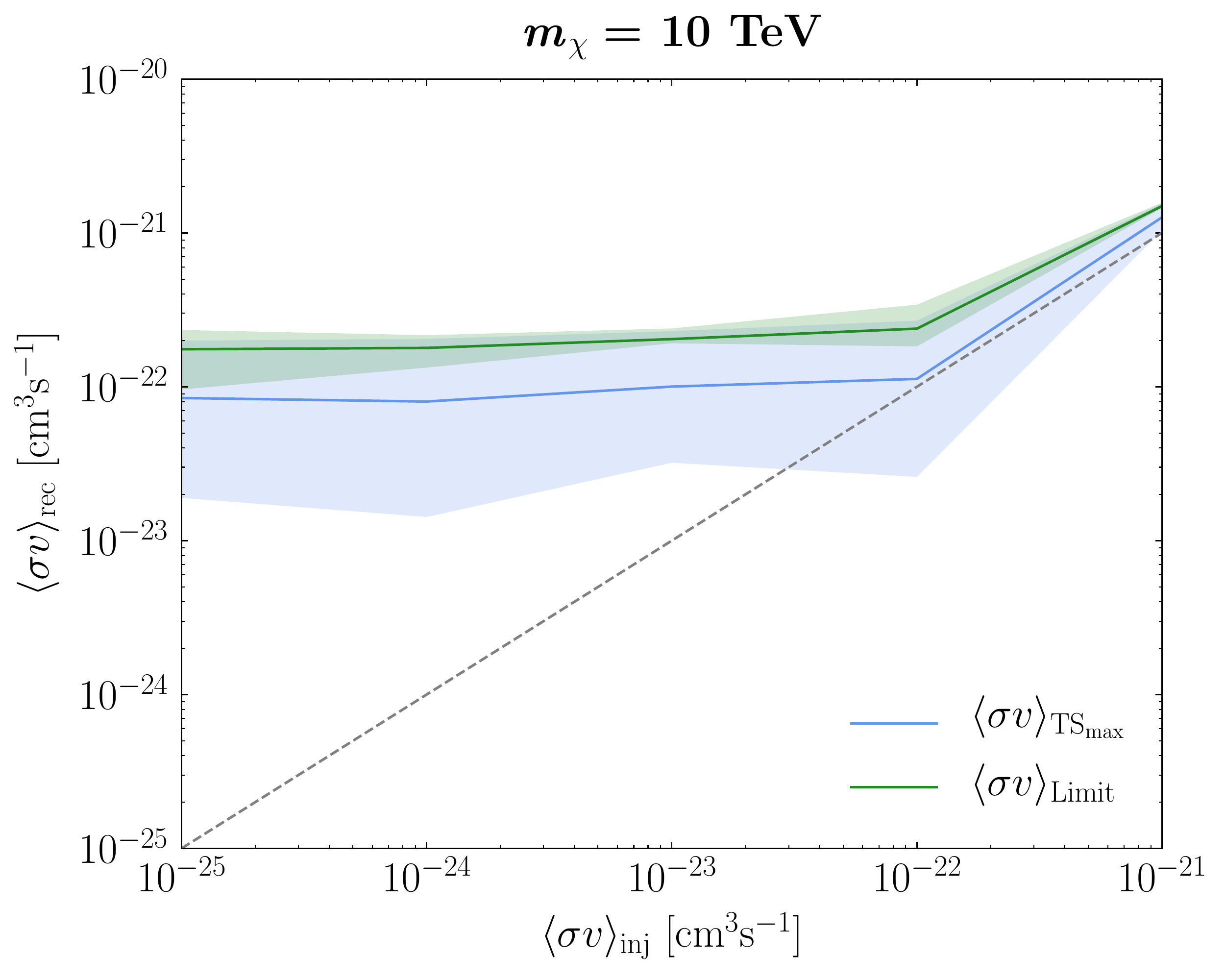} \hspace{0.01cm}
   \caption{ The same as Fig.~\ref{fig:DSinjsiglocs} for $m_\chi = 10$ TeV and for an observer at Location~1, except that the $J$-factors are fixed to their true values. Contrasting with the analogous plot in Fig.~\ref{fig:DSinjsiglocs}, we see that the $J$-factor uncertainties induce a small bias, at the $1$$\sigma$ level, towards higher recovered cross sections. }
   \label{fig:noJuncertainty}
\end{figure}

In Fig.~\ref{fig:DSinjsiglocs} of the main text,  the best-fit cross sections are systematically $\sim1$$\sigma$ higher than their injected values at high $\langle \sigma v \rangle_\text{inj}$ where DM detection is significant.  While this could be consistent with statistical fluctuations in the $J$-factor distributions, it  likely results from the fact that the assumed $J$-factor distributions, and the methods we have for calculating the central values and uncertainties, are not an exact representation of the actual $N$-body data.  To demonstrate this, Fig.~\ref{fig:noJuncertainty} shows the $m_\chi = 10$~TeV injected signal plot for an observer at Location~1, where all $J$-factors are fixed to their true values.  In this case, the best-fit cross section exactly matches the injected cross section at high values of $\langle \sigma v \rangle_\text{inj}$, confirming that it is indeed the assumed $J$-factor distributions that induce the bias seen in Fig.~\ref{fig:DSinjsiglocs}.

\subsection{Profile Likelihood Approximation}
\begin{figure*}[t]
   \centering
   \includegraphics[width=0.45\textwidth]{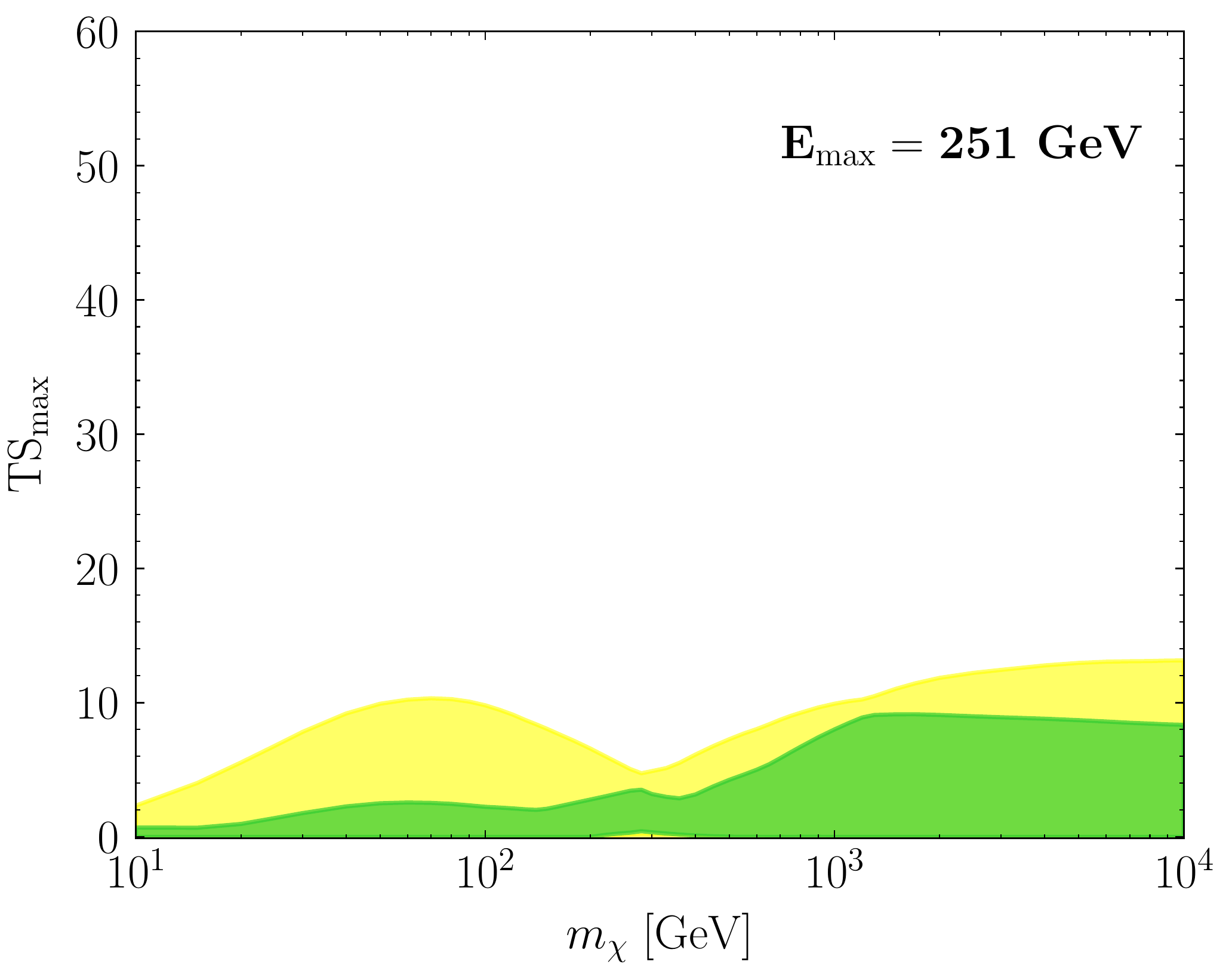}
   \includegraphics[width=0.45\textwidth]{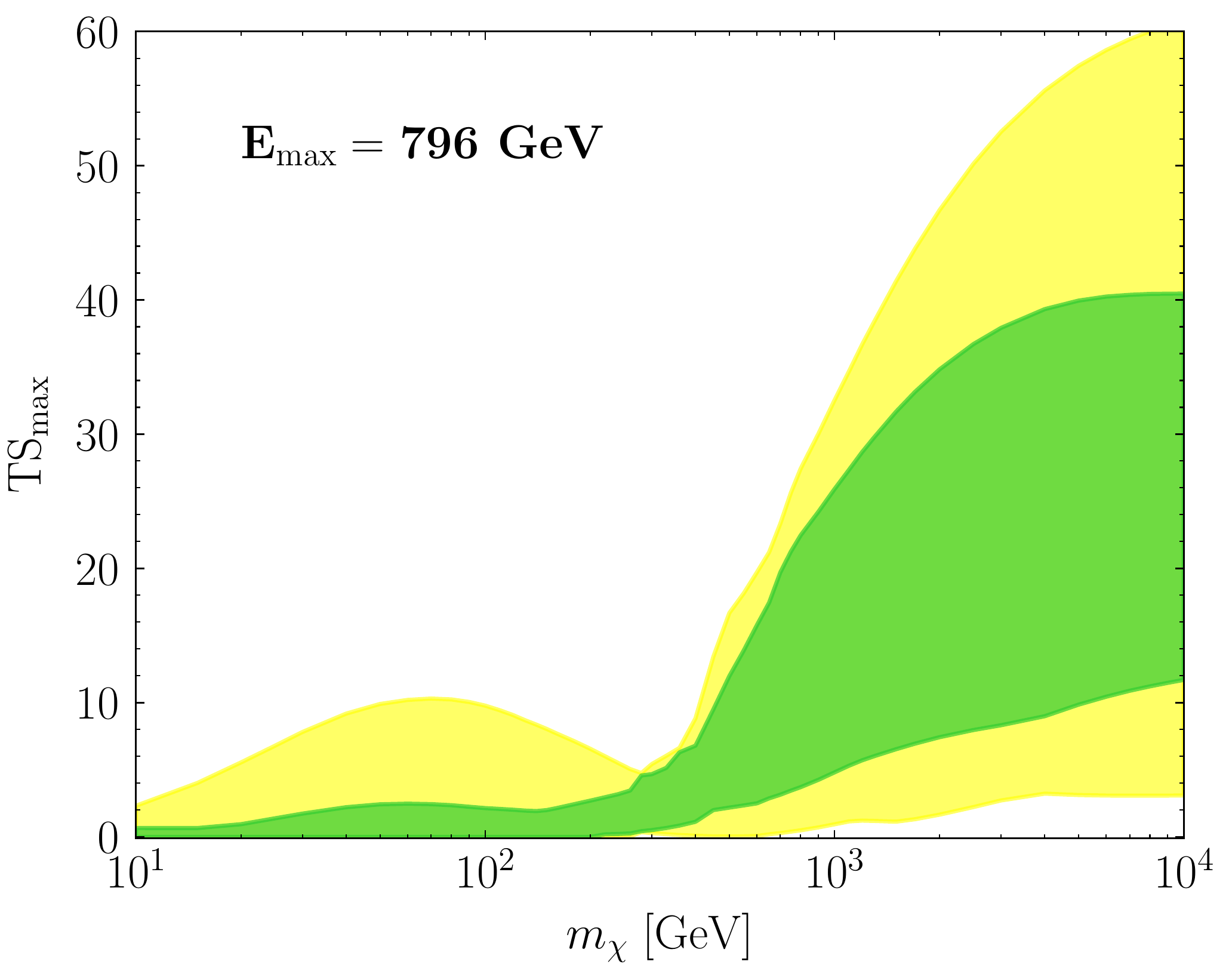}   
   \caption{ Maximum test statistic, TS$_\text{max}$, for the stacked analysis comparing the model with and without DM annihilating to $b \bar b$.  The green~(yellow) bands show the 68\%~(95\%) containment over multiple random sky locations.  The results are shown restricting the analysis to energies below 251 (left) and 796 (right)~GeV.  In both cases, the minimum energy is 500 MeV.}
   \label{fig:LikelihoodApprox}
\end{figure*}

As described in Sec.~\ref{sec:stats}, we use an approximation  to the full profile likelihood procedure to make the analysis computationally tractable.  This approximation comes into play when removing the nuisance parameters associated with the astrophysical templates:
\begin{equation}
\mathcal{L}_i^r(d_i^r | \psi_i, J^r) = \max_{\{\lambda_i^r\}}\,\mathcal{L}_i^r \left( d_i^r | \boldsymbol{\theta}_i^r, J^r \right)\,,
\label{eq:firstprofileSM}
\end{equation}
following the notation of Sec.~\ref{sec:stats}.  To briefly review, the full implementation of the profile likelihood requires maximizing the $\lambda_i^r$ for all $\psi_i$ and $J^r$.  Instead, we set the $\lambda_i^r$ to their maximum values, as obtained in an initial scan where all the template normalizations are floated. In general, we find that this approximation works very well, except at high photon energies ($\gtrsim 250$~GeV) and at low photon energy ($\lesssim 500$~MeV).  We now describe the challenges incurred in more detail.

At the highest energies, the number of photons becomes statistics-limited, and it is likely that there are very few photons in a given ROI. If one of these photons happens to fall  near the expected halo center, then the DM template will pick up flux in the initial scan, but all the other (astrophysical) templates will not.  In this case, the best-fit values for the astrophysical templates can be near zero, even though much larger normalizations for these templates would still be consistent with the data.  The problem arises when we set the $\lambda_i^r$ to their values from the initial scan and determine the DM intensities, because there will be evidence for a signal where there should be none, since the astrophysical templates are not allowed to adjust from their near-zero values.  This problem is not present in the full profile likelihood method, because in that case, one maximizes the likelihood over the $\lambda_i^r$ when constructing the likelihood profiles as functions of the DM intensities.  In particular, in the full profile likelihood method one obtains a better fit at low DM intensities, compared to that obtained in our approximation, because the astrophysical template normalizations are allowed to be higher.  We stress that this is not a concern at moderate energies, where the photon counts in each ROI are large enough such that the normalizations of the astrophysical templates are always well-determined. 

The behavior described above can be observed directly  in the mock data (with no injected signal).  For examples, the inset plot with $m_\chi = 10$ TeV  in Fig.~\ref{fig:DSinjsiglocs} shows the maximum test statistic as a function of the injected cross section.  At low injected cross sections, the data is described by the null hypothesis and so the TS$_\text{max}$ should follow a chi-square distribution.  In particular, this means that the 84$^\text{th}$ percentile, which is given by the upper  boundary of the red bands in Fig.~\ref{fig:DSinjsiglocs}, should asymptote to a value $\sim$$0.99$, while the lower part of the band should be consistent with zero.  
The discrepancy between the MC results and the chi-square expectation are most pronounced at high masses where the high energy bins are relatively more important.  

\begin{figure*}[t]
   \centering
   \includegraphics[width=0.45\textwidth]{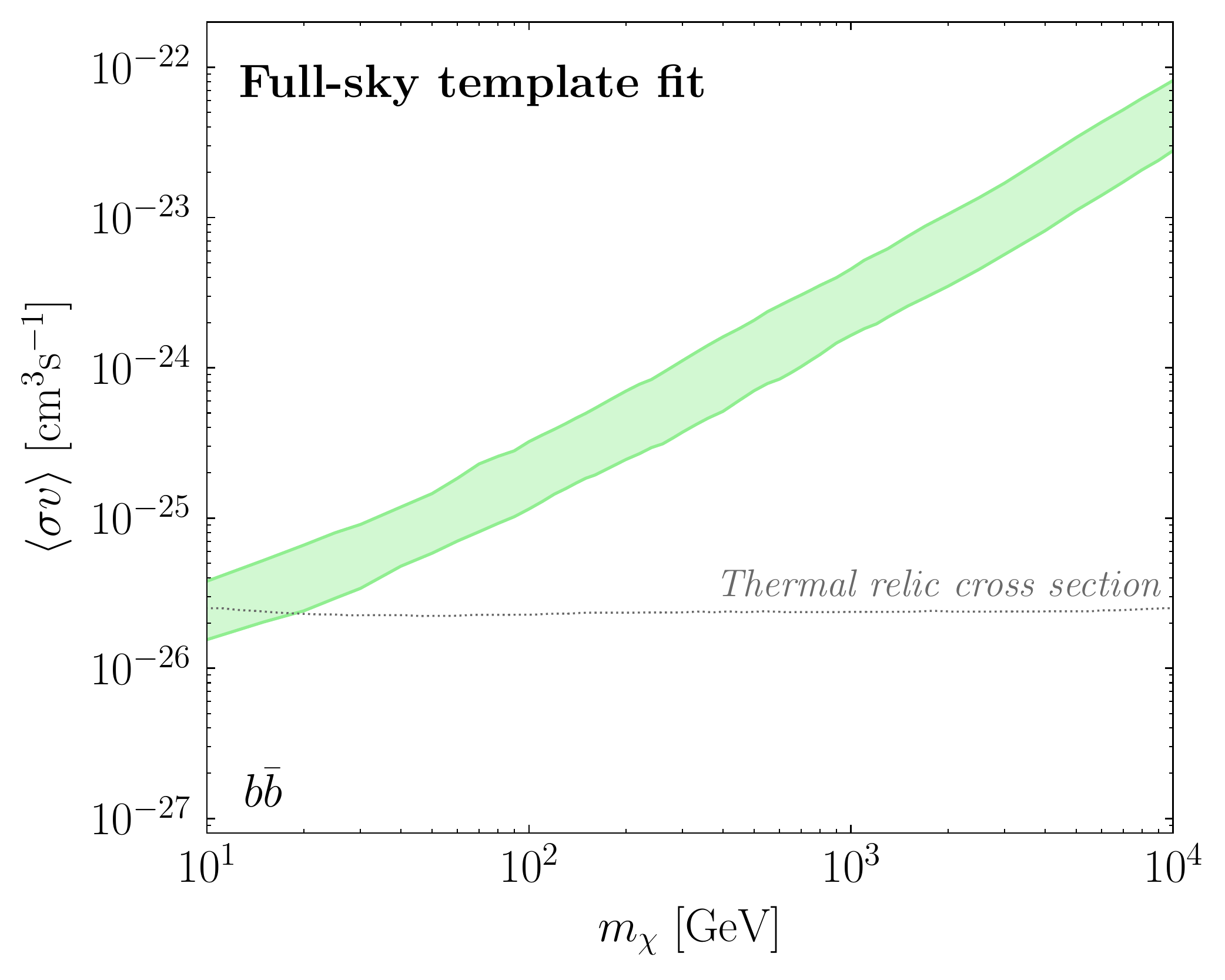}
     \includegraphics[width=0.45\textwidth]{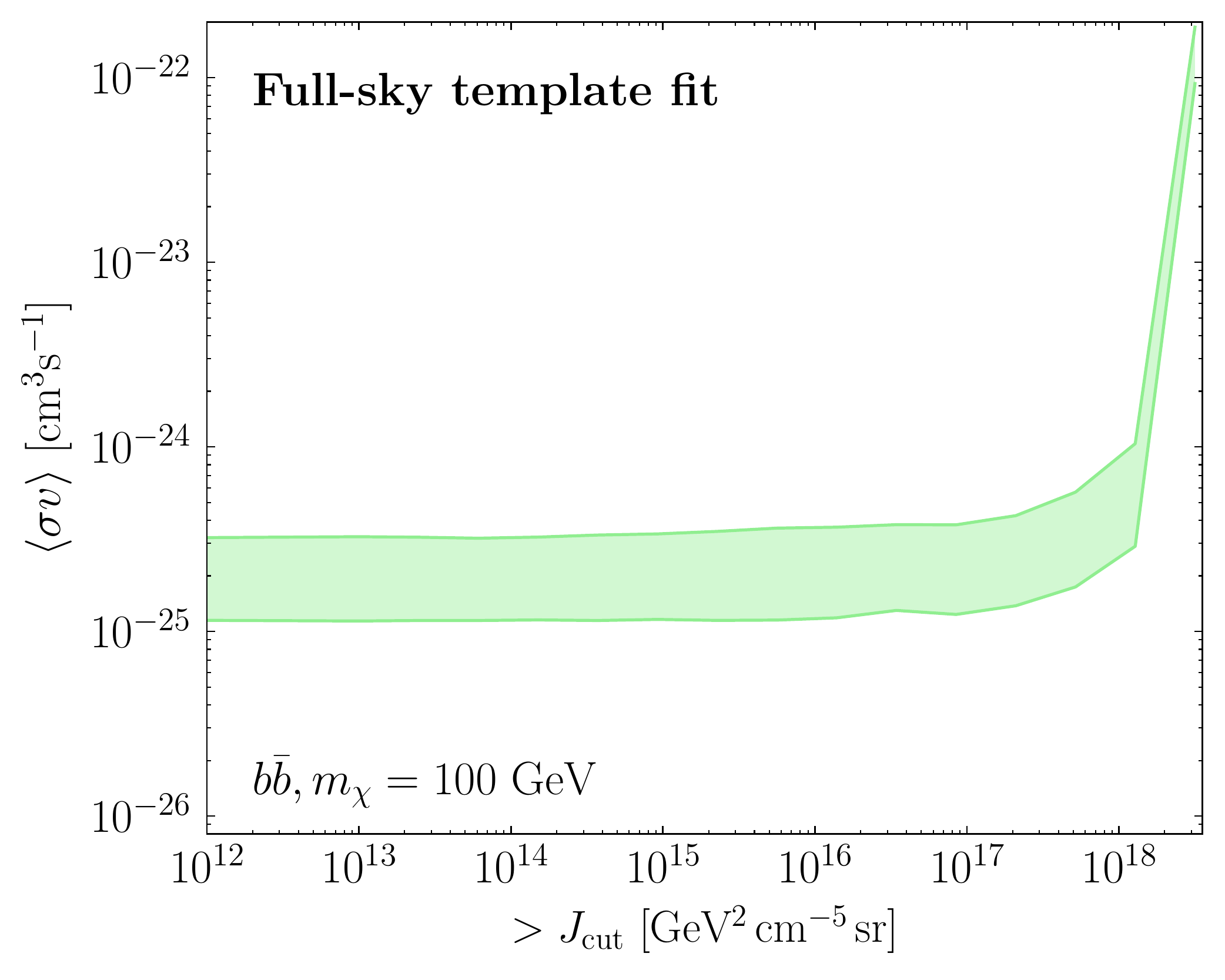}
   \caption{ (Left)  The predicted 68\% containment region over 100 MCs for the limit on the annihilation cross section for the full-sky template fitting method.  In this method, all of the $J$-factors from the individual halos are summed together.  Then, the likelihood profile is constructed for the single template.  Note that here we have assumed knowledge of the true $J$-factors, unlike in the previous figures, and have not marginalized over the corresponding uncertainties.
   (Right)  Projected limits at $m_\chi = 100$~GeV, using the single template-fitting method, as a function of the $J$-factor cut.  Only halos with $J$-factors higher than $> J_\text{cut}$, indicated on the $x$ axis, are included in the analysis.  These results show that the limit is dominated by a small subset of halos with high $J$-factors.
   }
   \label{fig:DarkSkyallhalossensitivity}
\end{figure*}

One method to alleviate this tension is to impose a maximum energy cut-off.   The right panel of Fig.~\ref{fig:LikelihoodApprox} shows the maximum test statistic, TS$_\text{max}$, of the stacked analysis (for the $b\bar{b}$ channel) for energies below 796~GeV, as a function of DM mass, for data that is a Poisson draw of the background models only.  The green~(yellow) bands show the 68\%~(95\%) spread over 20~MC realizations of the mock data.  We see that TS$_\text{max}$ increases with mass and is inconsistent with 0 at the 95\% level for $m_\chi \gtrsim 500$~GeV.  Note that the 97.5 percentile for TS$_\text{max}$ should be approximately 3.84 assuming a chi-square distribution, though it is over an order of magnitude larger in the MC results at high masses.  To avoid these issues when applying the analysis to real data, we plan to restrict to energies below $\sim$$250$~GeV. The left panel of Fig.~\ref{fig:LikelihoodApprox} shows the TS$_\text{max}$ distribution for energies below 251~GeV, as obtained from 20 MCs of the mock data.  In this case, the simulated TS$_\text{max}$ distribution does a much better, though not perfect, job of approximating the expectation from the chi-square distribution across the $m_\chi$ range.

A related issue is seen at low DM masses because the approximation we take to the profile likelihood method also breaks down in the lowest energy bins.  In this case, we find that the TS$_\text{max}$ distribution under the null hypothesis does not extend to values as high as would be expected from the chi-square distribution.  In particular, with our default minimum energy of 200 MeV, we find that the 84 and 97.5 percentiles of the TS$_\text{max}$ distribution under the null hypothesis for 10 GeV DM are $\sim$0.14 and $\sim$0.25, respectively.  On the other hand, increasing the minimum energy to 500 MeV brings these values to $\sim$0.70 and $\sim$2.33, respectively, which are in much closer agreement with the expectations from the chi-square distribution.  In Fig.~\ref{fig:LikelihoodApprox}, we implemented the 500 MeV minimum energy and we also plan on using this minimum energy when analyzing the real data.  We suspect that the issues with the lowest energy bins are related to the large PSF at low energies, which can extend the emission from the extragalactic DM halos on the size of the ROI.  This induces degeneracies between the DM template and the astrophysical templates.

One takeaway from these exercises is that the approximation to the full profile likelihood method discussed here has its limitations.  As a result, it would always be best to perform the full likelihood analysis on the data, if the necessary  computational resources are available.   
In lieu of that, one can adjust the energy binning to give results that well-approximate the expectations from the exact procedure.

\section{Limits Using a Full-Sky Dark Matter Template}
\label{app:inflimitimpact}

The main body of this paper presents the results of a stacked analysis in which we analyze individual targets and combine their likelihoods.  Here, we show the results of an analysis where we float the expected emission from all the halos together as a single template.  All resolved point sources are also floated together in this case. The left panel of~Fig.~\ref{fig:DarkSkyallhalossensitivity} shows the projected sensitivity in this case for $b\bar b$ annihilation, \emph{assuming the truth values for the $J$-factors}.  The green band shows the statistical variation of 100 MC iterations of the mock data.  The right panel shows how the limit for $m_\chi = 100$~GeV changes as only halos above a certain $J$-factor threshold are included in the template. Clearly, the sensitivity is dominated by a relatively small number of halos.  This suggests that the normalization of the DM annihilation flux template is being set primarily by the fluctuations of the brightest halos.  We conclude therefore that there is no benefit to doing a full-sky template analysis where the normalizations of all the halos are floated together.  Instead, these results suggest that it is far more advantageous to perform a stacked analysis of the individual galaxy groups, as  explored in this paper. 

\bibliography{fermi_darksky}

\begin{thebibliography}{94}%
\makeatletter
\providecommand \@ifxundefined [1]{%
 \@ifx{#1\undefined}
}%
\providecommand \@ifnum [1]{%
 \ifnum #1\expandafter \@firstoftwo
 \else \expandafter \@secondoftwo
 \fi
}%
\providecommand \@ifx [1]{%
 \ifx #1\expandafter \@firstoftwo
 \else \expandafter \@secondoftwo
 \fi
}%
\providecommand \natexlab [1]{#1}%
\providecommand \enquote  [1]{``#1''}%
\providecommand \bibnamefont  [1]{#1}%
\providecommand \bibfnamefont [1]{#1}%
\providecommand \citenamefont [1]{#1}%
\providecommand \href@noop [0]{\@secondoftwo}%
\providecommand \href [0]{\begingroup \@sanitize@url \@href}%
\providecommand \@href[1]{\@@startlink{#1}\@@href}%
\providecommand \@@href[1]{\endgroup#1\@@endlink}%
\providecommand \@sanitize@url [0]{\catcode `\\12\catcode `\$12\catcode
  `\&12\catcode `\#12\catcode `\^12\catcode `\_12\catcode `\%12\relax}%
\providecommand \@@startlink[1]{}%
\providecommand \@@endlink[0]{}%
\providecommand \url  [0]{\begingroup\@sanitize@url \@url }%
\providecommand \@url [1]{\endgroup\@href {#1}{\urlprefix }}%
\providecommand \urlprefix  [0]{URL }%
\providecommand \Eprint [0]{\href }%
\providecommand \doibase [0]{http://dx.doi.org/}%
\providecommand \selectlanguage [0]{\@gobble}%
\providecommand \bibinfo  [0]{\@secondoftwo}%
\providecommand \bibfield  [0]{\@secondoftwo}%
\providecommand \translation [1]{[#1]}%
\providecommand \BibitemOpen [0]{}%
\providecommand \bibitemStop [0]{}%
\providecommand \bibitemNoStop [0]{.\EOS\space}%
\providecommand \EOS [0]{\spacefactor3000\relax}%
\providecommand \BibitemShut  [1]{\csname bibitem#1\endcsname}%
\let\auto@bib@innerbib\@empty
%</preamble>
\bibitem [{\citenamefont {Albert}\ \emph {et~al.}(2017)\citenamefont {Albert}
  \emph {et~al.}}]{Fermi-LAT:2016uux}%
  \BibitemOpen
  \bibfield  {author} {\bibinfo {author} {\bibfnamefont {A.}~\bibnamefont
  {Albert}} \emph {et~al.} (\bibinfo {collaboration} {DES, Fermi-LAT}),\
  }\bibfield  {title} {\enquote {\bibinfo {title} {{Searching for Dark Matter
  Annihilation in Recently Discovered Milky Way Satellites with Fermi-LAT}},}\
  }\href {\doibase 10.3847/1538-4357/834/2/110} {\bibfield  {journal} {\bibinfo
   {journal} {Astrophys. J.}\ }\textbf {\bibinfo {volume} {834}},\ \bibinfo
  {pages} {110} (\bibinfo {year} {2017})},\ \Eprint
  {http://arxiv.org/abs/1611.03184} {arXiv:1611.03184 [astro-ph.HE]}
  \BibitemShut {NoStop}%
%%CITATION = ARXIV:1611.03184;%%
\bibitem [{\citenamefont {Ackermann}\ \emph
  {et~al.}(2015{\natexlab{a}})\citenamefont {Ackermann} \emph
  {et~al.}}]{Ackermann:2015zua}%
  \BibitemOpen
  \bibfield  {author} {\bibinfo {author} {\bibfnamefont {M.}~\bibnamefont
  {Ackermann}} \emph {et~al.} (\bibinfo {collaboration} {Fermi-LAT}),\
  }\bibfield  {title} {\enquote {\bibinfo {title} {{Searching for Dark Matter
  Annihilation from Milky Way Dwarf Spheroidal Galaxies with Six Years of Fermi
  Large Area Telescope Data}},}\ }\href {\doibase
  10.1103/PhysRevLett.115.231301} {\bibfield  {journal} {\bibinfo  {journal}
  {Phys. Rev. Lett.}\ }\textbf {\bibinfo {volume} {115}},\ \bibinfo {pages}
  {231301} (\bibinfo {year} {2015}{\natexlab{a}})},\ \Eprint
  {http://arxiv.org/abs/1503.02641} {arXiv:1503.02641 [astro-ph.HE]}
  \BibitemShut {NoStop}%
%%CITATION = ARXIV:1503.02641;%%
\bibitem [{\citenamefont {Bengtsson}\ \emph {et~al.}(1990)\citenamefont
  {Bengtsson}, \citenamefont {Salati},\ and\ \citenamefont
  {Silk}}]{Bengtsson:1990xf}%
  \BibitemOpen
  \bibfield  {author} {\bibinfo {author} {\bibfnamefont {Hans~Uno}\
  \bibnamefont {Bengtsson}}, \bibinfo {author} {\bibfnamefont {Pierre}\
  \bibnamefont {Salati}}, \ and\ \bibinfo {author} {\bibfnamefont {Joseph}\
  \bibnamefont {Silk}},\ }\bibfield  {title} {\enquote {\bibinfo {title}
  {{Quark Flavors and the gamma-ray Spectrum From Halo Dark Matter
  Annihilations}},}\ }\href {\doibase 10.1016/0550-3213(90)90241-5} {\bibfield
  {journal} {\bibinfo  {journal} {Nucl. Phys.}\ }\textbf {\bibinfo {volume}
  {B346}},\ \bibinfo {pages} {129--148} (\bibinfo {year} {1990})}\BibitemShut
  {NoStop}%
%%CITATION = NUPHA,B346,129;%%
\bibitem [{\citenamefont {Bergstrom}\ \emph {et~al.}(2001)\citenamefont
  {Bergstrom}, \citenamefont {Edsjo},\ and\ \citenamefont
  {Ullio}}]{Bergstrom:2001jj}%
  \BibitemOpen
  \bibfield  {author} {\bibinfo {author} {\bibfnamefont {Lars}\ \bibnamefont
  {Bergstrom}}, \bibinfo {author} {\bibfnamefont {Joakim}\ \bibnamefont
  {Edsjo}}, \ and\ \bibinfo {author} {\bibfnamefont {Piero}\ \bibnamefont
  {Ullio}},\ }\bibfield  {title} {\enquote {\bibinfo {title} {{Spectral
  gamma-ray signatures of cosmological dark matter annihilation}},}\ }\href
  {\doibase 10.1103/PhysRevLett.87.251301} {\bibfield  {journal} {\bibinfo
  {journal} {Phys. Rev. Lett.}\ }\textbf {\bibinfo {volume} {87}},\ \bibinfo
  {pages} {251301} (\bibinfo {year} {2001})},\ \Eprint
  {http://arxiv.org/abs/astro-ph/0105048} {arXiv:astro-ph/0105048 [astro-ph]}
  \BibitemShut {NoStop}%
%%CITATION = ASTRO-PH/0105048;%%
\bibitem [{\citenamefont {Ullio}\ \emph {et~al.}(2002)\citenamefont {Ullio},
  \citenamefont {Bergstrom}, \citenamefont {Edsjo},\ and\ \citenamefont
  {Lacey}}]{Ullio:2002pj}%
  \BibitemOpen
  \bibfield  {author} {\bibinfo {author} {\bibfnamefont {Piero}\ \bibnamefont
  {Ullio}}, \bibinfo {author} {\bibfnamefont {Lars}\ \bibnamefont {Bergstrom}},
  \bibinfo {author} {\bibfnamefont {Joakim}\ \bibnamefont {Edsjo}}, \ and\
  \bibinfo {author} {\bibfnamefont {Cedric~G.}\ \bibnamefont {Lacey}},\
  }\bibfield  {title} {\enquote {\bibinfo {title} {{Cosmological dark matter
  annihilations into gamma-rays - a closer look}},}\ }\href {\doibase
  10.1103/PhysRevD.66.123502} {\bibfield  {journal} {\bibinfo  {journal} {Phys.
  Rev.}\ }\textbf {\bibinfo {volume} {D66}},\ \bibinfo {pages} {123502}
  (\bibinfo {year} {2002})},\ \Eprint {http://arxiv.org/abs/astro-ph/0207125}
  {arXiv:astro-ph/0207125 [astro-ph]} \BibitemShut {NoStop}%
%%CITATION = ASTRO-PH/0207125;%%
\bibitem [{\citenamefont {Bottino}\ \emph {et~al.}(2004)\citenamefont
  {Bottino}, \citenamefont {Donato}, \citenamefont {Fornengo},\ and\
  \citenamefont {Scopel}}]{Bottino:2004qi}%
  \BibitemOpen
  \bibfield  {author} {\bibinfo {author} {\bibfnamefont {A.}~\bibnamefont
  {Bottino}}, \bibinfo {author} {\bibfnamefont {F.}~\bibnamefont {Donato}},
  \bibinfo {author} {\bibfnamefont {N.}~\bibnamefont {Fornengo}}, \ and\
  \bibinfo {author} {\bibfnamefont {S.}~\bibnamefont {Scopel}},\ }\bibfield
  {title} {\enquote {\bibinfo {title} {{Indirect signals from light neutralinos
  in supersymmetric models without gaugino mass unification}},}\ }\href
  {\doibase 10.1103/PhysRevD.70.015005} {\bibfield  {journal} {\bibinfo
  {journal} {Phys. Rev.}\ }\textbf {\bibinfo {volume} {D70}},\ \bibinfo {pages}
  {015005} (\bibinfo {year} {2004})},\ \Eprint
  {http://arxiv.org/abs/hep-ph/0401186} {arXiv:hep-ph/0401186 [hep-ph]}
  \BibitemShut {NoStop}%
%%CITATION = HEP-PH/0401186;%%
\bibitem [{\citenamefont {Bertone}\ \emph {et~al.}(2005)\citenamefont
  {Bertone}, \citenamefont {Hooper},\ and\ \citenamefont
  {Silk}}]{Bertone:2004pz}%
  \BibitemOpen
  \bibfield  {author} {\bibinfo {author} {\bibfnamefont {Gianfranco}\
  \bibnamefont {Bertone}}, \bibinfo {author} {\bibfnamefont {Dan}\ \bibnamefont
  {Hooper}}, \ and\ \bibinfo {author} {\bibfnamefont {Joseph}\ \bibnamefont
  {Silk}},\ }\bibfield  {title} {\enquote {\bibinfo {title} {{Particle dark
  matter: Evidence, candidates and constraints}},}\ }\href {\doibase
  10.1016/j.physrep.2004.08.031} {\bibfield  {journal} {\bibinfo  {journal}
  {Phys. Rept.}\ }\textbf {\bibinfo {volume} {405}},\ \bibinfo {pages}
  {279--390} (\bibinfo {year} {2005})},\ \Eprint
  {http://arxiv.org/abs/hep-ph/0404175} {arXiv:hep-ph/0404175 [hep-ph]}
  \BibitemShut {NoStop}%
%%CITATION = HEP-PH/0404175;%%
\bibitem [{\citenamefont {Bringmann}\ and\ \citenamefont
  {Weniger}(2012)}]{Bringmann:2012ez}%
  \BibitemOpen
  \bibfield  {author} {\bibinfo {author} {\bibfnamefont {Torsten}\ \bibnamefont
  {Bringmann}}\ and\ \bibinfo {author} {\bibfnamefont {Christoph}\ \bibnamefont
  {Weniger}},\ }\bibfield  {title} {\enquote {\bibinfo {title} {{Gamma Ray
  Signals from Dark Matter: Concepts, Status and Prospects}},}\ }\href
  {\doibase 10.1016/j.dark.2012.10.005} {\bibfield  {journal} {\bibinfo
  {journal} {Phys. Dark Univ.}\ }\textbf {\bibinfo {volume} {1}},\ \bibinfo
  {pages} {194--217} (\bibinfo {year} {2012})},\ \Eprint
  {http://arxiv.org/abs/1208.5481} {arXiv:1208.5481 [hep-ph]} \BibitemShut
  {NoStop}%
%%CITATION = ARXIV:1208.5481;%%
\bibitem [{\citenamefont {Ajello}\ \emph {et~al.}(2015)\citenamefont {Ajello}
  \emph {et~al.}}]{Ajello:2015mfa}%
  \BibitemOpen
  \bibfield  {author} {\bibinfo {author} {\bibfnamefont {M.}~\bibnamefont
  {Ajello}} \emph {et~al.},\ }\bibfield  {title} {\enquote {\bibinfo {title}
  {{The Origin of the Extragalactic Gamma-Ray Background and Implications for
  Dark-Matter Annihilation}},}\ }\href {\doibase 10.1088/2041-8205/800/2/L27}
  {\bibfield  {journal} {\bibinfo  {journal} {Astrophys. J.}\ }\textbf
  {\bibinfo {volume} {800}},\ \bibinfo {pages} {L27} (\bibinfo {year}
  {2015})},\ \Eprint {http://arxiv.org/abs/1501.05301} {arXiv:1501.05301
  [astro-ph.HE]} \BibitemShut {NoStop}%
%%CITATION = ARXIV:1501.05301;%%
\bibitem [{\citenamefont {Di~Mauro}\ and\ \citenamefont
  {Donato}(2015)}]{DiMauro:2015tfa}%
  \BibitemOpen
  \bibfield  {author} {\bibinfo {author} {\bibfnamefont {Mattia}\ \bibnamefont
  {Di~Mauro}}\ and\ \bibinfo {author} {\bibfnamefont {Fiorenza}\ \bibnamefont
  {Donato}},\ }\bibfield  {title} {\enquote {\bibinfo {title} {{Composition of
  the Fermi-LAT isotropic gamma-ray background intensity: Emission from
  extragalactic point sources and dark matter annihilations}},}\ }\href
  {\doibase 10.1103/PhysRevD.91.123001} {\bibfield  {journal} {\bibinfo
  {journal} {Phys. Rev.}\ }\textbf {\bibinfo {volume} {D91}},\ \bibinfo {pages}
  {123001} (\bibinfo {year} {2015})},\ \Eprint
  {http://arxiv.org/abs/1501.05316} {arXiv:1501.05316 [astro-ph.HE]}
  \BibitemShut {NoStop}%
%%CITATION = ARXIV:1501.05316;%%
\bibitem [{\citenamefont {Ackermann}\ \emph
  {et~al.}(2015{\natexlab{b}})\citenamefont {Ackermann} \emph
  {et~al.}}]{Ackermann:2015tah}%
  \BibitemOpen
  \bibfield  {author} {\bibinfo {author} {\bibfnamefont {M.}~\bibnamefont
  {Ackermann}} \emph {et~al.} (\bibinfo {collaboration} {Fermi-LAT}),\
  }\bibfield  {title} {\enquote {\bibinfo {title} {{Limits on Dark Matter
  Annihilation Signals from the Fermi LAT 4-year Measurement of the Isotropic
  Gamma-Ray Background}},}\ }\href {\doibase 10.1088/1475-7516/2015/09/008}
  {\bibfield  {journal} {\bibinfo  {journal} {JCAP}\ }\textbf {\bibinfo
  {volume} {1509}},\ \bibinfo {pages} {008} (\bibinfo {year}
  {2015}{\natexlab{b}})},\ \Eprint {http://arxiv.org/abs/1501.05464}
  {arXiv:1501.05464 [astro-ph.CO]} \BibitemShut {NoStop}%
%%CITATION = ARXIV:1501.05464;%%
\bibitem [{\citenamefont {Feng}\ \emph {et~al.}(2017)\citenamefont {Feng},
  \citenamefont {Cooray},\ and\ \citenamefont {Keating}}]{Feng:2016fkl}%
  \BibitemOpen
  \bibfield  {author} {\bibinfo {author} {\bibfnamefont {Chang}\ \bibnamefont
  {Feng}}, \bibinfo {author} {\bibfnamefont {Asantha}\ \bibnamefont {Cooray}},
  \ and\ \bibinfo {author} {\bibfnamefont {Brian}\ \bibnamefont {Keating}},\
  }\bibfield  {title} {\enquote {\bibinfo {title} {{Planck Lensing and Cosmic
  Infrared Background Cross-Correlation with Fermi-LAT: Tracing Dark Matter
  Signals in the Gamma-Ray Background}},}\ }\href {\doibase
  10.3847/1538-4357/836/1/127} {\bibfield  {journal} {\bibinfo  {journal}
  {Astrophys. J.}\ }\textbf {\bibinfo {volume} {836}},\ \bibinfo {pages} {127}
  (\bibinfo {year} {2017})},\ \Eprint {http://arxiv.org/abs/1608.04351}
  {arXiv:1608.04351 [astro-ph.CO]} \BibitemShut {NoStop}%
%%CITATION = ARXIV:1608.04351;%%
\bibitem [{\citenamefont {Ackermann}\ \emph {et~al.}(2012)\citenamefont
  {Ackermann} \emph {et~al.}}]{Ackermann:2012uf}%
  \BibitemOpen
  \bibfield  {author} {\bibinfo {author} {\bibfnamefont {M.}~\bibnamefont
  {Ackermann}} \emph {et~al.} (\bibinfo {collaboration} {Fermi-LAT}),\
  }\bibfield  {title} {\enquote {\bibinfo {title} {{Anisotropies in the diffuse
  gamma-ray background measured by the Fermi LAT}},}\ }\href {\doibase
  10.1103/PhysRevD.85.109901, 10.1103/PhysRevD.85.083007} {\bibfield  {journal}
  {\bibinfo  {journal} {Phys. Rev.}\ }\textbf {\bibinfo {volume} {D85}},\
  \bibinfo {pages} {083007} (\bibinfo {year} {2012})},\ \Eprint
  {http://arxiv.org/abs/1202.2856} {1202.2856} \BibitemShut {NoStop}%
%%CITATION = ARXIV:1202.2856;%%
\bibitem [{\citenamefont {Fornasa}\ \emph {et~al.}(2016)\citenamefont {Fornasa}
  \emph {et~al.}}]{Fornasa:2016ohl}%
  \BibitemOpen
  \bibfield  {author} {\bibinfo {author} {\bibfnamefont {Mattia}\ \bibnamefont
  {Fornasa}} \emph {et~al.},\ }\bibfield  {title} {\enquote {\bibinfo {title}
  {{Angular power spectrum of the diffuse gamma-ray emission as measured by the
  Fermi Large Area Telescope and constraints on its dark matter
  interpretation}},}\ }\href {\doibase 10.1103/PhysRevD.94.123005} {\bibfield
  {journal} {\bibinfo  {journal} {Phys. Rev.}\ }\textbf {\bibinfo {volume}
  {D94}},\ \bibinfo {pages} {123005} (\bibinfo {year} {2016})},\ \Eprint
  {http://arxiv.org/abs/1608.07289} {arXiv:1608.07289 [astro-ph.HE]}
  \BibitemShut {NoStop}%
%%CITATION = ARXIV:1608.07289;%%
\bibitem [{\citenamefont {Ando}\ \emph {et~al.}(2007)\citenamefont {Ando},
  \citenamefont {Komatsu}, \citenamefont {Narumoto},\ and\ \citenamefont
  {Totani}}]{Ando:2006cr}%
  \BibitemOpen
  \bibfield  {author} {\bibinfo {author} {\bibfnamefont {Shin'ichiro}\
  \bibnamefont {Ando}}, \bibinfo {author} {\bibfnamefont {Eiichiro}\
  \bibnamefont {Komatsu}}, \bibinfo {author} {\bibfnamefont {Takuro}\
  \bibnamefont {Narumoto}}, \ and\ \bibinfo {author} {\bibfnamefont {Tomonori}\
  \bibnamefont {Totani}},\ }\bibfield  {title} {\enquote {\bibinfo {title}
  {{Dark matter annihilation or unresolved astrophysical sources? Anisotropy
  probe of the origin of cosmic gamma-ray background}},}\ }\href {\doibase
  10.1103/PhysRevD.75.063519} {\bibfield  {journal} {\bibinfo  {journal} {Phys.
  Rev.}\ }\textbf {\bibinfo {volume} {D75}},\ \bibinfo {pages} {063519}
  (\bibinfo {year} {2007})},\ \Eprint {http://arxiv.org/abs/astro-ph/0612467}
  {arXiv:astro-ph/0612467 [astro-ph]} \BibitemShut {NoStop}%
%%CITATION = ASTRO-PH/0612467;%%
\bibitem [{\citenamefont {Ando}\ and\ \citenamefont
  {Komatsu}(2013)}]{Ando:2013ff}%
  \BibitemOpen
  \bibfield  {author} {\bibinfo {author} {\bibfnamefont {Shin'ichiro}\
  \bibnamefont {Ando}}\ and\ \bibinfo {author} {\bibfnamefont {Eiichiro}\
  \bibnamefont {Komatsu}},\ }\bibfield  {title} {\enquote {\bibinfo {title}
  {{Constraints on the annihilation cross section of dark matter particles from
  anisotropies in the diffuse gamma-ray background measured with Fermi-LAT}},}\
  }\href {\doibase 10.1103/PhysRevD.87.123539} {\bibfield  {journal} {\bibinfo
  {journal} {Phys. Rev.}\ }\textbf {\bibinfo {volume} {D87}},\ \bibinfo {pages}
  {123539} (\bibinfo {year} {2013})},\ \Eprint {http://arxiv.org/abs/1301.5901}
  {arXiv:1301.5901 [astro-ph.CO]} \BibitemShut {NoStop}%
%%CITATION = ARXIV:1301.5901;%%
\bibitem [{\citenamefont {Branchini}\ \emph {et~al.}(2017)\citenamefont
  {Branchini}, \citenamefont {Camera}, \citenamefont {Cuoco}, \citenamefont
  {Fornengo}, \citenamefont {Regis}, \citenamefont {Viel},\ and\ \citenamefont
  {Xia}}]{Branchini:2016glc}%
  \BibitemOpen
  \bibfield  {author} {\bibinfo {author} {\bibfnamefont {Enzo}\ \bibnamefont
  {Branchini}}, \bibinfo {author} {\bibfnamefont {Stefano}\ \bibnamefont
  {Camera}}, \bibinfo {author} {\bibfnamefont {Alessandro}\ \bibnamefont
  {Cuoco}}, \bibinfo {author} {\bibfnamefont {Nicolao}\ \bibnamefont
  {Fornengo}}, \bibinfo {author} {\bibfnamefont {Marco}\ \bibnamefont {Regis}},
  \bibinfo {author} {\bibfnamefont {Matteo}\ \bibnamefont {Viel}}, \ and\
  \bibinfo {author} {\bibfnamefont {Jun-Qing}\ \bibnamefont {Xia}},\ }\bibfield
   {title} {\enquote {\bibinfo {title} {{Cross-correlating the $\gamma$-ray sky
  with Catalogs of Galaxy Clusters}},}\ }\href {\doibase
  10.3847/1538-4365/228/1/8} {\bibfield  {journal} {\bibinfo  {journal}
  {Astrophys. J. Suppl.}\ }\textbf {\bibinfo {volume} {228}},\ \bibinfo {pages}
  {8} (\bibinfo {year} {2017})},\ \Eprint {http://arxiv.org/abs/1612.05788}
  {arXiv:1612.05788 [astro-ph.CO]} \BibitemShut {NoStop}%
%%CITATION = ARXIV:1612.05788;%%
\bibitem [{\citenamefont {Xia}\ \emph {et~al.}(2011)\citenamefont {Xia},
  \citenamefont {Cuoco}, \citenamefont {Branchini}, \citenamefont {Fornasa},\
  and\ \citenamefont {Viel}}]{Xia:2011ax}%
  \BibitemOpen
  \bibfield  {author} {\bibinfo {author} {\bibfnamefont {Jun-Qing}\
  \bibnamefont {Xia}}, \bibinfo {author} {\bibfnamefont {Alessandro}\
  \bibnamefont {Cuoco}}, \bibinfo {author} {\bibfnamefont {Enzo}\ \bibnamefont
  {Branchini}}, \bibinfo {author} {\bibfnamefont {Mattia}\ \bibnamefont
  {Fornasa}}, \ and\ \bibinfo {author} {\bibfnamefont {Matteo}\ \bibnamefont
  {Viel}},\ }\bibfield  {title} {\enquote {\bibinfo {title} {{A
  cross-correlation study of the Fermi-LAT $\gamma$-ray diffuse extragalactic
  signal}},}\ }\href {\doibase 10.1111/j.1365-2966.2011.19200.x} {\bibfield
  {journal} {\bibinfo  {journal} {Mon. Not. Roy. Astron. Soc.}\ }\textbf
  {\bibinfo {volume} {416}},\ \bibinfo {pages} {2247--2264} (\bibinfo {year}
  {2011})},\ \Eprint {http://arxiv.org/abs/1103.4861} {arXiv:1103.4861
  [astro-ph.CO]} \BibitemShut {NoStop}%
%%CITATION = ARXIV:1103.4861;%%
\bibitem [{\citenamefont {Ando}(2014)}]{Ando:2014aoa}%
  \BibitemOpen
  \bibfield  {author} {\bibinfo {author} {\bibfnamefont {Shin'ichiro}\
  \bibnamefont {Ando}},\ }\bibfield  {title} {\enquote {\bibinfo {title}
  {{Power spectrum tomography of dark matter annihilation with local galaxy
  distribution}},}\ }\href {\doibase 10.1088/1475-7516/2014/10/061} {\bibfield
  {journal} {\bibinfo  {journal} {JCAP}\ }\textbf {\bibinfo {volume} {1410}},\
  \bibinfo {pages} {061} (\bibinfo {year} {2014})},\ \Eprint
  {http://arxiv.org/abs/1407.8502} {arXiv:1407.8502 [astro-ph.CO]} \BibitemShut
  {NoStop}%
%%CITATION = ARXIV:1407.8502;%%
\bibitem [{\citenamefont {Ando}\ \emph {et~al.}(2014)\citenamefont {Ando},
  \citenamefont {Benoit-L{\'e}vy},\ and\ \citenamefont
  {Komatsu}}]{Ando:2013xwa}%
  \BibitemOpen
  \bibfield  {author} {\bibinfo {author} {\bibfnamefont {Shin'ichiro}\
  \bibnamefont {Ando}}, \bibinfo {author} {\bibfnamefont {Aur{\'e}lien}\
  \bibnamefont {Benoit-L{\'e}vy}}, \ and\ \bibinfo {author} {\bibfnamefont
  {Eiichiro}\ \bibnamefont {Komatsu}},\ }\bibfield  {title} {\enquote {\bibinfo
  {title} {{Mapping dark matter in the gamma-ray sky with galaxy catalogs}},}\
  }\href {\doibase 10.1103/PhysRevD.90.023514} {\bibfield  {journal} {\bibinfo
  {journal} {Phys. Rev.}\ }\textbf {\bibinfo {volume} {D90}},\ \bibinfo {pages}
  {023514} (\bibinfo {year} {2014})},\ \Eprint {http://arxiv.org/abs/1312.4403}
  {arXiv:1312.4403 [astro-ph.CO]} \BibitemShut {NoStop}%
%%CITATION = ARXIV:1312.4403;%%
\bibitem [{\citenamefont {Xia}\ \emph {et~al.}(2015)\citenamefont {Xia},
  \citenamefont {Cuoco}, \citenamefont {Branchini},\ and\ \citenamefont
  {Viel}}]{Xia:2015wka}%
  \BibitemOpen
  \bibfield  {author} {\bibinfo {author} {\bibfnamefont {Jun-Qing}\
  \bibnamefont {Xia}}, \bibinfo {author} {\bibfnamefont {Alessandro}\
  \bibnamefont {Cuoco}}, \bibinfo {author} {\bibfnamefont {Enzo}\ \bibnamefont
  {Branchini}}, \ and\ \bibinfo {author} {\bibfnamefont {Matteo}\ \bibnamefont
  {Viel}},\ }\bibfield  {title} {\enquote {\bibinfo {title} {{Tomography of the
  Fermi-lat $\gamma$-ray Diffuse Extragalactic Signal via Cross Correlations
  With Galaxy Catalogs}},}\ }\href {\doibase 10.1088/0067-0049/217/1/15}
  {\bibfield  {journal} {\bibinfo  {journal} {Astrophys. J. Suppl.}\ }\textbf
  {\bibinfo {volume} {217}},\ \bibinfo {pages} {15} (\bibinfo {year} {2015})},\
  \Eprint {http://arxiv.org/abs/1503.05918} {arXiv:1503.05918 [astro-ph.CO]}
  \BibitemShut {NoStop}%
%%CITATION = ARXIV:1503.05918;%%
\bibitem [{\citenamefont {Regis}\ \emph {et~al.}(2015)\citenamefont {Regis},
  \citenamefont {Xia}, \citenamefont {Cuoco}, \citenamefont {Branchini},
  \citenamefont {Fornengo},\ and\ \citenamefont {Viel}}]{Regis:2015zka}%
  \BibitemOpen
  \bibfield  {author} {\bibinfo {author} {\bibfnamefont {Marco}\ \bibnamefont
  {Regis}}, \bibinfo {author} {\bibfnamefont {Jun-Qing}\ \bibnamefont {Xia}},
  \bibinfo {author} {\bibfnamefont {Alessandro}\ \bibnamefont {Cuoco}},
  \bibinfo {author} {\bibfnamefont {Enzo}\ \bibnamefont {Branchini}}, \bibinfo
  {author} {\bibfnamefont {Nicolao}\ \bibnamefont {Fornengo}}, \ and\ \bibinfo
  {author} {\bibfnamefont {Matteo}\ \bibnamefont {Viel}},\ }\bibfield  {title}
  {\enquote {\bibinfo {title} {{Particle dark matter searches outside the Local
  Group}},}\ }\href {\doibase 10.1103/PhysRevLett.114.241301} {\bibfield
  {journal} {\bibinfo  {journal} {Phys. Rev. Lett.}\ }\textbf {\bibinfo
  {volume} {114}},\ \bibinfo {pages} {241301} (\bibinfo {year} {2015})},\
  \Eprint {http://arxiv.org/abs/1503.05922} {arXiv:1503.05922 [astro-ph.CO]}
  \BibitemShut {NoStop}%
%%CITATION = ARXIV:1503.05922;%%
\bibitem [{\citenamefont {Cuoco}\ \emph {et~al.}(2015)\citenamefont {Cuoco},
  \citenamefont {Xia}, \citenamefont {Regis}, \citenamefont {Branchini},
  \citenamefont {Fornengo},\ and\ \citenamefont {Viel}}]{Cuoco:2015rfa}%
  \BibitemOpen
  \bibfield  {author} {\bibinfo {author} {\bibfnamefont {Alessandro}\
  \bibnamefont {Cuoco}}, \bibinfo {author} {\bibfnamefont {Jun-Qing}\
  \bibnamefont {Xia}}, \bibinfo {author} {\bibfnamefont {Marco}\ \bibnamefont
  {Regis}}, \bibinfo {author} {\bibfnamefont {Enzo}\ \bibnamefont {Branchini}},
  \bibinfo {author} {\bibfnamefont {Nicolao}\ \bibnamefont {Fornengo}}, \ and\
  \bibinfo {author} {\bibfnamefont {Matteo}\ \bibnamefont {Viel}},\ }\bibfield
  {title} {\enquote {\bibinfo {title} {{Dark Matter Searches in the Gamma-ray
  Extragalactic Background via Cross-correlations With Galaxy Catalogs}},}\
  }\href {\doibase 10.1088/0067-0049/221/2/29} {\bibfield  {journal} {\bibinfo
  {journal} {Astrophys. J. Suppl.}\ }\textbf {\bibinfo {volume} {221}},\
  \bibinfo {pages} {29} (\bibinfo {year} {2015})},\ \Eprint
  {http://arxiv.org/abs/1506.01030} {arXiv:1506.01030 [astro-ph.HE]}
  \BibitemShut {NoStop}%
%%CITATION = ARXIV:1506.01030;%%
\bibitem [{\citenamefont {Ando}\ and\ \citenamefont
  {Ishiwata}(2016)}]{Ando:2016ang}%
  \BibitemOpen
  \bibfield  {author} {\bibinfo {author} {\bibfnamefont {Shin'ichiro}\
  \bibnamefont {Ando}}\ and\ \bibinfo {author} {\bibfnamefont {Koji}\
  \bibnamefont {Ishiwata}},\ }\bibfield  {title} {\enquote {\bibinfo {title}
  {{Constraining particle dark matter using local galaxy distribution}},}\
  }\href {\doibase 10.1088/1475-7516/2016/06/045} {\bibfield  {journal}
  {\bibinfo  {journal} {JCAP}\ }\textbf {\bibinfo {volume} {1606}},\ \bibinfo
  {pages} {045} (\bibinfo {year} {2016})},\ \Eprint
  {http://arxiv.org/abs/1604.02263} {arXiv:1604.02263 [hep-ph]} \BibitemShut
  {NoStop}%
%%CITATION = ARXIV:1604.02263;%%
\bibitem [{\citenamefont {Camera}\ \emph {et~al.}(2015)\citenamefont {Camera},
  \citenamefont {Fornasa}, \citenamefont {Fornengo},\ and\ \citenamefont
  {Regis}}]{Camera:2014rja}%
  \BibitemOpen
  \bibfield  {author} {\bibinfo {author} {\bibfnamefont {Stefano}\ \bibnamefont
  {Camera}}, \bibinfo {author} {\bibfnamefont {Mattia}\ \bibnamefont
  {Fornasa}}, \bibinfo {author} {\bibfnamefont {Nicolao}\ \bibnamefont
  {Fornengo}}, \ and\ \bibinfo {author} {\bibfnamefont {Marco}\ \bibnamefont
  {Regis}},\ }\bibfield  {title} {\enquote {\bibinfo {title}
  {{Tomographic-spectral approach for dark matter detection in the
  cross-correlation between cosmic shear and diffuse $\gamma$-ray emission}},}\
  }\href {\doibase 10.1088/1475-7516/2015/06/029} {\bibfield  {journal}
  {\bibinfo  {journal} {JCAP}\ }\textbf {\bibinfo {volume} {1506}},\ \bibinfo
  {pages} {029} (\bibinfo {year} {2015})},\ \Eprint
  {http://arxiv.org/abs/1411.4651} {arXiv:1411.4651 [astro-ph.CO]} \BibitemShut
  {NoStop}%
%%CITATION = ARXIV:1411.4651;%%
\bibitem [{\citenamefont {Tr{\"o}ster}\ \emph {et~al.}(2016)\citenamefont
  {Tr{\"o}ster} \emph {et~al.}}]{Troster:2016sgf}%
  \BibitemOpen
  \bibfield  {author} {\bibinfo {author} {\bibfnamefont {Tilman}\ \bibnamefont
  {Tr{\"o}ster}} \emph {et~al.},\ }\bibfield  {title} {\enquote {\bibinfo
  {title} {{Cross-correlation of weak lensing and gamma rays: implications for
  the nature of dark matter}},}\ }\href {\doibase 10.1093/mnras/stx365} {\
  (\bibinfo {year} {2016}),\ 10.1093/mnras/stx365},\ \bibinfo {note} {[Mon.
  Not. Roy. Astron. Soc.467,2706(2017)]},\ \Eprint
  {http://arxiv.org/abs/1611.03554} {arXiv:1611.03554 [astro-ph.CO]}
  \BibitemShut {NoStop}%
%%CITATION = ARXIV:1611.03554;%%
\bibitem [{\citenamefont {Choi}\ \emph {et~al.}(2016)\citenamefont {Choi},
  \citenamefont {Heymans}, \citenamefont {Blake}, \citenamefont {Hildebrandt},
  \citenamefont {Duncan}, \citenamefont {Erben}, \citenamefont {Nakajima},
  \citenamefont {Van~Waerbeke},\ and\ \citenamefont {Viola}}]{Choi:2015mnp}%
  \BibitemOpen
  \bibfield  {author} {\bibinfo {author} {\bibfnamefont {Ami}\ \bibnamefont
  {Choi}}, \bibinfo {author} {\bibfnamefont {Catherine}\ \bibnamefont
  {Heymans}}, \bibinfo {author} {\bibfnamefont {Chris}\ \bibnamefont {Blake}},
  \bibinfo {author} {\bibfnamefont {Hendrik}\ \bibnamefont {Hildebrandt}},
  \bibinfo {author} {\bibfnamefont {Christopher A.~J.}\ \bibnamefont {Duncan}},
  \bibinfo {author} {\bibfnamefont {Thomas}\ \bibnamefont {Erben}}, \bibinfo
  {author} {\bibfnamefont {Reiko}\ \bibnamefont {Nakajima}}, \bibinfo {author}
  {\bibfnamefont {Ludovic}\ \bibnamefont {Van~Waerbeke}}, \ and\ \bibinfo
  {author} {\bibfnamefont {Massimo}\ \bibnamefont {Viola}},\ }\bibfield
  {title} {\enquote {\bibinfo {title} {{CFHTLenS and RCSLenS: Testing
  Photometric Redshift Distributions Using Angular Cross-Correlations with
  Spectroscopic Galaxy Surveys}},}\ }\href {\doibase 10.1093/mnras/stw2241}
  {\bibfield  {journal} {\bibinfo  {journal} {Mon. Not. Roy. Astron. Soc.}\
  }\textbf {\bibinfo {volume} {463}},\ \bibinfo {pages} {3737--3754} (\bibinfo
  {year} {2016})},\ \Eprint {http://arxiv.org/abs/1512.03626} {arXiv:1512.03626
  [astro-ph.CO]} \BibitemShut {NoStop}%
%%CITATION = ARXIV:1512.03626;%%
\bibitem [{\citenamefont {Camera}\ \emph {et~al.}(2013)\citenamefont {Camera},
  \citenamefont {Fornasa}, \citenamefont {Fornengo},\ and\ \citenamefont
  {Regis}}]{Camera:2012cj}%
  \BibitemOpen
  \bibfield  {author} {\bibinfo {author} {\bibfnamefont {Stefano}\ \bibnamefont
  {Camera}}, \bibinfo {author} {\bibfnamefont {Mattia}\ \bibnamefont
  {Fornasa}}, \bibinfo {author} {\bibfnamefont {Nicolao}\ \bibnamefont
  {Fornengo}}, \ and\ \bibinfo {author} {\bibfnamefont {Marco}\ \bibnamefont
  {Regis}},\ }\bibfield  {title} {\enquote {\bibinfo {title} {{A Novel Approach
  in the Weakly Interacting Massive Particle Quest: Cross-correlation of
  Gamma-Ray Anisotropies and Cosmic Shear}},}\ }\href {\doibase
  10.1088/2041-8205/771/1/L5} {\bibfield  {journal} {\bibinfo  {journal}
  {Astrophys. J.}\ }\textbf {\bibinfo {volume} {771}},\ \bibinfo {pages} {L5}
  (\bibinfo {year} {2013})},\ \Eprint {http://arxiv.org/abs/1212.5018}
  {arXiv:1212.5018 [astro-ph.CO]} \BibitemShut {NoStop}%
%%CITATION = ARXIV:1212.5018;%%
\bibitem [{\citenamefont {Shirasaki}\ \emph {et~al.}(2015)\citenamefont
  {Shirasaki}, \citenamefont {Horiuchi},\ and\ \citenamefont
  {Yoshida}}]{Shirasaki:2015nqp}%
  \BibitemOpen
  \bibfield  {author} {\bibinfo {author} {\bibfnamefont {Masato}\ \bibnamefont
  {Shirasaki}}, \bibinfo {author} {\bibfnamefont {Shunsaku}\ \bibnamefont
  {Horiuchi}}, \ and\ \bibinfo {author} {\bibfnamefont {Naoki}\ \bibnamefont
  {Yoshida}},\ }\bibfield  {title} {\enquote {\bibinfo {title}
  {{Cross-Correlation of the Extragalactic Gamma-ray Background with Luminous
  Red Galaxies}},}\ }\href {\doibase 10.1103/PhysRevD.92.123540} {\bibfield
  {journal} {\bibinfo  {journal} {Phys. Rev.}\ }\textbf {\bibinfo {volume}
  {D92}},\ \bibinfo {pages} {123540} (\bibinfo {year} {2015})},\ \Eprint
  {http://arxiv.org/abs/1511.07092} {arXiv:1511.07092 [astro-ph.CO]}
  \BibitemShut {NoStop}%
%%CITATION = ARXIV:1511.07092;%%
\bibitem [{\citenamefont {Shirasaki}\ \emph {et~al.}(2014)\citenamefont
  {Shirasaki}, \citenamefont {Horiuchi},\ and\ \citenamefont
  {Yoshida}}]{Shirasaki:2014noa}%
  \BibitemOpen
  \bibfield  {author} {\bibinfo {author} {\bibfnamefont {Masato}\ \bibnamefont
  {Shirasaki}}, \bibinfo {author} {\bibfnamefont {Shunsaku}\ \bibnamefont
  {Horiuchi}}, \ and\ \bibinfo {author} {\bibfnamefont {Naoki}\ \bibnamefont
  {Yoshida}},\ }\bibfield  {title} {\enquote {\bibinfo {title}
  {{Cross-Correlation of Cosmic Shear and Extragalactic Gamma-ray Background:
  Constraints on the Dark Matter Annihilation Cross-Section}},}\ }\href
  {\doibase 10.1103/PhysRevD.90.063502} {\bibfield  {journal} {\bibinfo
  {journal} {Phys. Rev.}\ }\textbf {\bibinfo {volume} {D90}},\ \bibinfo {pages}
  {063502} (\bibinfo {year} {2014})},\ \Eprint {http://arxiv.org/abs/1404.5503}
  {arXiv:1404.5503 [astro-ph.CO]} \BibitemShut {NoStop}%
%%CITATION = ARXIV:1404.5503;%%
\bibitem [{\citenamefont {Shirasaki}\ \emph {et~al.}(2016)\citenamefont
  {Shirasaki}, \citenamefont {Macias}, \citenamefont {Horiuchi}, \citenamefont
  {Shirai},\ and\ \citenamefont {Yoshida}}]{Shirasaki:2016kol}%
  \BibitemOpen
  \bibfield  {author} {\bibinfo {author} {\bibfnamefont {Masato}\ \bibnamefont
  {Shirasaki}}, \bibinfo {author} {\bibfnamefont {Oscar}\ \bibnamefont
  {Macias}}, \bibinfo {author} {\bibfnamefont {Shunsaku}\ \bibnamefont
  {Horiuchi}}, \bibinfo {author} {\bibfnamefont {Satoshi}\ \bibnamefont
  {Shirai}}, \ and\ \bibinfo {author} {\bibfnamefont {Naoki}\ \bibnamefont
  {Yoshida}},\ }\bibfield  {title} {\enquote {\bibinfo {title} {{Cosmological
  constraints on dark matter annihilation and decay: Cross-correlation analysis
  of the extragalactic $\gamma$-ray background and cosmic shear}},}\ }\href
  {\doibase 10.1103/PhysRevD.94.063522} {\bibfield  {journal} {\bibinfo
  {journal} {Phys. Rev.}\ }\textbf {\bibinfo {volume} {D94}},\ \bibinfo {pages}
  {063522} (\bibinfo {year} {2016})},\ \Eprint
  {http://arxiv.org/abs/1607.02187} {arXiv:1607.02187 [astro-ph.CO]}
  \BibitemShut {NoStop}%
%%CITATION = ARXIV:1607.02187;%%
\bibitem [{\citenamefont {Fornengo}\ \emph {et~al.}(2015)\citenamefont
  {Fornengo}, \citenamefont {Perotto}, \citenamefont {Regis},\ and\
  \citenamefont {Camera}}]{Fornengo:2014cya}%
  \BibitemOpen
  \bibfield  {author} {\bibinfo {author} {\bibfnamefont {Nicolao}\ \bibnamefont
  {Fornengo}}, \bibinfo {author} {\bibfnamefont {Laurence}\ \bibnamefont
  {Perotto}}, \bibinfo {author} {\bibfnamefont {Marco}\ \bibnamefont {Regis}},
  \ and\ \bibinfo {author} {\bibfnamefont {Stefano}\ \bibnamefont {Camera}},\
  }\bibfield  {title} {\enquote {\bibinfo {title} {{Evidence of
  Cross-correlation between the CMB Lensing and the $\Gamma$-ray sky}},}\
  }\href {\doibase 10.1088/2041-8205/802/1/L1} {\bibfield  {journal} {\bibinfo
  {journal} {Astrophys. J.}\ }\textbf {\bibinfo {volume} {802}},\ \bibinfo
  {pages} {L1} (\bibinfo {year} {2015})},\ \Eprint
  {http://arxiv.org/abs/1410.4997} {arXiv:1410.4997 [astro-ph.CO]} \BibitemShut
  {NoStop}%
%%CITATION = ARXIV:1410.4997;%%
\bibitem [{\citenamefont {Ackermann}\ \emph {et~al.}(2017)\citenamefont
  {Ackermann} \emph {et~al.}}]{Ackermann:2017nya}%
  \BibitemOpen
  \bibfield  {author} {\bibinfo {author} {\bibfnamefont {M.}~\bibnamefont
  {Ackermann}} \emph {et~al.} (\bibinfo {collaboration} {Fermi-LAT}),\
  }\bibfield  {title} {\enquote {\bibinfo {title} {{Observations of M31 and M33
  with the Fermi Large Area Telescope: A Galactic Center Excess in
  Andromeda?}}}\ }\href {\doibase 10.3847/1538-4357/aa5c3d} {\bibfield
  {journal} {\bibinfo  {journal} {Astrophys. J.}\ }\textbf {\bibinfo {volume}
  {836}},\ \bibinfo {pages} {208} (\bibinfo {year} {2017})},\ \Eprint
  {http://arxiv.org/abs/1702.08602} {arXiv:1702.08602 [astro-ph.HE]}
  \BibitemShut {NoStop}%
%%CITATION = ARXIV:1702.08602;%%
\bibitem [{\citenamefont {Ackermann}\ \emph
  {et~al.}(2015{\natexlab{c}})\citenamefont {Ackermann} \emph
  {et~al.}}]{Ackermann:2015fdi}%
  \BibitemOpen
  \bibfield  {author} {\bibinfo {author} {\bibfnamefont {M.}~\bibnamefont
  {Ackermann}} \emph {et~al.} (\bibinfo {collaboration} {Fermi-LAT}),\
  }\bibfield  {title} {\enquote {\bibinfo {title} {{Search for extended
  gamma-ray emission from the Virgo galaxy cluster with Fermi-LAT}},}\ }\href
  {\doibase 10.1088/0004-637X/812/2/159} {\bibfield  {journal} {\bibinfo
  {journal} {Astrophys. J.}\ }\textbf {\bibinfo {volume} {812}},\ \bibinfo
  {pages} {159} (\bibinfo {year} {2015}{\natexlab{c}})},\ \Eprint
  {http://arxiv.org/abs/1510.00004} {arXiv:1510.00004 [astro-ph.HE]}
  \BibitemShut {NoStop}%
%%CITATION = ARXIV:1510.00004;%%
\bibitem [{\citenamefont {Ackermann}\ \emph {et~al.}(2010)\citenamefont
  {Ackermann} \emph {et~al.}}]{Ackermann:2010rg}%
  \BibitemOpen
  \bibfield  {author} {\bibinfo {author} {\bibfnamefont {M.}~\bibnamefont
  {Ackermann}} \emph {et~al.},\ }\bibfield  {title} {\enquote {\bibinfo {title}
  {{Constraints on Dark Matter Annihilation in Clusters of Galaxies with the
  Fermi Large Area Telescope}},}\ }\href {\doibase
  10.1088/1475-7516/2010/05/025} {\bibfield  {journal} {\bibinfo  {journal}
  {JCAP}\ }\textbf {\bibinfo {volume} {1005}},\ \bibinfo {pages} {025}
  (\bibinfo {year} {2010})},\ \Eprint {http://arxiv.org/abs/1002.2239}
  {arXiv:1002.2239 [astro-ph.CO]} \BibitemShut {NoStop}%
%%CITATION = ARXIV:1002.2239;%%
\bibitem [{\citenamefont {Ando}\ and\ \citenamefont
  {Nagai}(2012)}]{Ando:2012vu}%
  \BibitemOpen
  \bibfield  {author} {\bibinfo {author} {\bibfnamefont {Shinichiro}\
  \bibnamefont {Ando}}\ and\ \bibinfo {author} {\bibfnamefont {Daisuke}\
  \bibnamefont {Nagai}},\ }\bibfield  {title} {\enquote {\bibinfo {title}
  {{Fermi-LAT constraints on dark matter annihilation cross section from
  observations of the Fornax cluster}},}\ }\href {\doibase
  10.1088/1475-7516/2012/07/017} {\bibfield  {journal} {\bibinfo  {journal}
  {JCAP}\ }\textbf {\bibinfo {volume} {1207}},\ \bibinfo {pages} {017}
  (\bibinfo {year} {2012})},\ \Eprint {http://arxiv.org/abs/1201.0753}
  {arXiv:1201.0753 [astro-ph.HE]} \BibitemShut {NoStop}%
%%CITATION = ARXIV:1201.0753;%%
\bibitem [{\citenamefont {Ackermann}\ \emph
  {et~al.}(2014{\natexlab{a}})\citenamefont {Ackermann} \emph
  {et~al.}}]{Ackermann:2013iaq}%
  \BibitemOpen
  \bibfield  {author} {\bibinfo {author} {\bibfnamefont {M.}~\bibnamefont
  {Ackermann}} \emph {et~al.} (\bibinfo {collaboration} {Fermi-LAT}),\
  }\bibfield  {title} {\enquote {\bibinfo {title} {{Search for cosmic-ray
  induced gamma-ray emission in Galaxy Clusters}},}\ }\href {\doibase
  10.1088/0004-637X/787/1/18} {\bibfield  {journal} {\bibinfo  {journal}
  {Astrophys. J.}\ }\textbf {\bibinfo {volume} {787}},\ \bibinfo {pages} {18}
  (\bibinfo {year} {2014}{\natexlab{a}})},\ \Eprint
  {http://arxiv.org/abs/1308.5654} {arXiv:1308.5654 [astro-ph.HE]} \BibitemShut
  {NoStop}%
%%CITATION = ARXIV:1308.5654;%%
\bibitem [{\citenamefont {Anderson}\ \emph {et~al.}(2016)\citenamefont
  {Anderson}, \citenamefont {Zimmer}, \citenamefont {Conrad}, \citenamefont
  {Gustafsson}, \citenamefont {S{\'a}nchez-Conde},\ and\ \citenamefont
  {Caputo}}]{Anderson:2015dpc}%
  \BibitemOpen
  \bibfield  {author} {\bibinfo {author} {\bibfnamefont {B.}~\bibnamefont
  {Anderson}}, \bibinfo {author} {\bibfnamefont {S.}~\bibnamefont {Zimmer}},
  \bibinfo {author} {\bibfnamefont {J.}~\bibnamefont {Conrad}}, \bibinfo
  {author} {\bibfnamefont {M.}~\bibnamefont {Gustafsson}}, \bibinfo {author}
  {\bibfnamefont {M.}~\bibnamefont {S{\'a}nchez-Conde}}, \ and\ \bibinfo
  {author} {\bibfnamefont {R.}~\bibnamefont {Caputo}},\ }\bibfield  {title}
  {\enquote {\bibinfo {title} {{Search for Gamma-Ray Lines towards Galaxy
  Clusters with the Fermi-LAT}},}\ }\href {\doibase
  10.1088/1475-7516/2016/02/026} {\bibfield  {journal} {\bibinfo  {journal}
  {JCAP}\ }\textbf {\bibinfo {volume} {1602}},\ \bibinfo {pages} {026}
  (\bibinfo {year} {2016})},\ \Eprint {http://arxiv.org/abs/1511.00014}
  {arXiv:1511.00014 [astro-ph.HE]} \BibitemShut {NoStop}%
%%CITATION = ARXIV:1511.00014;%%
\bibitem [{\citenamefont {Ackermann}\ \emph {et~al.}(2016)\citenamefont
  {Ackermann} \emph {et~al.}}]{Rephaeli:2015nca}%
  \BibitemOpen
  \bibfield  {author} {\bibinfo {author} {\bibfnamefont {M.}~\bibnamefont
  {Ackermann}} \emph {et~al.} (\bibinfo {collaboration} {Fermi-LAT}),\
  }\bibfield  {title} {\enquote {\bibinfo {title} {{Search for gamma-ray
  emission from the Coma Cluster with six years of Fermi-LAT data}},}\ }\href
  {\doibase 10.3847/0004-637X/819/2/149} {\bibfield  {journal} {\bibinfo
  {journal} {Astrophys. J.}\ }\textbf {\bibinfo {volume} {819}},\ \bibinfo
  {pages} {149} (\bibinfo {year} {2016})},\ \Eprint
  {http://arxiv.org/abs/1507.08995} {arXiv:1507.08995 [astro-ph.HE]}
  \BibitemShut {NoStop}%
%%CITATION = ARXIV:1507.08995;%%
\bibitem [{\citenamefont {{Ahnen}}\ \emph {et~al.}(2016)\citenamefont {{Ahnen}}
  \emph {et~al.}}]{2016A&A...589A..33A}%
  \BibitemOpen
  \bibfield  {author} {\bibinfo {author} {\bibfnamefont {M.~L.}\ \bibnamefont
  {{Ahnen}}} \emph {et~al.},\ }\bibfield  {title} {\enquote {\bibinfo {title}
  {{Deep observation of the NGC 1275 region with MAGIC: search of diffuse
  {$\gamma$}-ray emission from cosmic rays in the Perseus cluster}},}\ }\href
  {\doibase 10.1051/0004-6361/201527846} {\bibfield  {journal} {\bibinfo
  {journal} {{Astron. Astrophys.}}\ }\textbf {\bibinfo {volume} {589}},\
  \bibinfo {eid} {A33} (\bibinfo {year} {2016})},\ \Eprint
  {http://arxiv.org/abs/1602.03099} {arXiv:1602.03099 [astro-ph.HE]}
  \BibitemShut {NoStop}%
\bibitem [{\citenamefont {Liang}\ \emph {et~al.}(2016)\citenamefont {Liang},
  \citenamefont {Shen}, \citenamefont {Li}, \citenamefont {Fan}, \citenamefont
  {Huang}, \citenamefont {Lei}, \citenamefont {Feng}, \citenamefont {Liang},\
  and\ \citenamefont {Chang}}]{Liang:2016pvm}%
  \BibitemOpen
  \bibfield  {author} {\bibinfo {author} {\bibfnamefont {Yun-Feng}\
  \bibnamefont {Liang}}, \bibinfo {author} {\bibfnamefont {Zhao-Qiang}\
  \bibnamefont {Shen}}, \bibinfo {author} {\bibfnamefont {Xiang}\ \bibnamefont
  {Li}}, \bibinfo {author} {\bibfnamefont {Yi-Zhong}\ \bibnamefont {Fan}},
  \bibinfo {author} {\bibfnamefont {Xiaoyuan}\ \bibnamefont {Huang}}, \bibinfo
  {author} {\bibfnamefont {Shi-Jun}\ \bibnamefont {Lei}}, \bibinfo {author}
  {\bibfnamefont {Lei}\ \bibnamefont {Feng}}, \bibinfo {author} {\bibfnamefont
  {En-Wei}\ \bibnamefont {Liang}}, \ and\ \bibinfo {author} {\bibfnamefont
  {Jin}\ \bibnamefont {Chang}},\ }\bibfield  {title} {\enquote {\bibinfo
  {title} {{Search for a gamma-ray line feature from a group of nearby galaxy
  clusters with Fermi LAT Pass 8 data}},}\ }\href {\doibase
  10.1103/PhysRevD.93.103525} {\bibfield  {journal} {\bibinfo  {journal} {Phys.
  Rev.}\ }\textbf {\bibinfo {volume} {D93}},\ \bibinfo {pages} {103525}
  (\bibinfo {year} {2016})},\ \Eprint {http://arxiv.org/abs/1602.06527}
  {arXiv:1602.06527 [astro-ph.HE]} \BibitemShut {NoStop}%
%%CITATION = ARXIV:1602.06527;%%
\bibitem [{\citenamefont {Huchra}\ \emph {et~al.}(2012)\citenamefont {Huchra}
  \emph {et~al.}}]{Huchra:2011ii}%
  \BibitemOpen
  \bibfield  {author} {\bibinfo {author} {\bibfnamefont {John~P.}\ \bibnamefont
  {Huchra}} \emph {et~al.},\ }\bibfield  {title} {\enquote {\bibinfo {title}
  {{The 2MASS Redshift Survey - Description and Data Release}},}\ }\href
  {\doibase 10.1088/0067-0049/199/2/26} {\bibfield  {journal} {\bibinfo
  {journal} {Astrophys. J. Suppl.}\ }\textbf {\bibinfo {volume} {199}},\
  \bibinfo {pages} {26} (\bibinfo {year} {2012})},\ \Eprint
  {http://arxiv.org/abs/1108.0669} {arXiv:1108.0669 [astro-ph.CO]} \BibitemShut
  {NoStop}%
%%CITATION = ARXIV:1108.0669;%%
\bibitem [{\citenamefont {Tully}(2015)}]{Tully:2015opa}%
  \BibitemOpen
  \bibfield  {author} {\bibinfo {author} {\bibfnamefont {R.~Brent}\
  \bibnamefont {Tully}},\ }\bibfield  {title} {\enquote {\bibinfo {title}
  {{Galaxy Groups: A 2MASS Catalog}},}\ }\href {\doibase
  10.1088/0004-6256/149/5/171} {\bibfield  {journal} {\bibinfo  {journal}
  {Astron. J.}\ }\textbf {\bibinfo {volume} {149}},\ \bibinfo {pages} {171}
  (\bibinfo {year} {2015})},\ \Eprint {http://arxiv.org/abs/1503.03134}
  {arXiv:1503.03134 [astro-ph.CO]} \BibitemShut {NoStop}%
%%CITATION = ARXIV:1503.03134;%%
\bibitem [{\citenamefont {Kourkchi}\ and\ \citenamefont
  {Tully}(2017)}]{2017ApJ...843...16K}%
  \BibitemOpen
  \bibfield  {author} {\bibinfo {author} {\bibfnamefont {Ehsan}\ \bibnamefont
  {Kourkchi}}\ and\ \bibinfo {author} {\bibfnamefont {R.~Brent}\ \bibnamefont
  {Tully}},\ }\bibfield  {title} {\enquote {\bibinfo {title} {{Galaxy Groups
  Within 3500 km s$^{-1}$}},}\ }\href {\doibase 10.3847/1538-4357/aa76db}
  {\bibfield  {journal} {\bibinfo  {journal} {Astrophys. J.}\ }\textbf
  {\bibinfo {volume} {843}},\ \bibinfo {pages} {16} (\bibinfo {year} {2017})},\
  \Eprint {http://arxiv.org/abs/1705.08068} {arXiv:1705.08068 [astro-ph.GA]}
  \BibitemShut {NoStop}%
%%CITATION = ARXIV:1503.03134;%%
\bibitem [{\citenamefont {Lu}\ \emph {et~al.}(2016)\citenamefont {Lu},
  \citenamefont {Yang}, \citenamefont {Shi}, \citenamefont {Mo}, \citenamefont
  {Tweed}, \citenamefont {Wang}, \citenamefont {Zhang}, \citenamefont {Li},\
  and\ \citenamefont {Lim}}]{Lu:2016vmu}%
  \BibitemOpen
  \bibfield  {author} {\bibinfo {author} {\bibfnamefont {Yi}~\bibnamefont
  {Lu}}, \bibinfo {author} {\bibfnamefont {Xiaohu}\ \bibnamefont {Yang}},
  \bibinfo {author} {\bibfnamefont {Feng}\ \bibnamefont {Shi}}, \bibinfo
  {author} {\bibfnamefont {H.~J.}\ \bibnamefont {Mo}}, \bibinfo {author}
  {\bibfnamefont {Dylan}\ \bibnamefont {Tweed}}, \bibinfo {author}
  {\bibfnamefont {Huiyuan}\ \bibnamefont {Wang}}, \bibinfo {author}
  {\bibfnamefont {Youcai}\ \bibnamefont {Zhang}}, \bibinfo {author}
  {\bibfnamefont {Shijie}\ \bibnamefont {Li}}, \ and\ \bibinfo {author}
  {\bibfnamefont {S.~H.}\ \bibnamefont {Lim}},\ }\bibfield  {title} {\enquote
  {\bibinfo {title} {{Galaxy groups in the 2MASS Redshift Survey}},}\ }\href
  {\doibase 10.3847/0004-637X/832/1/39} {\bibfield  {journal} {\bibinfo
  {journal} {Astrophys. J.}\ }\textbf {\bibinfo {volume} {832}},\ \bibinfo
  {pages} {39} (\bibinfo {year} {2016})},\ \Eprint
  {http://arxiv.org/abs/1607.03982} {arXiv:1607.03982 [astro-ph.GA]}
  \BibitemShut {NoStop}%
%%CITATION = ARXIV:1607.03982;%%
\bibitem [{\citenamefont {Adams}\ \emph {et~al.}(2016)\citenamefont {Adams},
  \citenamefont {Bergstrom},\ and\ \citenamefont {Spolyar}}]{Adams:2016alz}%
  \BibitemOpen
  \bibfield  {author} {\bibinfo {author} {\bibfnamefont {Douglas~Quincy}\
  \bibnamefont {Adams}}, \bibinfo {author} {\bibfnamefont {Lars}\ \bibnamefont
  {Bergstrom}}, \ and\ \bibinfo {author} {\bibfnamefont {Douglas}\ \bibnamefont
  {Spolyar}},\ }\bibfield  {title} {\enquote {\bibinfo {title} {{Improved
  Constraints on Dark Matter Annihilation to a Line using Fermi-LAT
  observations of Galaxy Clusters}},}\ }\href@noop {} {\  (\bibinfo {year}
  {2016})},\ \Eprint {http://arxiv.org/abs/1606.09642} {arXiv:1606.09642
  [astro-ph.CO]} \BibitemShut {NoStop}%
%%CITATION = ARXIV:1606.09642;%%
\bibitem [{\citenamefont {Lisanti}\ \emph {et~al.}(2017)\citenamefont
  {Lisanti}, \citenamefont {Mishra-Sharma}, \citenamefont {Rodd},\ and\
  \citenamefont {Safdi}}]{companion}%
  \BibitemOpen
  \bibfield  {author} {\bibinfo {author} {\bibfnamefont {Mariangela}\
  \bibnamefont {Lisanti}}, \bibinfo {author} {\bibfnamefont {Siddharth}\
  \bibnamefont {Mishra-Sharma}}, \bibinfo {author} {\bibfnamefont
  {Nicholas~L.}\ \bibnamefont {Rodd}}, \ and\ \bibinfo {author} {\bibfnamefont
  {Benjamin~R.}\ \bibnamefont {Safdi}},\ }\bibfield  {title} {\enquote
  {\bibinfo {title} {{A Search for Dark Matter Annihilation in Galaxy
  Groups}},}\ }\href@noop {} {\  (\bibinfo {year} {2017})},\ \Eprint
  {http://arxiv.org/abs/1708.09385} {arXiv:1708.09385 [astro-ph.CO]}
  \BibitemShut {NoStop}%
%%CITATION = ARXIV:1708.09385;%%
\bibitem [{\citenamefont {Skillman}\ \emph {et~al.}(2014)\citenamefont
  {Skillman}, \citenamefont {Warren}, \citenamefont {Turk}, \citenamefont
  {Wechsler}, \citenamefont {Holz},\ and\ \citenamefont
  {Sutter}}]{Skillman:2014qca}%
  \BibitemOpen
  \bibfield  {author} {\bibinfo {author} {\bibfnamefont {Samuel~W.}\
  \bibnamefont {Skillman}}, \bibinfo {author} {\bibfnamefont {Michael~S.}\
  \bibnamefont {Warren}}, \bibinfo {author} {\bibfnamefont {Matthew~J.}\
  \bibnamefont {Turk}}, \bibinfo {author} {\bibfnamefont {Risa~H.}\
  \bibnamefont {Wechsler}}, \bibinfo {author} {\bibfnamefont {Daniel~E.}\
  \bibnamefont {Holz}}, \ and\ \bibinfo {author} {\bibfnamefont {P.~M.}\
  \bibnamefont {Sutter}},\ }\bibfield  {title} {\enquote {\bibinfo {title}
  {{Dark Sky Simulations: Early Data Release}},}\ }\href@noop {} {\  (\bibinfo
  {year} {2014})},\ \Eprint {http://arxiv.org/abs/1407.2600} {arXiv:1407.2600
  [astro-ph.CO]} \BibitemShut {NoStop}%
%%CITATION = ARXIV:1407.2600;%%
\bibitem [{\citenamefont {Lehmann}\ \emph {et~al.}(2017)\citenamefont
  {Lehmann}, \citenamefont {Mao}, \citenamefont {Becker}, \citenamefont
  {Skillman},\ and\ \citenamefont {Wechsler}}]{Lehmann:2015ioa}%
  \BibitemOpen
  \bibfield  {author} {\bibinfo {author} {\bibfnamefont {Benjamin~V.}\
  \bibnamefont {Lehmann}}, \bibinfo {author} {\bibfnamefont {Yao-Yuan}\
  \bibnamefont {Mao}}, \bibinfo {author} {\bibfnamefont {Matthew~R.}\
  \bibnamefont {Becker}}, \bibinfo {author} {\bibfnamefont {Samuel~W.}\
  \bibnamefont {Skillman}}, \ and\ \bibinfo {author} {\bibfnamefont {Risa~H.}\
  \bibnamefont {Wechsler}},\ }\bibfield  {title} {\enquote {\bibinfo {title}
  {{The Concentration Dependence of the Galaxy-Halo Connection: Modeling
  Assembly Bias with Abundance Matching}},}\ }\href {\doibase
  10.3847/1538-4357/834/1/37} {\bibfield  {journal} {\bibinfo  {journal}
  {Astrophys. J.}\ }\textbf {\bibinfo {volume} {834}},\ \bibinfo {pages} {37}
  (\bibinfo {year} {2017})},\ \Eprint {http://arxiv.org/abs/1510.05651}
  {arXiv:1510.05651 [astro-ph.CO]} \BibitemShut {NoStop}%
%%CITATION = ARXIV:1510.05651;%%
\bibitem [{\citenamefont {Bilicki}\ \emph {et~al.}(2013)\citenamefont
  {Bilicki}, \citenamefont {Jarrett}, \citenamefont {Peacock}, \citenamefont
  {Cluver},\ and\ \citenamefont {Steward}}]{Bilicki:2013sza}%
  \BibitemOpen
  \bibfield  {author} {\bibinfo {author} {\bibfnamefont {Maciej}\ \bibnamefont
  {Bilicki}}, \bibinfo {author} {\bibfnamefont {Thomas~H.}\ \bibnamefont
  {Jarrett}}, \bibinfo {author} {\bibfnamefont {John~A.}\ \bibnamefont
  {Peacock}}, \bibinfo {author} {\bibfnamefont {Michelle~E.}\ \bibnamefont
  {Cluver}}, \ and\ \bibinfo {author} {\bibfnamefont {Louise}\ \bibnamefont
  {Steward}},\ }\bibfield  {title} {\enquote {\bibinfo {title} {{2MASS
  Photometric Redshift catalog: a comprehensive three-dimensional census of the
  whole sky}},}\ }\href {\doibase 10.1088/0067-0049/210/1/9} {\bibfield
  {journal} {\bibinfo  {journal} {Astrophys. J. Suppl.}\ }\textbf {\bibinfo
  {volume} {210}},\ \bibinfo {pages} {9} (\bibinfo {year} {2013})},\ \Eprint
  {http://arxiv.org/abs/1311.5246} {arXiv:1311.5246 [astro-ph.CO]} \BibitemShut
  {NoStop}%
%%CITATION = ARXIV:1311.5246;%%
\bibitem [{\citenamefont {Warren}(2013)}]{Warren:2013vma}%
  \BibitemOpen
  \bibfield  {author} {\bibinfo {author} {\bibfnamefont {Michael~S.}\
  \bibnamefont {Warren}},\ }\bibfield  {title} {\enquote {\bibinfo {title}
  {{2HOT: An Improved Parallel Hashed Oct-Tree N-Body Algorithm for
  Cosmological Simulation}},}\ \ }(\bibinfo {year} {2013})\ \Eprint
  {http://arxiv.org/abs/1310.4502} {arXiv:1310.4502 [astro-ph.IM]} \BibitemShut
  {NoStop}%
%%CITATION = ARXIV:1310.4502;%%
\bibitem [{\citenamefont {Behroozi}\ \emph {et~al.}(2013)\citenamefont
  {Behroozi}, \citenamefont {Wechsler},\ and\ \citenamefont
  {Wu}}]{Behroozi:2011ju}%
  \BibitemOpen
  \bibfield  {author} {\bibinfo {author} {\bibfnamefont {Peter~S.}\
  \bibnamefont {Behroozi}}, \bibinfo {author} {\bibfnamefont {Risa~H.}\
  \bibnamefont {Wechsler}}, \ and\ \bibinfo {author} {\bibfnamefont {Hao-Yi}\
  \bibnamefont {Wu}},\ }\bibfield  {title} {\enquote {\bibinfo {title} {{The
  Rockstar Phase-Space Temporal Halo Finder and the Velocity Offsets of Cluster
  Cores}},}\ }\href {\doibase 10.1088/0004-637X/762/2/109} {\bibfield
  {journal} {\bibinfo  {journal} {Astrophys. J.}\ }\textbf {\bibinfo {volume}
  {762}},\ \bibinfo {pages} {109} (\bibinfo {year} {2013})},\ \Eprint
  {http://arxiv.org/abs/1110.4372} {arXiv:1110.4372 [astro-ph.CO]} \BibitemShut
  {NoStop}%
%%CITATION = ARXIV:1110.4372;%%
\bibitem [{\citenamefont {Behroozi}\ \emph {et~al.}(2010)\citenamefont
  {Behroozi}, \citenamefont {Conroy},\ and\ \citenamefont
  {Wechsler}}]{Behroozi:2010rx}%
  \BibitemOpen
  \bibfield  {author} {\bibinfo {author} {\bibfnamefont {Peter~S.}\
  \bibnamefont {Behroozi}}, \bibinfo {author} {\bibfnamefont {Charlie}\
  \bibnamefont {Conroy}}, \ and\ \bibinfo {author} {\bibfnamefont {Risa~H.}\
  \bibnamefont {Wechsler}},\ }\bibfield  {title} {\enquote {\bibinfo {title}
  {{A Comprehensive Analysis of Uncertainties Affecting the Stellar Mass-Halo
  Mass Relation for 0<z<4}},}\ }\href {\doibase 10.1088/0004-637X/717/1/379}
  {\bibfield  {journal} {\bibinfo  {journal} {Astrophys. J.}\ }\textbf
  {\bibinfo {volume} {717}},\ \bibinfo {pages} {379--403} (\bibinfo {year}
  {2010})},\ \Eprint {http://arxiv.org/abs/1001.0015} {arXiv:1001.0015
  [astro-ph.CO]} \BibitemShut {NoStop}%
%%CITATION = ARXIV:1001.0015;%%
\bibitem [{\citenamefont {Reddick}\ \emph {et~al.}(2013)\citenamefont
  {Reddick}, \citenamefont {Wechsler}, \citenamefont {Tinker},\ and\
  \citenamefont {Behroozi}}]{Reddick:2012qy}%
  \BibitemOpen
  \bibfield  {author} {\bibinfo {author} {\bibfnamefont {Rachel~M.}\
  \bibnamefont {Reddick}}, \bibinfo {author} {\bibfnamefont {Risa~H.}\
  \bibnamefont {Wechsler}}, \bibinfo {author} {\bibfnamefont {Jeremy~L.}\
  \bibnamefont {Tinker}}, \ and\ \bibinfo {author} {\bibfnamefont {Peter~S.}\
  \bibnamefont {Behroozi}},\ }\bibfield  {title} {\enquote {\bibinfo {title}
  {{The Connection between Galaxies and Dark Matter Structures in the Local
  Universe}},}\ }\href {\doibase 10.1088/0004-637X/771/1/30} {\bibfield
  {journal} {\bibinfo  {journal} {Astrophys. J.}\ }\textbf {\bibinfo {volume}
  {771}},\ \bibinfo {pages} {30} (\bibinfo {year} {2013})},\ \Eprint
  {http://arxiv.org/abs/1207.2160} {arXiv:1207.2160 [astro-ph.CO]} \BibitemShut
  {NoStop}%
%%CITATION = ARXIV:1207.2160;%%
\bibitem [{\citenamefont {Gorski}\ \emph {et~al.}(2005)\citenamefont {Gorski},
  \citenamefont {Hivon}, \citenamefont {Banday}, \citenamefont {Wandelt},
  \citenamefont {Hansen}, \citenamefont {Reinecke},\ and\ \citenamefont
  {Bartelman}}]{Gorski:2004by}%
  \BibitemOpen
  \bibfield  {author} {\bibinfo {author} {\bibfnamefont {K.~M.}\ \bibnamefont
  {Gorski}}, \bibinfo {author} {\bibfnamefont {Eric}\ \bibnamefont {Hivon}},
  \bibinfo {author} {\bibfnamefont {A.~J.}\ \bibnamefont {Banday}}, \bibinfo
  {author} {\bibfnamefont {B.~D.}\ \bibnamefont {Wandelt}}, \bibinfo {author}
  {\bibfnamefont {F.~K.}\ \bibnamefont {Hansen}}, \bibinfo {author}
  {\bibfnamefont {M.}~\bibnamefont {Reinecke}}, \ and\ \bibinfo {author}
  {\bibfnamefont {M.}~\bibnamefont {Bartelman}},\ }\bibfield  {title} {\enquote
  {\bibinfo {title} {{HEALPix - A Framework for high resolution discretization,
  and fast analysis of data distributed on the sphere}},}\ }\href {\doibase
  10.1086/427976} {\bibfield  {journal} {\bibinfo  {journal} {Astrophys. J.}\
  }\textbf {\bibinfo {volume} {622}},\ \bibinfo {pages} {759--771} (\bibinfo
  {year} {2005})},\ \Eprint {http://arxiv.org/abs/astro-ph/0409513}
  {arXiv:astro-ph/0409513 [astro-ph]} \BibitemShut {NoStop}%
%%CITATION = ASTRO-PH/0409513;%%
\bibitem [{\citenamefont {Navarro}\ \emph {et~al.}(1996)\citenamefont
  {Navarro}, \citenamefont {Frenk},\ and\ \citenamefont
  {White}}]{Navarro:1995iw}%
  \BibitemOpen
  \bibfield  {author} {\bibinfo {author} {\bibfnamefont {Julio~F.}\
  \bibnamefont {Navarro}}, \bibinfo {author} {\bibfnamefont {Carlos~S.}\
  \bibnamefont {Frenk}}, \ and\ \bibinfo {author} {\bibfnamefont {Simon D.~M.}\
  \bibnamefont {White}},\ }\bibfield  {title} {\enquote {\bibinfo {title} {{The
  Structure of cold dark matter halos}},}\ }\href {\doibase 10.1086/177173}
  {\bibfield  {journal} {\bibinfo  {journal} {Astrophys. J.}\ }\textbf
  {\bibinfo {volume} {462}},\ \bibinfo {pages} {563--575} (\bibinfo {year}
  {1996})},\ \Eprint {http://arxiv.org/abs/astro-ph/9508025}
  {arXiv:astro-ph/9508025 [astro-ph]} \BibitemShut {NoStop}%
%%CITATION = ASTRO-PH/9508025;%%
\bibitem [{\citenamefont {Cirelli}\ \emph {et~al.}(2011)\citenamefont
  {Cirelli}, \citenamefont {Corcella}, \citenamefont {Hektor}, \citenamefont
  {Hutsi}, \citenamefont {Kadastik}, \citenamefont {Panci}, \citenamefont
  {Raidal}, \citenamefont {Sala},\ and\ \citenamefont
  {Strumia}}]{Cirelli:2010xx}%
  \BibitemOpen
  \bibfield  {author} {\bibinfo {author} {\bibfnamefont {Marco}\ \bibnamefont
  {Cirelli}}, \bibinfo {author} {\bibfnamefont {Gennaro}\ \bibnamefont
  {Corcella}}, \bibinfo {author} {\bibfnamefont {Andi}\ \bibnamefont {Hektor}},
  \bibinfo {author} {\bibfnamefont {Gert}\ \bibnamefont {Hutsi}}, \bibinfo
  {author} {\bibfnamefont {Mario}\ \bibnamefont {Kadastik}}, \bibinfo {author}
  {\bibfnamefont {Paolo}\ \bibnamefont {Panci}}, \bibinfo {author}
  {\bibfnamefont {Martti}\ \bibnamefont {Raidal}}, \bibinfo {author}
  {\bibfnamefont {Filippo}\ \bibnamefont {Sala}}, \ and\ \bibinfo {author}
  {\bibfnamefont {Alessandro}\ \bibnamefont {Strumia}},\ }\bibfield  {title}
  {\enquote {\bibinfo {title} {{PPPC 4 DM ID: A Poor Particle Physicist
  Cookbook for Dark Matter Indirect Detection}},}\ }\href {\doibase
  10.1088/1475-7516/2012/10/E01, 10.1088/1475-7516/2011/03/051} {\bibfield
  {journal} {\bibinfo  {journal} {JCAP}\ }\textbf {\bibinfo {volume} {1103}},\
  \bibinfo {pages} {051} (\bibinfo {year} {2011})},\ \bibinfo {note} {[Erratum:
  JCAP1210,E01(2012)]},\ \Eprint {http://arxiv.org/abs/1012.4515}
  {arXiv:1012.4515 [hep-ph]} \BibitemShut {NoStop}%
%%CITATION = ARXIV:1012.4515;%%
\bibitem [{\citenamefont {Elor}\ \emph {et~al.}(2015)\citenamefont {Elor},
  \citenamefont {Rodd},\ and\ \citenamefont {Slatyer}}]{Elor:2015tva}%
  \BibitemOpen
  \bibfield  {author} {\bibinfo {author} {\bibfnamefont {Gilly}\ \bibnamefont
  {Elor}}, \bibinfo {author} {\bibfnamefont {Nicholas~L.}\ \bibnamefont
  {Rodd}}, \ and\ \bibinfo {author} {\bibfnamefont {Tracy~R.}\ \bibnamefont
  {Slatyer}},\ }\bibfield  {title} {\enquote {\bibinfo {title} {{Multistep
  cascade annihilations of dark matter and the Galactic Center excess}},}\
  }\href {\doibase 10.1103/PhysRevD.91.103531} {\bibfield  {journal} {\bibinfo
  {journal} {Phys. Rev.}\ }\textbf {\bibinfo {volume} {D91}},\ \bibinfo {pages}
  {103531} (\bibinfo {year} {2015})},\ \Eprint
  {http://arxiv.org/abs/1503.01773} {arXiv:1503.01773 [hep-ph]} \BibitemShut
  {NoStop}%
%%CITATION = ARXIV:1503.01773;%%
\bibitem [{\citenamefont {Elor}\ \emph {et~al.}(2016)\citenamefont {Elor},
  \citenamefont {Rodd}, \citenamefont {Slatyer},\ and\ \citenamefont
  {Xue}}]{Elor:2015bho}%
  \BibitemOpen
  \bibfield  {author} {\bibinfo {author} {\bibfnamefont {Gilly}\ \bibnamefont
  {Elor}}, \bibinfo {author} {\bibfnamefont {Nicholas~L.}\ \bibnamefont
  {Rodd}}, \bibinfo {author} {\bibfnamefont {Tracy~R.}\ \bibnamefont
  {Slatyer}}, \ and\ \bibinfo {author} {\bibfnamefont {Wei}\ \bibnamefont
  {Xue}},\ }\bibfield  {title} {\enquote {\bibinfo {title} {{Model-Independent
  Indirect Detection Constraints on Hidden Sector Dark Matter}},}\ }\href
  {\doibase 10.1088/1475-7516/2016/06/024} {\bibfield  {journal} {\bibinfo
  {journal} {JCAP}\ }\textbf {\bibinfo {volume} {1606}},\ \bibinfo {pages}
  {024} (\bibinfo {year} {2016})},\ \Eprint {http://arxiv.org/abs/1511.08787}
  {arXiv:1511.08787 [hep-ph]} \BibitemShut {NoStop}%
%%CITATION = ARXIV:1511.08787;%%
\bibitem [{\citenamefont {Bartels}\ and\ \citenamefont
  {Ando}(2015)}]{Bartels:2015uba}%
  \BibitemOpen
  \bibfield  {author} {\bibinfo {author} {\bibfnamefont {Richard}\ \bibnamefont
  {Bartels}}\ and\ \bibinfo {author} {\bibfnamefont {Shin'ichiro}\ \bibnamefont
  {Ando}},\ }\bibfield  {title} {\enquote {\bibinfo {title} {{Boosting the
  annihilation boost: Tidal effects on dark matter subhalos and consistent
  luminosity modeling}},}\ }\href {\doibase 10.1103/PhysRevD.92.123508}
  {\bibfield  {journal} {\bibinfo  {journal} {Phys. Rev.}\ }\textbf {\bibinfo
  {volume} {D92}},\ \bibinfo {pages} {123508} (\bibinfo {year} {2015})},\
  \Eprint {http://arxiv.org/abs/1507.08656} {arXiv:1507.08656 [astro-ph.CO]}
  \BibitemShut {NoStop}%
%%CITATION = ARXIV:1507.08656;%%
\bibitem [{\citenamefont {Tully}\ \emph {et~al.}(2016)\citenamefont {Tully},
  \citenamefont {Courtois},\ and\ \citenamefont {Sorce}}]{Tully:2016ppz}%
  \BibitemOpen
  \bibfield  {author} {\bibinfo {author} {\bibfnamefont {R.~Brent}\
  \bibnamefont {Tully}}, \bibinfo {author} {\bibfnamefont {Helene~M.}\
  \bibnamefont {Courtois}}, \ and\ \bibinfo {author} {\bibfnamefont {Jenny~G.}\
  \bibnamefont {Sorce}},\ }\bibfield  {title} {\enquote {\bibinfo {title}
  {{Cosmicflows-3}},}\ }\href {\doibase 10.3847/0004-6256/152/2/50} {\bibfield
  {journal} {\bibinfo  {journal} {Astron. J.}\ }\textbf {\bibinfo {volume}
  {152}},\ \bibinfo {pages} {50} (\bibinfo {year} {2016})},\ \Eprint
  {http://arxiv.org/abs/1605.01765} {arXiv:1605.01765 [astro-ph.CO]}
  \BibitemShut {NoStop}%
%%CITATION = ARXIV:1605.01765;%%
\bibitem [{\citenamefont {Vale}\ and\ \citenamefont
  {Ostriker}(2006)}]{Vale:2005mw}%
  \BibitemOpen
  \bibfield  {author} {\bibinfo {author} {\bibfnamefont {Antonio}\ \bibnamefont
  {Vale}}\ and\ \bibinfo {author} {\bibfnamefont {J.~P.}\ \bibnamefont
  {Ostriker}},\ }\bibfield  {title} {\enquote {\bibinfo {title} {{The
  non-parametric model for linking galaxy luminosity with halo/subhalo
  mass}},}\ }\href {\doibase 10.1111/j.1365-2966.2006.10605.x} {\bibfield
  {journal} {\bibinfo  {journal} {Mon. Not. Roy. Astron. Soc.}\ }\textbf
  {\bibinfo {volume} {371}},\ \bibinfo {pages} {1173--1187} (\bibinfo {year}
  {2006})},\ \Eprint {http://arxiv.org/abs/astro-ph/0511816}
  {arXiv:astro-ph/0511816 [astro-ph]} \BibitemShut {NoStop}%
%%CITATION = ASTRO-PH/0511816;%%
\bibitem [{\citenamefont {Bryan}\ and\ \citenamefont
  {Norman}(1998)}]{Bryan:1997dn}%
  \BibitemOpen
  \bibfield  {author} {\bibinfo {author} {\bibfnamefont {G.~L.}\ \bibnamefont
  {Bryan}}\ and\ \bibinfo {author} {\bibfnamefont {M.~L.}\ \bibnamefont
  {Norman}},\ }\bibfield  {title} {\enquote {\bibinfo {title} {{Statistical
  properties of x-ray clusters: Analytic and numerical comparisons}},}\ }\href
  {\doibase 10.1086/305262} {\bibfield  {journal} {\bibinfo  {journal}
  {Astrophys. J.}\ }\textbf {\bibinfo {volume} {495}},\ \bibinfo {pages} {80}
  (\bibinfo {year} {1998})},\ \Eprint {http://arxiv.org/abs/astro-ph/9710107}
  {arXiv:astro-ph/9710107 [astro-ph]} \BibitemShut {NoStop}%
%%CITATION = ASTRO-PH/9710107;%%
\bibitem [{\citenamefont {Correa}\ \emph {et~al.}(2015)\citenamefont {Correa},
  \citenamefont {Wyithe}, \citenamefont {Schaye},\ and\ \citenamefont
  {Duffy}}]{Correa:2015dva}%
  \BibitemOpen
  \bibfield  {author} {\bibinfo {author} {\bibfnamefont {Camila~A.}\
  \bibnamefont {Correa}}, \bibinfo {author} {\bibfnamefont {J.~Stuart~B.}\
  \bibnamefont {Wyithe}}, \bibinfo {author} {\bibfnamefont {Joop}\ \bibnamefont
  {Schaye}}, \ and\ \bibinfo {author} {\bibfnamefont {Alan~R.}\ \bibnamefont
  {Duffy}},\ }\bibfield  {title} {\enquote {\bibinfo {title} {{The accretion
  history of dark matter haloes -- III. A physical model for the
  concentration--mass relation}},}\ }\href {\doibase 10.1093/mnras/stv1363}
  {\bibfield  {journal} {\bibinfo  {journal} {Mon. Not. Roy. Astron. Soc.}\
  }\textbf {\bibinfo {volume} {452}},\ \bibinfo {pages} {1217--1232} (\bibinfo
  {year} {2015})},\ \Eprint {http://arxiv.org/abs/1502.00391} {arXiv:1502.00391
  [astro-ph.CO]} \BibitemShut {NoStop}%
%%CITATION = ARXIV:1502.00391;%%
\bibitem [{\citenamefont {Diemer}\ and\ \citenamefont
  {Kravtsov}(2015)}]{Diemer:2014gba}%
  \BibitemOpen
  \bibfield  {author} {\bibinfo {author} {\bibfnamefont {Benedikt}\
  \bibnamefont {Diemer}}\ and\ \bibinfo {author} {\bibfnamefont {Andrey~V.}\
  \bibnamefont {Kravtsov}},\ }\bibfield  {title} {\enquote {\bibinfo {title}
  {{A universal model for halo concentrations}},}\ }\href {\doibase
  10.1088/0004-637X/799/1/108} {\bibfield  {journal} {\bibinfo  {journal}
  {Astrophys. J.}\ }\textbf {\bibinfo {volume} {799}},\ \bibinfo {pages} {108}
  (\bibinfo {year} {2015})},\ \Eprint {http://arxiv.org/abs/1407.4730}
  {arXiv:1407.4730 [astro-ph.CO]} \BibitemShut {NoStop}%
%%CITATION = ARXIV:1407.4730;%%
\bibitem [{\citenamefont {Prada}\ \emph {et~al.}(2012)\citenamefont {Prada},
  \citenamefont {Klypin}, \citenamefont {Cuesta}, \citenamefont
  {Betancort-Rijo},\ and\ \citenamefont {Primack}}]{Prada:2011jf}%
  \BibitemOpen
  \bibfield  {author} {\bibinfo {author} {\bibfnamefont {Francisco}\
  \bibnamefont {Prada}}, \bibinfo {author} {\bibfnamefont {Anatoly~A.}\
  \bibnamefont {Klypin}}, \bibinfo {author} {\bibfnamefont {Antonio~J.}\
  \bibnamefont {Cuesta}}, \bibinfo {author} {\bibfnamefont {Juan~E.}\
  \bibnamefont {Betancort-Rijo}}, \ and\ \bibinfo {author} {\bibfnamefont
  {Joel}\ \bibnamefont {Primack}},\ }\bibfield  {title} {\enquote {\bibinfo
  {title} {{Halo concentrations in the standard LCDM cosmology}},}\ }\href
  {\doibase 10.1111/j.1365-2966.2012.21007.x} {\bibfield  {journal} {\bibinfo
  {journal} {Mon. Not. Roy. Astron. Soc.}\ }\textbf {\bibinfo {volume} {423}},\
  \bibinfo {pages} {3018--3030} (\bibinfo {year} {2012})},\ \Eprint
  {http://arxiv.org/abs/1104.5130} {arXiv:1104.5130 [astro-ph.CO]} \BibitemShut
  {NoStop}%
%%CITATION = ARXIV:1104.5130;%%
\bibitem [{\citenamefont {S{\'a}nchez-Conde}\ and\ \citenamefont
  {Prada}(2014)}]{Sanchez-Conde:2013yxa}%
  \BibitemOpen
  \bibfield  {author} {\bibinfo {author} {\bibfnamefont {Miguel~A.}\
  \bibnamefont {S{\'a}nchez-Conde}}\ and\ \bibinfo {author} {\bibfnamefont
  {Francisco}\ \bibnamefont {Prada}},\ }\bibfield  {title} {\enquote {\bibinfo
  {title} {{The flattening of the concentration--mass relation towards low halo
  masses and its implications for the annihilation signal boost}},}\ }\href
  {\doibase 10.1093/mnras/stu1014} {\bibfield  {journal} {\bibinfo  {journal}
  {Mon. Not. Roy. Astron. Soc.}\ }\textbf {\bibinfo {volume} {442}},\ \bibinfo
  {pages} {2271--2277} (\bibinfo {year} {2014})},\ \Eprint
  {http://arxiv.org/abs/1312.1729} {arXiv:1312.1729 [astro-ph.CO]} \BibitemShut
  {NoStop}%
%%CITATION = ARXIV:1312.1729;%%
\bibitem [{\citenamefont {Molin{\'e}}\ \emph {et~al.}(2017)\citenamefont
  {Molin{\'e}}, \citenamefont {S{\'a}nchez-Conde}, \citenamefont
  {Palomares-Ruiz},\ and\ \citenamefont {Prada}}]{Moline:2016pbm}%
  \BibitemOpen
  \bibfield  {author} {\bibinfo {author} {\bibfnamefont {{\'A}ngeles}\
  \bibnamefont {Molin{\'e}}}, \bibinfo {author} {\bibfnamefont {Miguel~A.}\
  \bibnamefont {S{\'a}nchez-Conde}}, \bibinfo {author} {\bibfnamefont {Sergio}\
  \bibnamefont {Palomares-Ruiz}}, \ and\ \bibinfo {author} {\bibfnamefont
  {Francisco}\ \bibnamefont {Prada}},\ }\bibfield  {title} {\enquote {\bibinfo
  {title} {{Characterization of subhalo structural properties and implications
  for dark matter annihilation signals}},}\ }\href {\doibase
  10.1093/mnras/stx026} {\bibfield  {journal} {\bibinfo  {journal} {Mon. Not.
  Roy. Astron. Soc.}\ }\textbf {\bibinfo {volume} {466}},\ \bibinfo {pages}
  {4974--4990} (\bibinfo {year} {2017})},\ \Eprint
  {http://arxiv.org/abs/1603.04057} {arXiv:1603.04057 [astro-ph.CO]}
  \BibitemShut {NoStop}%
%%CITATION = ARXIV:1603.04057;%%
\bibitem [{\citenamefont {Gao}\ \emph {et~al.}(2012)\citenamefont {Gao},
  \citenamefont {Frenk}, \citenamefont {Jenkins}, \citenamefont {Springel},\
  and\ \citenamefont {White}}]{Gao:2011rf}%
  \BibitemOpen
  \bibfield  {author} {\bibinfo {author} {\bibfnamefont {L.}~\bibnamefont
  {Gao}}, \bibinfo {author} {\bibfnamefont {C.~S.}\ \bibnamefont {Frenk}},
  \bibinfo {author} {\bibfnamefont {A.}~\bibnamefont {Jenkins}}, \bibinfo
  {author} {\bibfnamefont {V.}~\bibnamefont {Springel}}, \ and\ \bibinfo
  {author} {\bibfnamefont {S.~D.~M.}\ \bibnamefont {White}},\ }\bibfield
  {title} {\enquote {\bibinfo {title} {{Where will supersymmetric dark matter
  first be seen?}}}\ }\href {\doibase 10.1111/j.1365-2966.2011.19836.x}
  {\bibfield  {journal} {\bibinfo  {journal} {Mon. Not. Roy. Astron. Soc.}\
  }\textbf {\bibinfo {volume} {419}},\ \bibinfo {pages} {1721} (\bibinfo {year}
  {2012})},\ \Eprint {http://arxiv.org/abs/1107.1916} {arXiv:1107.1916
  [astro-ph.CO]} \BibitemShut {NoStop}%
%%CITATION = ARXIV:1107.1916;%%
\bibitem [{\citenamefont {Duffy}\ \emph {et~al.}(2008)\citenamefont {Duffy},
  \citenamefont {Schaye}, \citenamefont {Kay},\ and\ \citenamefont
  {Dalla~Vecchia}}]{Duffy:2008pz}%
  \BibitemOpen
  \bibfield  {author} {\bibinfo {author} {\bibfnamefont {Alan~R.}\ \bibnamefont
  {Duffy}}, \bibinfo {author} {\bibfnamefont {Joop}\ \bibnamefont {Schaye}},
  \bibinfo {author} {\bibfnamefont {Scott~T.}\ \bibnamefont {Kay}}, \ and\
  \bibinfo {author} {\bibfnamefont {Claudio}\ \bibnamefont {Dalla~Vecchia}},\
  }\bibfield  {title} {\enquote {\bibinfo {title} {{Dark matter halo
  concentrations in the Wilkinson Microwave Anisotropy Probe year 5
  cosmology}},}\ }\href {\doibase 10.1111/j.1745-3933.2008.00537.x} {\bibfield
  {journal} {\bibinfo  {journal} {Mon. Not. Roy. Astron. Soc.}\ }\textbf
  {\bibinfo {volume} {390}},\ \bibinfo {pages} {L64} (\bibinfo {year}
  {2008})},\ \bibinfo {note} {[Erratum: Mon. Not. Roy. Astron.
  Soc.415,L85(2011)]},\ \Eprint {http://arxiv.org/abs/0804.2486}
  {arXiv:0804.2486 [astro-ph]} \BibitemShut {NoStop}%
%%CITATION = ARXIV:0804.2486;%%
\bibitem [{\citenamefont {Klypin}\ \emph {et~al.}(2016)\citenamefont {Klypin},
  \citenamefont {Yepes}, \citenamefont {Gottlober}, \citenamefont {Prada},\
  and\ \citenamefont {Hess}}]{Klypin:2014kpa}%
  \BibitemOpen
  \bibfield  {author} {\bibinfo {author} {\bibfnamefont {Anatoly}\ \bibnamefont
  {Klypin}}, \bibinfo {author} {\bibfnamefont {Gustavo}\ \bibnamefont {Yepes}},
  \bibinfo {author} {\bibfnamefont {Stefan}\ \bibnamefont {Gottlober}},
  \bibinfo {author} {\bibfnamefont {Francisco}\ \bibnamefont {Prada}}, \ and\
  \bibinfo {author} {\bibfnamefont {Steffen}\ \bibnamefont {Hess}},\ }\bibfield
   {title} {\enquote {\bibinfo {title} {{MultiDark simulations: the story of
  dark matter halo concentrations and density profiles}},}\ }\href {\doibase
  10.1093/mnras/stw248} {\bibfield  {journal} {\bibinfo  {journal} {Mon. Not.
  Roy. Astron. Soc.}\ }\textbf {\bibinfo {volume} {457}},\ \bibinfo {pages}
  {4340--4359} (\bibinfo {year} {2016})},\ \Eprint
  {http://arxiv.org/abs/1411.4001} {arXiv:1411.4001 [astro-ph.CO]} \BibitemShut
  {NoStop}%
%%CITATION = ARXIV:1411.4001;%%
\bibitem [{\citenamefont {Dutton}\ and\ \citenamefont
  {Macci{\`o}}(2014)}]{Dutton:2014xda}%
  \BibitemOpen
  \bibfield  {author} {\bibinfo {author} {\bibfnamefont {Aaron~A.}\
  \bibnamefont {Dutton}}\ and\ \bibinfo {author} {\bibfnamefont {Andrea~V.}\
  \bibnamefont {Macci{\`o}}},\ }\bibfield  {title} {\enquote {\bibinfo {title}
  {{Cold dark matter haloes in the Planck era: evolution of structural
  parameters for Einasto and NFW profiles}},}\ }\href {\doibase
  10.1093/mnras/stu742} {\bibfield  {journal} {\bibinfo  {journal} {Mon. Not.
  Roy. Astron. Soc.}\ }\textbf {\bibinfo {volume} {441}},\ \bibinfo {pages}
  {3359--3374} (\bibinfo {year} {2014})},\ \Eprint
  {http://arxiv.org/abs/1402.7073} {arXiv:1402.7073 [astro-ph.CO]} \BibitemShut
  {NoStop}%
%%CITATION = ARXIV:1402.7073;%%
\bibitem [{\citenamefont {H{\"u}tten}\ \emph {et~al.}(2016)\citenamefont
  {H{\"u}tten}, \citenamefont {Combet}, \citenamefont {Maier},\ and\
  \citenamefont {Maurin}}]{Hutten:2016jko}%
  \BibitemOpen
  \bibfield  {author} {\bibinfo {author} {\bibfnamefont {M.}~\bibnamefont
  {H{\"u}tten}}, \bibinfo {author} {\bibfnamefont {C.}~\bibnamefont {Combet}},
  \bibinfo {author} {\bibfnamefont {G.}~\bibnamefont {Maier}}, \ and\ \bibinfo
  {author} {\bibfnamefont {D.}~\bibnamefont {Maurin}},\ }\bibfield  {title}
  {\enquote {\bibinfo {title} {{Dark matter substructure modelling and
  sensitivity of the Cherenkov Telescope Array to Galactic dark halos}},}\
  }\href {\doibase 10.1088/1475-7516/2016/09/047} {\bibfield  {journal}
  {\bibinfo  {journal} {JCAP}\ }\textbf {\bibinfo {volume} {1609}},\ \bibinfo
  {pages} {047} (\bibinfo {year} {2016})},\ \Eprint
  {http://arxiv.org/abs/1606.04898} {arXiv:1606.04898 [astro-ph.HE]}
  \BibitemShut {NoStop}%
%%CITATION = ARXIV:1606.04898;%%
\bibitem [{\citenamefont {Su}\ \emph {et~al.}(2010)\citenamefont {Su},
  \citenamefont {Slatyer},\ and\ \citenamefont {Finkbeiner}}]{Su:2010qj}%
  \BibitemOpen
  \bibfield  {author} {\bibinfo {author} {\bibfnamefont {Meng}\ \bibnamefont
  {Su}}, \bibinfo {author} {\bibfnamefont {Tracy~R.}\ \bibnamefont {Slatyer}},
  \ and\ \bibinfo {author} {\bibfnamefont {Douglas~P.}\ \bibnamefont
  {Finkbeiner}},\ }\bibfield  {title} {\enquote {\bibinfo {title} {{Giant
  Gamma-ray Bubbles from Fermi-LAT: AGN Activity or Bipolar Galactic Wind?}}}\
  }\href {\doibase 10.1088/0004-637X/724/2/1044} {\bibfield  {journal}
  {\bibinfo  {journal} {Astrophys. J.}\ }\textbf {\bibinfo {volume} {724}},\
  \bibinfo {pages} {1044--1082} (\bibinfo {year} {2010})},\ \Eprint
  {http://arxiv.org/abs/1005.5480} {arXiv:1005.5480 [astro-ph.HE]} \BibitemShut
  {NoStop}%
%%CITATION = ARXIV:1005.5480;%%
\bibitem [{\citenamefont {Acero}\ \emph {et~al.}(2015)\citenamefont {Acero}
  \emph {et~al.}}]{Acero:2015hja}%
  \BibitemOpen
  \bibfield  {author} {\bibinfo {author} {\bibfnamefont {F.}~\bibnamefont
  {Acero}} \emph {et~al.} (\bibinfo {collaboration} {Fermi-LAT}),\ }\bibfield
  {title} {\enquote {\bibinfo {title} {{Fermi Large Area Telescope Third Source
  Catalog}},}\ }\href {\doibase 10.1088/0067-0049/218/2/23} {\bibfield
  {journal} {\bibinfo  {journal} {Astrophys. J. Suppl.}\ }\textbf {\bibinfo
  {volume} {218}},\ \bibinfo {pages} {23} (\bibinfo {year} {2015})},\ \Eprint
  {http://arxiv.org/abs/1501.02003} {arXiv:1501.02003 [astro-ph.HE]}
  \BibitemShut {NoStop}%
%%CITATION = ARXIV:1501.02003;%%
\bibitem [{\citenamefont {Rolke}\ \emph {et~al.}(2005)\citenamefont {Rolke},
  \citenamefont {Lopez},\ and\ \citenamefont {Conrad}}]{Rolke:2004mj}%
  \BibitemOpen
  \bibfield  {author} {\bibinfo {author} {\bibfnamefont {Wolfgang~A.}\
  \bibnamefont {Rolke}}, \bibinfo {author} {\bibfnamefont {Angel~M.}\
  \bibnamefont {Lopez}}, \ and\ \bibinfo {author} {\bibfnamefont {Jan}\
  \bibnamefont {Conrad}},\ }\bibfield  {title} {\enquote {\bibinfo {title}
  {{Limits and confidence intervals in the presence of nuisance parameters}},}\
  }\href {\doibase 10.1016/j.nima.2005.05.068} {\bibfield  {journal} {\bibinfo
  {journal} {Nucl. Instrum. Meth.}\ }\textbf {\bibinfo {volume} {A551}},\
  \bibinfo {pages} {493--503} (\bibinfo {year} {2005})},\ \Eprint
  {http://arxiv.org/abs/physics/0403059} {arXiv:physics/0403059 [physics]}
  \BibitemShut {NoStop}%
%%CITATION = PHYSICS/0403059;%%
\bibitem [{\citenamefont {Mishra-Sharma}\ \emph {et~al.}(2017)\citenamefont
  {Mishra-Sharma}, \citenamefont {Rodd},\ and\ \citenamefont
  {Safdi}}]{Mishra-Sharma:2016gis}%
  \BibitemOpen
  \bibfield  {author} {\bibinfo {author} {\bibfnamefont {Siddharth}\
  \bibnamefont {Mishra-Sharma}}, \bibinfo {author} {\bibfnamefont
  {Nicholas~L.}\ \bibnamefont {Rodd}}, \ and\ \bibinfo {author} {\bibfnamefont
  {Benjamin~R.}\ \bibnamefont {Safdi}},\ }\bibfield  {title} {\enquote
  {\bibinfo {title} {{NPTFit: A code package for Non-Poissonian Template
  Fitting}},}\ }\href {\doibase 10.3847/1538-3881/aa6d5f} {\bibfield  {journal}
  {\bibinfo  {journal} {Astron. J.}\ }\textbf {\bibinfo {volume} {153}},\
  \bibinfo {pages} {253} (\bibinfo {year} {2017})},\ \Eprint
  {http://arxiv.org/abs/1612.03173} {arXiv:1612.03173 [astro-ph.HE]}
  \BibitemShut {NoStop}%
%%CITATION = ARXIV:1612.03173;%%
\bibitem [{\citenamefont {Feroz}\ \emph {et~al.}(2009)\citenamefont {Feroz},
  \citenamefont {Hobson},\ and\ \citenamefont {Bridges}}]{Feroz:2008xx}%
  \BibitemOpen
  \bibfield  {author} {\bibinfo {author} {\bibfnamefont {F.}~\bibnamefont
  {Feroz}}, \bibinfo {author} {\bibfnamefont {M.~P.}\ \bibnamefont {Hobson}}, \
  and\ \bibinfo {author} {\bibfnamefont {M.}~\bibnamefont {Bridges}},\
  }\bibfield  {title} {\enquote {\bibinfo {title} {{MultiNest: an efficient and
  robust Bayesian inference tool for cosmology and particle physics}},}\ }\href
  {\doibase 10.1111/j.1365-2966.2009.14548.x} {\bibfield  {journal} {\bibinfo
  {journal} {Mon. Not. Roy. Astron. Soc.}\ }\textbf {\bibinfo {volume} {398}},\
  \bibinfo {pages} {1601--1614} (\bibinfo {year} {2009})},\ \Eprint
  {http://arxiv.org/abs/0809.3437} {arXiv:0809.3437 [astro-ph]} \BibitemShut
  {NoStop}%
%%CITATION = ARXIV:0809.3437;%%
\bibitem [{\citenamefont {Buchner}\ \emph {et~al.}(2014)\citenamefont
  {Buchner}, \citenamefont {Georgakakis}, \citenamefont {Nandra}, \citenamefont
  {Hsu}, \citenamefont {Rangel}, \citenamefont {Brightman}, \citenamefont
  {Merloni}, \citenamefont {Salvato}, \citenamefont {Donley},\ and\
  \citenamefont {Kocevski}}]{Buchner:2014nha}%
  \BibitemOpen
  \bibfield  {author} {\bibinfo {author} {\bibfnamefont {J.}~\bibnamefont
  {Buchner}}, \bibinfo {author} {\bibfnamefont {A.}~\bibnamefont
  {Georgakakis}}, \bibinfo {author} {\bibfnamefont {K.}~\bibnamefont {Nandra}},
  \bibinfo {author} {\bibfnamefont {L.}~\bibnamefont {Hsu}}, \bibinfo {author}
  {\bibfnamefont {C.}~\bibnamefont {Rangel}}, \bibinfo {author} {\bibfnamefont
  {M.}~\bibnamefont {Brightman}}, \bibinfo {author} {\bibfnamefont
  {A.}~\bibnamefont {Merloni}}, \bibinfo {author} {\bibfnamefont
  {M.}~\bibnamefont {Salvato}}, \bibinfo {author} {\bibfnamefont
  {J.}~\bibnamefont {Donley}}, \ and\ \bibinfo {author} {\bibfnamefont
  {D.}~\bibnamefont {Kocevski}},\ }\bibfield  {title} {\enquote {\bibinfo
  {title} {{X-ray spectral modelling of the AGN obscuring region in the CDFS:
  Bayesian model selection and catalogue}},}\ }\href {\doibase
  10.1051/0004-6361/201322971} {\bibfield  {journal} {\bibinfo  {journal}
  {Astron. Astrophys.}\ }\textbf {\bibinfo {volume} {564}},\ \bibinfo {pages}
  {A125} (\bibinfo {year} {2014})},\ \Eprint {http://arxiv.org/abs/1402.0004}
  {arXiv:1402.0004 [astro-ph.HE]} \BibitemShut {NoStop}%
%%CITATION = ARXIV:1402.0004;%%
\bibitem [{\citenamefont {James}\ and\ \citenamefont
  {Roos}(1975)}]{James:1975dr}%
  \BibitemOpen
  \bibfield  {author} {\bibinfo {author} {\bibfnamefont {F.}~\bibnamefont
  {James}}\ and\ \bibinfo {author} {\bibfnamefont {M.}~\bibnamefont {Roos}},\
  }\bibfield  {title} {\enquote {\bibinfo {title} {{Minuit: A System for
  Function Minimization and Analysis of the Parameter Errors and
  Correlations}},}\ }\href {\doibase 10.1016/0010-4655(75)90039-9} {\bibfield
  {journal} {\bibinfo  {journal} {Comput. Phys. Commun.}\ }\textbf {\bibinfo
  {volume} {10}},\ \bibinfo {pages} {343--367} (\bibinfo {year}
  {1975})}\BibitemShut {NoStop}%
%%CITATION = CPHCB,10,343;%%
\bibitem [{\citenamefont {Steigman}\ \emph {et~al.}(2012)\citenamefont
  {Steigman}, \citenamefont {Dasgupta},\ and\ \citenamefont
  {Beacom}}]{Steigman:2012nb}%
  \BibitemOpen
  \bibfield  {author} {\bibinfo {author} {\bibfnamefont {Gary}\ \bibnamefont
  {Steigman}}, \bibinfo {author} {\bibfnamefont {Basudeb}\ \bibnamefont
  {Dasgupta}}, \ and\ \bibinfo {author} {\bibfnamefont {John~F.}\ \bibnamefont
  {Beacom}},\ }\bibfield  {title} {\enquote {\bibinfo {title} {{Precise Relic
  WIMP Abundance and its Impact on Searches for Dark Matter Annihilation}},}\
  }\href {\doibase 10.1103/PhysRevD.86.023506} {\bibfield  {journal} {\bibinfo
  {journal} {Phys. Rev.}\ }\textbf {\bibinfo {volume} {D86}},\ \bibinfo {pages}
  {023506} (\bibinfo {year} {2012})},\ \Eprint {http://arxiv.org/abs/1204.3622}
  {arXiv:1204.3622 [hep-ph]} \BibitemShut {NoStop}%
%%CITATION = ARXIV:1204.3622;%%
\bibitem [{\citenamefont {Ackermann}\ \emph
  {et~al.}(2015{\natexlab{d}})\citenamefont {Ackermann} \emph
  {et~al.}}]{Ackermann:2014usa}%
  \BibitemOpen
  \bibfield  {author} {\bibinfo {author} {\bibfnamefont {M.}~\bibnamefont
  {Ackermann}} \emph {et~al.} (\bibinfo {collaboration} {Fermi-LAT}),\
  }\bibfield  {title} {\enquote {\bibinfo {title} {{The spectrum of isotropic
  diffuse gamma-ray emission between 100 MeV and 820 GeV}},}\ }\href {\doibase
  10.1088/0004-637X/799/1/86} {\bibfield  {journal} {\bibinfo  {journal}
  {Astrophys. J.}\ }\textbf {\bibinfo {volume} {799}},\ \bibinfo {pages} {86}
  (\bibinfo {year} {2015}{\natexlab{d}})},\ \Eprint
  {http://arxiv.org/abs/1410.3696} {arXiv:1410.3696 [astro-ph.HE]} \BibitemShut
  {NoStop}%
%%CITATION = ARXIV:1410.3696;%%
\bibitem [{\citenamefont {Reiprich}\ and\ \citenamefont
  {Boehringer}(2002)}]{Reiprich:2001zv}%
  \BibitemOpen
  \bibfield  {author} {\bibinfo {author} {\bibfnamefont {Thomas~H.}\
  \bibnamefont {Reiprich}}\ and\ \bibinfo {author} {\bibfnamefont {Hans}\
  \bibnamefont {Boehringer}},\ }\bibfield  {title} {\enquote {\bibinfo {title}
  {{The Mass function of an X-ray flux-limited sample of galaxy clusters}},}\
  }\href {\doibase 10.1086/338753} {\bibfield  {journal} {\bibinfo  {journal}
  {Astrophys. J.}\ }\textbf {\bibinfo {volume} {567}},\ \bibinfo {pages}
  {716--740} (\bibinfo {year} {2002})},\ \Eprint
  {http://arxiv.org/abs/astro-ph/0111285} {arXiv:astro-ph/0111285 [astro-ph]}
  \BibitemShut {NoStop}%
%%CITATION = ASTRO-PH/0111285;%%
\bibitem [{\citenamefont {Chen}\ \emph {et~al.}(2007)\citenamefont {Chen},
  \citenamefont {Reiprich}, \citenamefont {Bohringer}, \citenamefont {Ikebe},\
  and\ \citenamefont {Zhang}}]{Chen:2007sz}%
  \BibitemOpen
  \bibfield  {author} {\bibinfo {author} {\bibfnamefont {Yong}\ \bibnamefont
  {Chen}}, \bibinfo {author} {\bibfnamefont {T.~H.}\ \bibnamefont {Reiprich}},
  \bibinfo {author} {\bibfnamefont {H.}~\bibnamefont {Bohringer}}, \bibinfo
  {author} {\bibfnamefont {Y.}~\bibnamefont {Ikebe}}, \ and\ \bibinfo {author}
  {\bibfnamefont {Y.~Y.}\ \bibnamefont {Zhang}},\ }\bibfield  {title} {\enquote
  {\bibinfo {title} {{Statistics of X-ray observables for the cooling-core and
  non-cooling core galaxy clusters}},}\ }\href {\doibase
  10.1051/0004-6361:20066471} {\bibfield  {journal} {\bibinfo  {journal}
  {Astron. Astrophys.}\ } (\bibinfo {year} {2007}),\
  10.1051/0004-6361:20066471},\ \bibinfo {note} {[Astron.
  Astrophys.466,805(2007)]},\ \Eprint {http://arxiv.org/abs/astro-ph/0702482}
  {arXiv:astro-ph/0702482 [astro-ph]} \BibitemShut {NoStop}%
%%CITATION = ASTRO-PH/0702482;%%
\bibitem [{\citenamefont {Ade}\ \emph {et~al.}(2016)\citenamefont {Ade} \emph
  {et~al.}}]{Ade:2015xua}%
  \BibitemOpen
  \bibfield  {author} {\bibinfo {author} {\bibfnamefont {P.~A.~R.}\
  \bibnamefont {Ade}} \emph {et~al.} (\bibinfo {collaboration} {Planck}),\
  }\bibfield  {title} {\enquote {\bibinfo {title} {{Planck 2015 results. XIII.
  Cosmological parameters}},}\ }\href {\doibase 10.1051/0004-6361/201525830}
  {\bibfield  {journal} {\bibinfo  {journal} {Astron. Astrophys.}\ }\textbf
  {\bibinfo {volume} {594}},\ \bibinfo {pages} {A13} (\bibinfo {year}
  {2016})},\ \Eprint {http://arxiv.org/abs/1502.01589} {arXiv:1502.01589
  [astro-ph.CO]} \BibitemShut {NoStop}%
%%CITATION = ARXIV:1502.01589;%%
\bibitem [{\citenamefont {{Astropy Collaboration}}\ \emph
  {et~al.}(2013)\citenamefont {{Astropy Collaboration}}, \citenamefont
  {{Robitaille}}, \citenamefont {{Tollerud}}, \citenamefont {{Greenfield}},
  \citenamefont {{Droettboom}}, \citenamefont {{Bray}}, \citenamefont
  {{Aldcroft}}, \citenamefont {{Davis}}, \citenamefont {{Ginsburg}},
  \citenamefont {{Price-Whelan}}, \citenamefont {{Kerzendorf}}, \citenamefont
  {{Conley}}, \citenamefont {{Crighton}}, \citenamefont {{Barbary}},
  \citenamefont {{Muna}}, \citenamefont {{Ferguson}}, \citenamefont
  {{Grollier}}, \citenamefont {{Parikh}}, \citenamefont {{Nair}}, \citenamefont
  {{Unther}}, \citenamefont {{Deil}}, \citenamefont {{Woillez}}, \citenamefont
  {{Conseil}}, \citenamefont {{Kramer}}, \citenamefont {{Turner}},
  \citenamefont {{Singer}}, \citenamefont {{Fox}}, \citenamefont {{Weaver}},
  \citenamefont {{Zabalza}}, \citenamefont {{Edwards}}, \citenamefont {{Azalee
  Bostroem}}, \citenamefont {{Burke}}, \citenamefont {{Casey}}, \citenamefont
  {{Crawford}}, \citenamefont {{Dencheva}}, \citenamefont {{Ely}},
  \citenamefont {{Jenness}}, \citenamefont {{Labrie}}, \citenamefont {{Lim}},
  \citenamefont {{Pierfederici}}, \citenamefont {{Pontzen}}, \citenamefont
  {{Ptak}}, \citenamefont {{Refsdal}}, \citenamefont {{Servillat}},\ and\
  \citenamefont {{Streicher}}}]{2013A&A...558A..33A}%
  \BibitemOpen
  \bibfield  {author} {\bibinfo {author} {\bibnamefont {{Astropy
  Collaboration}}}, \bibinfo {author} {\bibfnamefont {T.~P.}\ \bibnamefont
  {{Robitaille}}}, \bibinfo {author} {\bibfnamefont {E.~J.}\ \bibnamefont
  {{Tollerud}}}, \bibinfo {author} {\bibfnamefont {P.}~\bibnamefont
  {{Greenfield}}}, \bibinfo {author} {\bibfnamefont {M.}~\bibnamefont
  {{Droettboom}}}, \bibinfo {author} {\bibfnamefont {E.}~\bibnamefont
  {{Bray}}}, \bibinfo {author} {\bibfnamefont {T.}~\bibnamefont {{Aldcroft}}},
  \bibinfo {author} {\bibfnamefont {M.}~\bibnamefont {{Davis}}}, \bibinfo
  {author} {\bibfnamefont {A.}~\bibnamefont {{Ginsburg}}}, \bibinfo {author}
  {\bibfnamefont {A.~M.}\ \bibnamefont {{Price-Whelan}}}, \bibinfo {author}
  {\bibfnamefont {W.~E.}\ \bibnamefont {{Kerzendorf}}}, \bibinfo {author}
  {\bibfnamefont {A.}~\bibnamefont {{Conley}}}, \bibinfo {author}
  {\bibfnamefont {N.}~\bibnamefont {{Crighton}}}, \bibinfo {author}
  {\bibfnamefont {K.}~\bibnamefont {{Barbary}}}, \bibinfo {author}
  {\bibfnamefont {D.}~\bibnamefont {{Muna}}}, \bibinfo {author} {\bibfnamefont
  {H.}~\bibnamefont {{Ferguson}}}, \bibinfo {author} {\bibfnamefont
  {F.}~\bibnamefont {{Grollier}}}, \bibinfo {author} {\bibfnamefont {M.~M.}\
  \bibnamefont {{Parikh}}}, \bibinfo {author} {\bibfnamefont {P.~H.}\
  \bibnamefont {{Nair}}}, \bibinfo {author} {\bibfnamefont {H.~M.}\
  \bibnamefont {{Unther}}}, \bibinfo {author} {\bibfnamefont {C.}~\bibnamefont
  {{Deil}}}, \bibinfo {author} {\bibfnamefont {J.}~\bibnamefont {{Woillez}}},
  \bibinfo {author} {\bibfnamefont {S.}~\bibnamefont {{Conseil}}}, \bibinfo
  {author} {\bibfnamefont {R.}~\bibnamefont {{Kramer}}}, \bibinfo {author}
  {\bibfnamefont {J.~E.~H.}\ \bibnamefont {{Turner}}}, \bibinfo {author}
  {\bibfnamefont {L.}~\bibnamefont {{Singer}}}, \bibinfo {author}
  {\bibfnamefont {R.}~\bibnamefont {{Fox}}}, \bibinfo {author} {\bibfnamefont
  {B.~A.}\ \bibnamefont {{Weaver}}}, \bibinfo {author} {\bibfnamefont
  {V.}~\bibnamefont {{Zabalza}}}, \bibinfo {author} {\bibfnamefont {Z.~I.}\
  \bibnamefont {{Edwards}}}, \bibinfo {author} {\bibfnamefont {K.}~\bibnamefont
  {{Azalee Bostroem}}}, \bibinfo {author} {\bibfnamefont {D.~J.}\ \bibnamefont
  {{Burke}}}, \bibinfo {author} {\bibfnamefont {A.~R.}\ \bibnamefont
  {{Casey}}}, \bibinfo {author} {\bibfnamefont {S.~M.}\ \bibnamefont
  {{Crawford}}}, \bibinfo {author} {\bibfnamefont {N.}~\bibnamefont
  {{Dencheva}}}, \bibinfo {author} {\bibfnamefont {J.}~\bibnamefont {{Ely}}},
  \bibinfo {author} {\bibfnamefont {T.}~\bibnamefont {{Jenness}}}, \bibinfo
  {author} {\bibfnamefont {K.}~\bibnamefont {{Labrie}}}, \bibinfo {author}
  {\bibfnamefont {P.~L.}\ \bibnamefont {{Lim}}}, \bibinfo {author}
  {\bibfnamefont {F.}~\bibnamefont {{Pierfederici}}}, \bibinfo {author}
  {\bibfnamefont {A.}~\bibnamefont {{Pontzen}}}, \bibinfo {author}
  {\bibfnamefont {A.}~\bibnamefont {{Ptak}}}, \bibinfo {author} {\bibfnamefont
  {B.}~\bibnamefont {{Refsdal}}}, \bibinfo {author} {\bibfnamefont
  {M.}~\bibnamefont {{Servillat}}}, \ and\ \bibinfo {author} {\bibfnamefont
  {O.}~\bibnamefont {{Streicher}}},\ }\bibfield  {title} {\enquote {\bibinfo
  {title} {{Astropy: A community Python package for astronomy}},}\ }\href
  {\doibase 10.1051/0004-6361/201322068} {\bibfield  {journal} {\bibinfo
  {journal} {AAP}\ }\textbf {\bibinfo {volume} {558}},\ \bibinfo {eid} {A33}
  (\bibinfo {year} {2013})},\ \Eprint {http://arxiv.org/abs/1307.6212}
  {arXiv:1307.6212 [astro-ph.IM]} \BibitemShut {NoStop}%
\bibitem [{\citenamefont {P\'erez}\ and\ \citenamefont
  {Granger}(2007)}]{PER-GRA:2007}%
  \BibitemOpen
  \bibfield  {author} {\bibinfo {author} {\bibfnamefont {Fernando}\
  \bibnamefont {P\'erez}}\ and\ \bibinfo {author} {\bibfnamefont {Brian~E.}\
  \bibnamefont {Granger}},\ }\bibfield  {title} {\enquote {\bibinfo {title}
  {{IP}ython: a system for interactive scientific computing},}\ }\href
  {\doibase 10.1109/MCSE.2007.53} {\bibfield  {journal} {\bibinfo  {journal}
  {Computing in Science and Engineering}\ }\textbf {\bibinfo {volume} {9}},\
  \bibinfo {pages} {21--29} (\bibinfo {year} {2007})}\BibitemShut {NoStop}%
\bibitem [{\citenamefont {Abdo}\ \emph {et~al.}(2010)\citenamefont {Abdo} \emph
  {et~al.}}]{Abdo:2010ex}%
  \BibitemOpen
  \bibfield  {author} {\bibinfo {author} {\bibfnamefont {A.~A.}\ \bibnamefont
  {Abdo}} \emph {et~al.} (\bibinfo {collaboration} {Fermi-LAT}),\ }\bibfield
  {title} {\enquote {\bibinfo {title} {{Observations of Milky Way Dwarf
  Spheroidal galaxies with the Fermi-LAT detector and constraints on Dark
  Matter models}},}\ }\href {\doibase 10.1088/0004-637X/712/1/147} {\bibfield
  {journal} {\bibinfo  {journal} {Astrophys. J.}\ }\textbf {\bibinfo {volume}
  {712}},\ \bibinfo {pages} {147--158} (\bibinfo {year} {2010})},\ \Eprint
  {http://arxiv.org/abs/1001.4531} {arXiv:1001.4531 [astro-ph.CO]} \BibitemShut
  {NoStop}%
%%CITATION = ARXIV:1001.4531;%%
\bibitem [{\citenamefont {Charbonnier}\ \emph {et~al.}(2011)\citenamefont
  {Charbonnier} \emph {et~al.}}]{Charbonnier:2011ft}%
  \BibitemOpen
  \bibfield  {author} {\bibinfo {author} {\bibfnamefont {A.}~\bibnamefont
  {Charbonnier}} \emph {et~al.},\ }\bibfield  {title} {\enquote {\bibinfo
  {title} {{Dark matter profiles and annihilation in dwarf spheroidal galaxies:
  prospectives for present and future gamma-ray observatories - I. The
  classical dSphs}},}\ }\href {\doibase 10.1111/j.1365-2966.2011.19387.x}
  {\bibfield  {journal} {\bibinfo  {journal} {Mon. Not. Roy. Astron. Soc.}\
  }\textbf {\bibinfo {volume} {418}},\ \bibinfo {pages} {1526--1556} (\bibinfo
  {year} {2011})},\ \Eprint {http://arxiv.org/abs/1104.0412} {arXiv:1104.0412
  [astro-ph.HE]} \BibitemShut {NoStop}%
%%CITATION = ARXIV:1104.0412;%%
\bibitem [{\citenamefont {Charbonnier}\ \emph {et~al.}(2012)\citenamefont
  {Charbonnier}, \citenamefont {Combet},\ and\ \citenamefont
  {Maurin}}]{Charbonnier:2012gf}%
  \BibitemOpen
  \bibfield  {author} {\bibinfo {author} {\bibfnamefont {A.}~\bibnamefont
  {Charbonnier}}, \bibinfo {author} {\bibfnamefont {C.}~\bibnamefont {Combet}},
  \ and\ \bibinfo {author} {\bibfnamefont {D.}~\bibnamefont {Maurin}},\
  }\bibfield  {title} {\enquote {\bibinfo {title} {{CLUMPY: a code for
  gamma-ray signals from dark matter structures}},}\ }\href {\doibase
  10.1016/j.cpc.2011.10.017} {\bibfield  {journal} {\bibinfo  {journal}
  {Comput. Phys. Commun.}\ }\textbf {\bibinfo {volume} {183}},\ \bibinfo
  {pages} {656--668} (\bibinfo {year} {2012})},\ \Eprint
  {http://arxiv.org/abs/1201.4728} {arXiv:1201.4728 [astro-ph.HE]} \BibitemShut
  {NoStop}%
%%CITATION = ARXIV:1201.4728;%%
\bibitem [{\citenamefont {Evans}\ \emph {et~al.}(2016)\citenamefont {Evans},
  \citenamefont {Sanders},\ and\ \citenamefont
  {Geringer-Sameth}}]{Evans:2016xwx}%
  \BibitemOpen
  \bibfield  {author} {\bibinfo {author} {\bibfnamefont {N.~W.}\ \bibnamefont
  {Evans}}, \bibinfo {author} {\bibfnamefont {J.~L.}\ \bibnamefont {Sanders}},
  \ and\ \bibinfo {author} {\bibfnamefont {A.}~\bibnamefont
  {Geringer-Sameth}},\ }\bibfield  {title} {\enquote {\bibinfo {title} {{Simple
  J-Factors and D-Factors for Indirect Dark Matter Detection}},}\ }\href
  {\doibase 10.1103/PhysRevD.93.103512} {\bibfield  {journal} {\bibinfo
  {journal} {Phys. Rev.}\ }\textbf {\bibinfo {volume} {D93}},\ \bibinfo {pages}
  {103512} (\bibinfo {year} {2016})},\ \Eprint
  {http://arxiv.org/abs/1604.05599} {arXiv:1604.05599 [astro-ph.GA]}
  \BibitemShut {NoStop}%
%%CITATION = ARXIV:1604.05599;%%
\bibitem [{\citenamefont {Burkert}(1996)}]{Burkert:1995yz}%
  \BibitemOpen
  \bibfield  {author} {\bibinfo {author} {\bibfnamefont {A.}~\bibnamefont
  {Burkert}},\ }\bibfield  {title} {\enquote {\bibinfo {title} {{The Structure
  of dark matter halos in dwarf galaxies}},}\ }\bibfield  {booktitle} {\emph
  {\bibinfo {booktitle} {{IAU Symposium 171: New Light on Galaxy Evolution
  Heidelberg, Germany, June 26-30, 1995}}},\ }\href {\doibase 10.1086/309560}
  {\bibfield  {journal} {\bibinfo  {journal} {IAU Symp.}\ }\textbf {\bibinfo
  {volume} {171}},\ \bibinfo {pages} {175} (\bibinfo {year} {1996})},\ \bibinfo
  {note} {[Astrophys. J.447,L25(1995)]},\ \Eprint
  {http://arxiv.org/abs/astro-ph/9504041} {arXiv:astro-ph/9504041 [astro-ph]}
  \BibitemShut {NoStop}%
%%CITATION = ASTRO-PH/9504041;%%
\bibitem [{\citenamefont {Ackermann}\ \emph {et~al.}(2011)\citenamefont
  {Ackermann} \emph {et~al.}}]{Ackermann:2011wa}%
  \BibitemOpen
  \bibfield  {author} {\bibinfo {author} {\bibfnamefont {M.}~\bibnamefont
  {Ackermann}} \emph {et~al.} (\bibinfo {collaboration} {Fermi-LAT}),\
  }\bibfield  {title} {\enquote {\bibinfo {title} {{Constraining Dark Matter
  Models from a Combined Analysis of Milky Way Satellites with the Fermi Large
  Area Telescope}},}\ }\href {\doibase 10.1103/PhysRevLett.107.241302}
  {\bibfield  {journal} {\bibinfo  {journal} {Phys. Rev. Lett.}\ }\textbf
  {\bibinfo {volume} {107}},\ \bibinfo {pages} {241302} (\bibinfo {year}
  {2011})},\ \Eprint {http://arxiv.org/abs/1108.3546} {arXiv:1108.3546
  [astro-ph.HE]} \BibitemShut {NoStop}%
%%CITATION = ARXIV:1108.3546;%%
\bibitem [{\citenamefont {Ackermann}\ \emph
  {et~al.}(2014{\natexlab{b}})\citenamefont {Ackermann} \emph
  {et~al.}}]{Ackermann:2013yva}%
  \BibitemOpen
  \bibfield  {author} {\bibinfo {author} {\bibfnamefont {M.}~\bibnamefont
  {Ackermann}} \emph {et~al.} (\bibinfo {collaboration} {Fermi-LAT}),\
  }\bibfield  {title} {\enquote {\bibinfo {title} {{Dark matter constraints
  from observations of 25 Milky Way satellite galaxies with the Fermi Large
  Area Telescope}},}\ }\href {\doibase 10.1103/PhysRevD.89.042001} {\bibfield
  {journal} {\bibinfo  {journal} {Phys. Rev.}\ }\textbf {\bibinfo {volume}
  {D89}},\ \bibinfo {pages} {042001} (\bibinfo {year} {2014}{\natexlab{b}})},\
  \Eprint {http://arxiv.org/abs/1310.0828} {arXiv:1310.0828 [astro-ph.HE]}
  \BibitemShut {NoStop}%
%%CITATION = ARXIV:1310.0828;%%
\end{thebibliography}%

\end{document}